\documentclass[a4paper,11pt]{article}

\usepackage{jheppub} 

\usepackage{lmodern}
\usepackage{amsmath,amssymb}
\usepackage{hyperref}
\usepackage{xcolor}
\usepackage{graphicx}
\usepackage{caption}
\usepackage{subcaption}
\usepackage{float}
\usepackage{subfloat}

\usepackage{tikz}
\usetikzlibrary{positioning,decorations.pathmorphing}

\let\oldFootnote\footnote
\newcommand\nextToken\relax

\renewcommand\footnote[1]{%
	\oldFootnote{#1}\futurelet\nextToken\isFootnote}

\newcommand\isFootnote{%
	\ifx\footnote\nextToken\textsuperscript{,}\fi}

\usepackage{enumerate}

\usetikzlibrary{decorations.markings}

\def\id{{1 \kern-.28em {\rm l}}}

\def\K3{{\bf K3}}
\def\journal#1&#2(#3){\unskip, \sl #1\ \bf #2 \rm(19#3) }
\def\andjournal#1&#2(#3){\sl #1~\bf #2 \rm (19#3) }

\def\bar{\overline}

\def\ie{{\it i.e.}}
\def\eg{{\it e.g.}}

\def\etc{{\it etc}}

\def\tilde{\widetilde}

\def\frac#1#2{{#1\over#2}}

\def\inbar{\,\vrule height1.5ex width.4pt depth0pt}
\def\IC{\relax\hbox{$\inbar\kern-.3em{\rm C}$}}
\def\IR{\relax{\rm I\kern-.18em R}}
\def\IP{\relax{\rm I\kern-.18em P}}

\catcode`\@=11
\def\slash#1{\mathord{\mathpalette\c@ncel{#1}}}
\def\underrel#1\over#2{\mathrel{\mathop{\kern\z@#1}\limits_{#2}}}

\catcode`\@=12

\def \sinh{{\rm sinh}}

\def\exp{{\rm exp}}

\def\ie{{\it i.e.}}
\def\eg{{\it e.g.}}

\title{ Holographic Complexity of LST and Single Trace $T\bar{T},\,J\bar{T}$ and $ T\bar{J}$ deformations}

\author{Gaurav Katoch, Swejyoti Mitra and Shubho R. Roy}
\emailAdd{gauravitation@gmail.com}
\emailAdd{swejyoti@gmail.com}
\emailAdd{roy.shubho@gmail.com}

\affiliation{Department of Physics,\\Indian Institute of Technology, Hyderabad\\
	Kandi, Sangareddy 502285, Medak, Telengana, India}

\abstract{This work is an extension of our previous work \cite{Chakraborty:2020fpt} where we exploited holography to compute the complexity characteristics of Little String Theory (LST), a nonlocal, nongravitational field theory which flows to a local 2d CFT in the IR under RG via an integrable irrelevant ($T\overline{T}$) deformation. 
Here we look at the more general LST obtained by UV deforming the 2d CFT by incorporating Lorentz violating irrelevant $J\overline{T}$ and $T\overline{J}$ deformations on top of $T\overline{T}$ deformation, in an effort to capture the novel signatures of Lorentz violation (on top of nonlocality) on quantum complexity.  In anticipation of the fact that the dual field theory is Lorentz violating, we compute the volume complexity in two different Lorentz frames and the comparison is drawn between the results. It turns out that for this system the nonlocality and Lorentz violation effects are inextricably intertwined in the UV divergence structure of the quantum complexity. The coefficients of the divergences carry the signature of Lorentz boost violation. We also compute the subregion complexity which displays a (Hagedorn) phase transition with the transition point being the same as that for the phase transition of entanglement entropy \cite{Chakraborty:2020udr}. These new results are consistent with our previous work \cite {Chakraborty:2020fpt}. Null warped AdS$_{3}$ is treated as special case of interest.}

\begin{document}
	\maketitle
	\flushbottom
	
\section{Introduction $\&$ Summary}
	
Our understanding of strongly coupled gauge theories has been revolutionized by the AdS/CFT \cite{Maldacena:1997re, Gubser:1998bc, Witten:1998qj, Aharony:1999ti} duality (or more generally gauge/gravity duality \cite{Itzhaki:1998dd}). Strongly coupled regimes of field theories, once considered beyond the reach of analytical control due to breakdown of coupling constant perturbation theory, are now routinely being investigated by going over to the dual, weakly coupled physical system in weakly curved spacetime (most of the time constructed from the ``\emph{bottom up}" without the need of any details of string theory/M-theory compactifications). Effectively one solves (in most cases numerically) a much easier classical gravity-matter system, \ie\ Einstein field equations coupled to classical matter. This ``\emph{holographic approach}" of solving strongly coupled fields theories (with or without gauge fields), has proliferated the use of GR/SUGRA tools in the fields of condensed matter many-body physics \cite{Sachdev:2010ch, McGreevy:2009xe, Hartnoll:2009sz} and QCD \cite{Erlich:2005qh, DaRold:2005mxj, Karch:2006pv}. In fact, the impact of gauge/gravity duality has been far more profound than simply providing a geometric computational toolkit for strongly coupled regimes of field theory. Thinking about how a dual field theory encodes various phenomena on the gravity side, such the emergent holographic (radial) direction, spatial connectivity in/of the bulk, presence of event horizons in the bulk, formation of gravitational singularities in the bulk \etc, has led to the realization of the significance of various concepts from the quantum information and computation (QIC) canon which are able to capture aspects of field theory not captured by traditional observables such as correlators of local operators, or even Wilson loop operators. To name a few such concepts: Information geometry and information metrics, Shanon or Von-Neumann entropy \cite{Ryu:2006bv, Hubeny:2007xt} and Renyi \cite{Dong:2016fnf} Entropy, Mutual Information, Tensor networks \cite{Swingle:2009bg}, Computational Complexity, Fidelity susceptibility, Quantum error correcting codes. Influx of these ideas from QIC has turned out to be a ground-breaking enterprise leading to novel insights which might even have resolved the information paradox \cite{Penington:2019npb, Almheiri:2019hni}. Combining insights from complementary approaches such as holography, integrability or supersymmetry based arguments, lattice based approaches and perturbative approaches, we have explored the landscape of local quantum field theories rather comprehensively. However, the landscape of nonlocal quantum field theories is still mostly unexplored. Nonlocal field theories arise in various contexts in high energy physics both as effective or emergent theories (e.g. \cite{Namsrai:1986md}) as well fundamental (UV complete) theories (e.g.\cite{Sen:2015uaa}), they can be finite \cite{Efimov:1969fd} (or super-renormalizable) and unitary. We are optimistic that holography will be as productive in demystifying many aspects/properties of nonlocal quantum field theories such as the LST as it has been for enhancing our understanding of strongly coupled regimes of local field theories.  Another fact is that holography beyond the traditional asymptotically AdS setting is also little explored. Our hope is that studying set ups such as the LST will help us get an handle on nonperturbative quantum gravity beyond pure AdS asymptotics to flat asymptotics.\\

Our present understanding of holography is that the bulk spacetime geometry is a representation or form of encoding of the entanglement structure of the dual field theory state \cite{VanRaamsdonk:2009ar, VanRaamsdonk:2010pw}. The well known Ryu-Takayanagi (RT) proposal \cite{Ryu:2006bv, Hubeny:2007xt} was one of the earliest major piece of evidence to point in this direction (along with Maldacena's construction \cite{Maldacena:2001kr} of the eternal Schwarzschild-AdS (SAdS) as a thermally entangled state of two CFTs). Since then an impressive list of quantum entanglement related CFT observables have been related to classical geometric features of the bulk (see \eg\  \cite{VanRaamsdonk:2016exw} for a review). However, entanglement entropy or other entanglement related concepts such as tensor networks or error-correcting codes are yet to capture the essential features of bulk geometry which lay hidden behind the black hole horizons. Take for instance the case of the Einstein-Rosen Bridge (ERB) behind the black hole horizons. Entanglement entropy saturates in a short time upon reaching thermalization whereas, ER bridge continues to grow linearly with time even after the dual field theory attains thermalization. To explain the ERB growth, Susskind  \cite{Susskind:2014rva} has imported another concept from quantum information theory and added it to the holographic dictionary, namely the \emph{computational complexity} of the dual CFT state. Complexity is the property associated with the states in the Hilbert space of states of a quantum mechanical system quantifying the difficulty of preparing a state (called the target state), starting from the given reference state. While this is a well defined quantity for discrete systems, like quantum circuits in information theory, it has turned out to be enormously hard to define complexity for the continuous systems described by a QFT. It is fair to say that a precise and unambiguous definition of complexity is still lacking for field theories. In the approach of Nielsen et. al.\cite{2005quant.ph..2070N, 2006Sci...311.1133N} a definition of circuit complexity in field theory has been proposed as the minimum number of unitary gates in the space of unitary operators which has a Finsler geometry. The complexity of a target state, with respect to a reference state, is defined to be the geodesic length in a Finsler manifold with suitable cost functions and penalty factors, which acts like Lagrangian in typical variational problem. These cost functions are further required to obey certain conditions such as continuity, positive definiteness and satisfying the triangle inequality \etc. Despite this attempt at achieving precision, there is still arbitrariness in the choice of cost functions which fixes the Finsler metric and complexity depends upon the choice of the metric. Several attempts have been made to define complexity in the continuum limit (see \eg\ \cite{Jefferson:2017sdb, Chapman:2017rqy, Khan:2018rzm, Yang:2018nda, Molina-Vilaplana:2018sfn, Hackl:2018ptj, Bhattacharyya:2018wym, Guo:2018kzl, Bhattacharyya:2018bbv, Yang:2018tpo, Camargo:2019isp, Balasubramanian:2019wgd, Bhattacharyya:2019kvj, Erdmenger:2020sup, Bueno:2019ajd, Chen:2020nlj, Flory:2020eot, Flory:2020dja} for an incomplete but representative list). However, it is fair to say that as yet there exists neither any universal and unanimous definition of complexity in the continuum limit nor does there exist a complete study of the possible universality classes. In the continuum limit, complexity, even  in principle, is a UV divergent quantity because it is defined to within a tolerance ($\epsilon$) with respect to the target state. Demanding more precision of reproducing the target state requires including more gates which leads to a dependence on the inverse tolerance which is a divergent term. Conventionally UV divergent or quantities explicitly dependent on the cutoff in QFT are considered unphysical as their value can be altered by simply altering the UV cutoff. But the characteristic UV cutoff dependence is a feature which seems to be indispensable while defining complexity in QFT. 
	
	There are two proposals in holography, each with its own distinct motivation, as to which bulk geometric feature represents the complexity of the dual boundary field theory quantum state. First one prescribes the field theory complexity to be proportional to the volume of the maximal volume spacelike hypersurface in the bulk which terminates at the exact boundary spatial slice on which the boundary quantum state is specified \cite{Susskind:2014rva}. This is the complexity-volume ($CV$) proposal. The second proposal \cite{Brown:2015bva, Brown:2015lvg} prescribes the complexity to be proportional to the on-shell bulk SUGRA action integral supported over the Wheeler-deWitt (WdW) patch of the boundary spatial slice on which the field theory state is specified\footnote{The WdW patch of a given spatial slice on the boundary is defined to be the bulk subregion covered by the union of all possible spacelike surfaces in the bulk which terminates on the same spatial slice at the boundary.}. This is the complexity-action ($CA$) proposal. Since the bulk is noncompact, both these bulk geometric duals of complexity are manifestly UV divergent, hence regularization is necessary as the lore goes in holography. In the CV proposal there is an ambiguity - in order to make the expression dimensionless one must include a \emph{length scale}, $L$, characteristic of the geometry for which there is no unique candidate. For the CA proposal, there are also couple of issues. Some boundaries of the WdW patch are codimension one null submanifolds \emph{with} edges or joints. The presence of such null boundaries and their joints (edges) requires the inclusion of carefully defined GHY boundary terms as discussed in \cite{Lehner:2016vdi}. In this paper, we take an alternative approach to this issue \cite{Brown:2015lvg, Parattu:2015gga, Bolognesi:2018ion}. Since we have to regulate the WdW patch in any case, we use a special regularization which deforms the WdW null boundary to timelike and in the process also smooths out the joints. At the end the regulator is removed. In this way we can compute the GHY terms and obtain a UV-regulated result in one go.
	
	In a recent work \cite{Chakraborty:2020fpt} we focused our attention on the decoupled regime of the theory of a stack of large number ($k\gg1$) of NS$5$ branes wrapping $T^4 \times S^1$, the so called Little String theory (LST) in $1+1$ dimensions. This system is unlike the theory of a stack of D$p$ branes, since the worldvolume theory living on the NS$5$ branes decouples from the bulk  at \emph{finite} value of the string length $l_s=\sqrt{\alpha'}$. This implies that this decoupled theory, namely LST living on the NS$5$ branes, still retains stringy nonlocality and is \emph{not} a local quantum field theory. In fact this decoupled theory living on the NS$5$ branes is to some extent intermediate between string theory (which is nonlocal theory containing massless gravitons) and a local QFT. The dual holographic background is then obtained by taking the near horizon geometry of the NS$5$ branes - it is a metrically flat spacetime with a linear dilaton $\mathbb{R}^{1,1}\times \mathbb{R}_\phi$ turned on all the way to spatial infinity. Such a holographic duality has been studied quite extensively in the past \cite{Aharony:1998ub,Kutasov:2001uf}. Now if one introduces $p\gg 1$ F$1$ strings wrapping a $S^1$ along the NS$5$ directions, the near horizon geometry of the F$1$ strings is given by AdS$_3$. Thus the full geometry interpolates between AdS$_3$ in the IR (which corresponds to the near horizon geometry of the F$1$ strings) to flat spacetime with a linear dilaton in the UV (which corresponds to the near horizon geometry of just the NS$5$ branes). Correspondingly, the boundary field theory interpolates between  a local CFT$_2$ dual to AdS$_3$ in the IR to LST in the UV. The interpolating geometry discussed above is often referred to in the literature as $\mathcal{M}_3$. In the wake of the recent developments in the subject of $T\bar{T}$ deformation \cite{Smirnov:2016lqw,Cavaglia:2016oda}, it was proposed in \cite{Giveon:2017nie} that there exists an analogous deformation of string theory in $AdS_3$ that shares many properties in common with the  double trace $T\bar{T}$ deformation.
This is often referred to as the single trace $T\bar{T}$ deformation in the literature which changes the UV asymptotics of the bulk geometry from $AdS_3$ to flat spacetime with a linear dilaton keeping fix the IR regime of the geometry. Analysis in \cite{Giveon:2017nie} shows that the dual background geometry interpolates between $AdS_3$ in the IR to flat spacetime with a linear dilaton in the UV.  Holography in this background (often referred to as $\mathcal{M}_3$) can be realized as a concrete example of holography in non-AdS background that is smoothly connected to $AdS_3$. Naturally this non-AdS holography set up has attracted a lot of attention and there has been a lot of studies where holography has been exploited to investigate various aspects of nonlocal field theories such as LST which admit gravity duals, e.g. \cite{Asrat:2017tzd,Chakraborty:2018kpr,Chakraborty:2018aji,Chakraborty:2020xyz,Chakraborty:2020udr,Chakraborty:2020yka}. In our recent work \cite{Chakraborty:2020fpt} we probed this theory using holographic complexity as a probe. There we computed the volume and action complexity, both at zero and finite temperature. The complexity expressions contained imprints of the stringy nonlocality on the UV divergence structure. To be specific, we encountered quadratic and logarithmic divergences, evidently not to be associated with local field theory in $1$ space dimension (where we expect a linear divergence) when the UV cutoff is smaller than the (Hagedorn) length scale, $\beta_H = 2\pi l_s \sqrt{k\,\lambda}$, set by the $T\overline{T}$ coupling $\lambda$). When the UV cutoff is held larger than the Hagedorn scale, complexity displays a linear UV divergence, much akin to a local field theory in $1$ space dimension. For completeness we computed the holographic complexity at finite temperature as well, however no unanticipated newer type of UV divergences were encountered in perturbation theory around zero temperature.\\

The purpose of this paper is to extend the our work in \cite{Chakraborty:2020fpt} to a more general linear combination
of irrelevant single trace deformations, namely the single trace $T\overline{T}$,$ J\overline{T}$ and $T\overline{J}$  of a CFT$_2$ which contains/involves conserved left (right)-moving current $J (\overline{J})$. These irrelevant deformations drive the UV theory to nonlocality, in the sense that the UV is not a local fixed point as the high energy density of states exhibits a exponential Hagedorn growth \cite{Chakraborty:2020xyz}. Moreover, the effect of turning on the irrelevant current $J (\overline{J})$ couplings is to explicitly break Lorentz boost symmetry in the UV. The dual gravity (string) background was introduced in \cite{Chakraborty:2019mdf, Chakraborty:2020cgo} which interpolates between AdS$_3$ in the IR to a linear dilaton background in the UV. From the string viewpoint, the UV is the near horizon limit of the stack of $k$ NS$5$ branes with $p$ F$1$ strings propagating in the world volume while incorporating NS-NS $H$-flux along the world volume directions violating Lorentz boost invariance \cite{Chakraborty:2019mdf, Chakraborty:2018vja, Apolo:2018qpq}. Our main motivation to investigate this set up is to capture the imprint of Lorentz boost symmetry violation in the holographic complexity, to be specific in the UV divergence structure of holographic complexity. In particular, we are interested in finding out whether the imprints of Lorentz symmetry violation and nonlocality on the UV divergence are separate or different kind. Also since the theory does not respect boost symmetry, we would like to know how the UV divergences in complexity change as we move from one Lorentz frame to another.  Another motivation  of the present work is to study subsystem holographic complexity  \cite{Alishahiha:2015rta, Ben-Ami:2016qex, Carmi:2016wjl} which we had omitted in our previous work \cite{Chakraborty:2020fpt}. Subsystem complexity, just like entanglement entropy of a subsystem's reduced density matrix is expected to display phase transitions as the subsystem size is tuned. In particular, in the work \cite{Chakraborty:2020udr}, which looked at entanglement entropy of this system, namely the $T\overline{T}$,$ J\overline{T}$ and $T\overline{J}$ deformed CFT$_2$, entanglement entropy undergoes a (Hagedorn) phase transition when the subsystem size is tuned to a critical spatial size determined by the strength of the irrelevant couplings.
\\
	
	The plan of the paper is as follows. In section \ref{sec2}, we give a briefly recap aspects of string theory in $AdS_3$, its single trace $T\overline{T}, J\overline{T}, T\overline{J}$ deformations and highlight interesting features of LST for the sake of completeness. We also review some features of the dual holographic background (bulk). In this regard we would like to point out that one may work with either a $3+1$-dimensional bulk as was done in the works on entanglement entropy \cite{Chakraborty:2020udr}, or equivalently one can perform a KK reduction on the $y$ circle fiber and work with an effective bulk background in $2+1$ dimensions \cite{Chakraborty:2019mdf}. Here we take the second approach because it affords us performing immediate comparision or checks with our previous work \cite{Chakraborty:2020fpt} at every step.   In section \ref{0T complexity}, we set out to compute the holographic complexity of the $T\overline{T}, J\overline{T}, T\overline{J}$ deformed CFT$_2$ by implementing the $CV$ prescription\footnote{Actually we used a generalized prescription of the volume complexity put forth in our previous work \cite{Chakraborty:2020fpt} in the string frame since a non-trivial dilaton field turned on in the bulk, and this modification is necessary to get the correct powers of $G_N$. Similar considerations led the authors in \cite{Klebanov:2007ws} to a generalization of the Ryu-Takayanagi formula for holographic entanglement entropy for bulk backgrounds supporting a non-trivial dilation in the string frame. } in two distinct (boundary) Lorentz frames, which we dub as the \emph{stationary frame} and \emph{static frame} (for reasons which will become obvious),  related to each other by a Lorentz boost. In either frame, the volume complexity diverges quadratically with a subleading logarithmic divergence. However, anticipated, due to lack of boost symmetry, the coefficients of the quadratic and logarithmic divergence differ in the two frames (and even the finite piece differs). We find that the Lorentz violation effects (governed by the parameter $\epsilon_\pm$) and nonlocality effects (governed by the parameter $\lambda$) are inextricably linked - the UV divergence structure depends on a \emph{single} parameter, namely $\mu = \lambda -  \left( \epsilon_+ + \epsilon_{-}\right)^2$ in the stationary frame, and the parameter $\lambda'\equiv\lambda-4\epsilon_{+}\epsilon_{-}$ in the static frame.  There is no way to cleanly separate the effects of nonlocality and Lorentz boost asymmetry. This is perhaps mildly disappointing since our hope was to be able to see the effects of nonlocality and Lorentz violation in separate or independent UV divergence structures. These results are consistent with the results obtained in \cite{Chakraborty:2020fpt} - setting $\epsilon_\pm=0$ reproduces the volume complexity of the LST dual to the $\mathcal{M}_3$ geometry.  The quadratic and logarithmic divergences of the volume complexity immediately reveals the nonlocal nature of the dual field theory (LST) as for a local theory the complexity is expected to scale with volume $\mathcal{V}$ (here length) and hence should diverge as lattice cell volume inverse $\mathcal{V}/\epsilon^{d}$. In either frames, the nonlocality scale is set by the respective Hagedorn length $\rho_H \propto \sqrt{k \mu} \,l_s$ in the stationary frame and $\beta'_H\propto \sqrt{k\,\lambda'}\,l_s$ in the static frame. When the lattice spacing is larger than the Hagedorn length scales in the respective frame ($\epsilon\gg \rho_H$ or $\epsilon\gg \beta'_H$), the complexity expression reduces to that of a local field theory with a linear divergence (volume scaling).  However if the lattice spacing is shorter than the Hagedorn length scale $\epsilon \ll \rho_H,$ or $\epsilon\ll\lambda'_H$, stringy physics takes over and the theory departs from behaving like a local field theory. Finally we note that the logarithmic divergent pieces (subleading divergence) in the complexity expressions in either frame which are accompanied by a dimensionless universal constant coefficient. This coefficient can be given the interpretation of the total number of  ``regularized/effective" degrees of freedom in the spacetime theory in the nonlocal stringy regime as opposed to the true degrees of freedom of LST which naively diverges \cite{Barbon:2008ut,Chakraborty:2018kpr}. Next in Sec. \ref{subregionC}, we proceed to evaluate the subregion complexity, in both the stationary (Sec. \ref{subregionC in stationary}) and static (Sec. \ref{subregionC in static}) frames. The exact results for subregion complexity are obtained numerically, and the results are displayed graphically, subregion complexity plotted as function of the subregion size, $\mathcal{C}_V$ vs $L$ for several different choices of the set of parameters $\lambda, \epsilon_\pm$. In either frame, the plots clearly show the Hagedorn phase transition - at a critical subregion size, $L_c = \frac{\pi\sqrt{ k \lambda\lambda' } l_s  }{2\sqrt{\mu}}$ in the stationary frame and $L'_c = \frac{1}{2} \pi  \sqrt{k \lambda} l_s $ in the static frame. For subregion sizes larger than the critical size, the subregion complexity grows linearly with subregion size (length), characteristic of a CFT$_2$ while for subregion sizes lower than the critical subregion length, subregion complexity grows quadratically with subsystem size (length), which is more like a nonlocal LST. The reason we identify this transition as the Hagedorn transition because the critical length, read off from the numerics (plot), is identical to the phase transition point of entanglement entropy \cite{Chakraborty:2020udr}! The fact that the critical length is different in the two frames related by a Lorentz boost simply reflects the boost asymmetry of the LST. In Sec. \ref{null WAdS3 CV} we explore a very interesting special point in the parameter space of the couplings, namely when $\lambda=\epsilon_+=0$ (or $\lambda=\epsilon_-=0$) which is dual to the null warped AdS$_3$ geometry (with nonvanishing dilaton and NS-NS $B$ field). Although this might appear to be a slight digression, we explore this case since this falls under the same broader umbrella of sting theory in AdS$_3$. Since this limit is singular, instead of naively taking this limit in the final complexity expression of the general case, and redo some of the intermediate steps.  The complexity is only well defined (real) when the UV cut off is restricted $\epsilon\leq \sqrt{k} \epsilon_- l_s$, a trait which lends support to the claims in the literature that the null warped AdS$_3$ spacetime is the holographic dual to field theory which \emph{does not} possess a UV completion. For the null WAdS$_3$, the UV divergence structure is also special, one obtains UV divergences to all orders! In other words the complexity is not an analytic function of the UV cut off. This alludes to the fact that the boundary theory is highly nonlocal (and does not possess boost symmetry either). We also compute the subregion complexity numerically for a boundary interval of length, $L$ and present our results graphically via subregion $C_V$ vs $L$ plot in Fig. \ref{subregionCV for WAdS3} for a (allowable) range of the warping parameter $\epsilon_-$. The subregion complexity monotonically increases with the subregion size and approaches the subregion complexity of a CFT$_2$ (i.e. pure AdS$_3$ linear regime) as $L$ is progressively increased. However, unlike what we found for the case of general values of the couplings $\lambda, \epsilon_\pm$, there is no Hagedorn like phase transition. These results were obtained in the stationary frame, and there is no static frame for this case since the associated boost transformation which takes one from the stationary to static frame, becomes singular. Next, in Sec. \ref{CA static} we set out to compute the action complexity for the LST (i.e. the $T\overline{T}, J\overline{T}, T\overline{J}$ deformed CFT$_2$). Here we realize that the construction of null surfaces bounding the so called Wheeler-de Witt (WdW) patch is simplest in the boosted frame in the boundary since it leads to a static metric in the bulk. So we exclusively stick to this coordinate system for the entire section/calculation. We leave the construction of lightsheets associated with the WdW patch and the subsequent evaluation of the action-complexity for the stationary frame for future work. While computing the WdW we are confronted with a choice, either to the use the $3+1$-d bulk geometry or to work with the $2+1$-d bulk geometry by dimensionally reducing over the $y$-fiber. Although we present the calculation performed in the dimensionally reduced $2+1$-d set up, pleasantly the action complexities obtained using the the $3+1$-d and $2+1$-d bulk actions agree \emph{provided we retain the total derivative terms in the lower dimensional action one gets while performing a dimensional reduction}. Usually such total derivative terms are omitted from the dimensionally reduced action as they do not contribute to the equations of motion, but they do contribute to (action) complexity. The action complexity results display the exact same divergence structures, quadratic and logarithm when $\epsilon\ll \beta'_H$. Modulo an overall constant (courtesy the ambiguity in the choice of the ``characteristic length-scale of the geometry" in the definition of the volume complexity), the leading quadratic divergence piece matches for both the volume and action complexities. However we find that the subleading logarithmic divergence, while having the same magnitude in both prescriptions, \emph{differs by a sign} in the volume and action complexity expressions. This is not a total surprise. Past studies have revealed that the coefficients of the subleading divergent pieces might be different \cite{Bolognesi:2018ion}  hinting to the fact that the two bulk/holographic prescriptions of complexity might actually correspond to different schemes of defining complexity in the boundary field theory. These are also consistent with the results of our previous paper \cite{Chakraborty:2020fpt}. As a final check, we extract the behavior of the action complexity in the deep IR limit (\ie\ $\epsilon\gg\beta'_H$) where it indeed reproduces the pure AdS$_3$ or  CFT$_2$ vacuum state complexity \cite{Reynolds:2016rvl, Carmi:2016wjl} (for both prescriptions). In Sec. \ref{nullWAdS3 CA}, we revisit the null Warped AdS$_3$ background (with dilaton and B-field) located at point in the coupling space, $\lambda=\epsilon_+=0$ and compute the action complexity of dual WCFT$_2$ using this bulk background. As remarked before, the static frame does not exist for this case, on cannot obtain the results by simply plugging $\lambda=\epsilon_+=0$ in the results of Sec. \ref{CA static}. We to tackle the calculation in the stationary frame itself where the construction of the WdW patch boundaries is more complicated than for a static geometry (but far simpler than that for the more general $\lambda, \epsilon_+\neq0$ case). We find that the action complexity null warped AdS$_3$ vanishes! We believe this is purely a dimensional accident, the action complexity for pure AdS$_{d+1}$ analogously vanishes \cite{Reynolds:2016rvl, Chakraborty:2020fpt}  due to an overall factor of $(d-2)$. Finally, in section \ref{DiscOut} we conclude by discussing our results and provide an outlook for future work. In the appendices, we gather some results for ready references in the main sections. In Appendix \ref{sigma model to 4d}  we recap the sigma model with the $T\overline{T}, J\overline{T}, T\overline{J}$ deformations and the $4$d target spacetime which follows and work out the action complexity terms for the $4$d geometry. Next in Appendix \ref{KK redone} we recap the KK reduction over the $y$ circle fiber following the conventions of \cite{Chakraborty:2019mdf}, and obtain the dimensionally reduced 3d metric, Dilaton, B-field and KK scalar and KK gauge fields (the KK gauge fields obtained after reducing the 4d NS-NS B-field were missing in \cite{Chakraborty:2019mdf}. Subsequently we demonstrate the action complexity integrals for the $3$d background work out to be the same as those from the $4$d background worked out in the previous section provided we \emph{retain the total derivative terms in the $3$d action}. In Appendix \ref{section:GHY} we compute the new GHY term contribution as a result of keeping the total derivative term in the $3$d lagrangian (action) and the net GHY contribution. In Appendix \ref{HEE} we compute the holographic entanglement entropy for the WCFT dual to null warped AdS$_3$, for a boundary interval of size $L$, thereby closing a gap in the literature. For null WCFT we find that the entanglement entropy is log divergent, just like that of an local CFT$_2$, but the coefficient of the log divergence (central charge) now depends on the warping parameter.\\

For other interesting works on complexity in the context of double trace $T\bar{T}$ deformed CFT see \cite{Akhavan:2018wla,Jafari:2019qns,Geng:2019yxo}.
	
	\section{Review of string theory in AdS$_3$, single trace $T\bar{T}$ and LST}\label{sec2}
	
We first consider critical superstring background AdS$_3\times \mathcal{M}$, with $\mathcal{M}$ being a compact seven dimensional spacelike manifold, which preserves $\mathcal{N}\geq2$ supersymmetry. A classic example of this kind of a set up consists of type $II$ strings on  AdS$_3\times S^3\times T^4$ preserving $(4,4)$ supersymmetry. The worldsheet theory of strings propagating in AdS$_3$ with NS-NS fluxes switched on but R-R fluxes turned off is a WZW nonlinear sigma model of the noncompact group manifold $SL(2,\mathbb{R})$. The worldsheet theory is symmetric under the holomorphic (left moving) and antiholomorphic (right moving) components of $sl(2,\mathbb{R})$ current algebra with level $k$. The AdS radius , $R_{AdS}$, is related to the level of the current algebra by the relation $R_{AdS}=\sqrt{k}l_s$, $l_s=\sqrt{\alpha'}$ being the string length.
	
	According to the AdS/CFT correspondence, string theory on (asymptotically) AdS$_3$ is dual to a two-dimensional CFT living on the conformal boundary of AdS$_3$. For supergravity approximation to be valid, we will have to work in the parameter regime $k\gg1$. In the presence of the NS-NS three form H-flux, the spacetime theory has the following properties:
	\begin{enumerate}
		\item{The spacetime theory has a normalizable $SL(2,\mathbb{C})$ invariant vacuum state:\begin{itemize}
				\item{The NS vacuum, which corresponds to global AdS$_3$ as the bulk.}
				\item{The R vacuum, that corresponds to massless spinless ($M=J=0$) BTZ as the bulk.}
		\end{itemize}}
		\item{The NS sector consists of a sequence of discrete states coming from the discrete series representation of $SL(2,\mathbb{R})$ followed by a continuum of long strings. The continuum starts above a gap of order $\frac{k}{2}$ \cite{Maldacena:2000hw}.}
		\item{The R-sector states contain a continuum above a gap of order $\frac{1}{k}$. Here the fate of the discrete series states is unclear.}
	\end{enumerate}
	In the discussion that follows, we focus exclusively on the long strings of the R-sector.\\
	
	In \cite{Seiberg:1999xz}, it was argued, that for string theory on AdS$_3\times\mathcal{M}$, the theory supported on a single long string is described by a sigma model on 
	\begin{eqnarray}\label{Mlong}
		\mathcal{M}^{(L)}_{6k}=\mathbb{R}_\phi\times \mathcal{M}~,
	\end{eqnarray}
	with central charge $6k$. The theory on $\mathbb{R}_\phi$ has a dilaton field $\Phi$ that is linear in the coordinate $\phi$ with a slope given by
	\begin{eqnarray}\label{Qlong}
		Q^{(L)}=(k-1)\sqrt{\frac{2}{k}}~.
	\end{eqnarray}
	Thus the theory on the long string worldsheet has an effective interaction strength given by $\exp({Q^{(L)}\phi})$ and as a result the dynamics of the long strings becomes strongly coupled as they approach spatial infinity (boundary). But there is a wide range of positions on the radial direction where the long strings are weakly coupled.  A natural question that one may ask at this point is: what is the full boundary theory dual to string theory in $AdS_3$. The answer to that question, for generic $k$, is unknown, but there are strong evidences to convince that the theory on the long strings is described by the symmetric product CFT 
	\begin{eqnarray}\label{longsym}
		(\mathcal{M}^{(L)}_{6k})^p/S_p~, \label{2.3}
	\end{eqnarray}
	where $p$ represents the number of fundamental (F1) strings that form the background.\\
	
	String theory in AdS$_3$ admits an operator $D(x,\bar{x})$ \cite{Kutasov:1999xu} (where $x$ and $\bar{x}$ are coordinates of the two-dimensional spacetime theory), in the long string sector that has many properties in common with the $T\bar{T}$ operator. For example $D(x,\bar{x})$ is a $(2,2)$ quasi-primary operator of the spacetime Virasoro and has the same OPE with the stress tensor as the $T\bar{T}$ operator. However, there is an important difference between the $T\bar{T}$ operator and the operator $D(x,\bar{x})$: 
	$T\bar{T}$ is a double trace whereas $D(x,\bar{x})$ is single trace.\footnote{Here single trace refers to the fact that $D(x,\bar{x})$ can be expressed as a single integral over the worldsheet of a certain worldsheet vertex operator. The operator $T\bar{T}$ on the other hand is double trace because it can be expressed as a product of two single trace operators in the sense just described.} In fact 
	\begin{eqnarray}\label{D}
		D(x,\bar{x})=\sum_{i=1}^pT_i\bar{T}_i~,
	\end{eqnarray}
	where $T_i\bar{T}_i$ can be thought of as the $T\bar{T}$ operator of the $i^{th}$ block  $\mathcal{M}^{(L)}_{6k}$ in the symmetric product CFT $(\mathcal{M}^{(L)}_{6k})^p/S_p$. For an elaborate discussion along this line see \cite{Chakraborty:2019mdf,Chakraborty:2018vja}
	
	Next, consider the deformation of the long string symmetric product by the operator $D(x,\bar{x})$. This deforms the $i^{th}$ block CFT $\mathcal{M}^{(L)}_{6k}$ by the operator $T_i\bar{T}_i$ and is subsequently symmetrized. Such a deformation is evidently irrelevant and it involves \emph{flowing up} the renormalization group (RG) trajectory. This $D(x,\bar{x})$ deformation of the spacetime theory translates to turning on the worldsheet a truly marginal deformation:
	\begin{eqnarray}\label{wsind}
		\int_{(\mathcal{M}^{(L)}_{6k})^p/S_p} d^2 xD(x,\bar{x})\sim\int_{\Sigma} d^2z J_{SL}^-\bar{J}_{SL}^-~,
	\end{eqnarray}
	where $z,\bar{z}$ are the complex coordinates of the worldsheet Riemann surface $\Sigma$, $J_{SL}^-$ and $\bar{J}_{SL}^-$ are respectively the left and right moving null $sl(2,\mathbb{R})$ currents of the worldsheet theory.     
	
	The above current-anti-current deformation of the worldsheet $\sigma-$model is exactly solvable, and standard worldsheet techniques yield the metric (in string frame), dilaton and the B-field as \cite{Forste:1994wp,Israel:2003ry}
	\begin{eqnarray}
		\begin{split}\label{M3}
			& ds^2=f^{-1}(-dt^2+dx^2)+kl_s^2\frac{dU^2}{U^2}~,\\
			& e^{2\Phi}=\frac{g_s^2}{kU^2}f^{-1}~,\\
			& dB=\frac{2i}{k^{3/2}\,l_s\,U^2}f^{-1}\epsilon_3~,
		\end{split}
	\end{eqnarray}
	where $f=\lambda+\frac{1}{kU^2}$, $\lambda$ is the dimensionless coupling \footnote{Note that without loss of generality, the value of $\lambda$ can be set to an appropriate value as discussed in \cite{Giveon:2017nie}.} of the marginal worldsheet deformation  and $g_s$ is the asymptotic string coupling in $AdS_3$ with $g_s^2=e^{2\Phi(U\to 0)}\equiv e^{2\Phi_0}$. This background is popularly known as $\mathcal{M}_3$. The background $\mathcal{M}_3$ \eqref{M3} interpolates between $AdS_3$ in the IR (\ie\ $U\ll 1/\sqrt{k\lambda}$) to flat spacetime with a linear dilaton, $\mathbb{R}^{1,1}\times\mathbb{R}_\phi$ in the UV (\ie\ $U\gg 1/\sqrt{k\lambda}$). The coupling $\lambda$ sets the scale at which the transition happens. 
	
	The deformed sigma model background \eqref{M3} can also be obtained as a solution to the equations of the motion of  three dimensional supergravity action \cite{Giveon:2005mi,Chakraborty:2020yka}
	\begin{eqnarray}\label{sugra}
		S= \frac{1}{16\pi G_N}\int d^3 X \sqrt{-g}e^{-2(\Phi-\Phi_0)}\left( R+4g^{\mu\nu}\partial_\mu\Phi\partial_\nu\Phi-\frac{1}{12}H^2-4\Lambda\right)~,
	\end{eqnarray} 
	where $G_N$ is the three-dimensional Newton's constant in $AdS_3$, $g_{\mu\nu}$ is the string frame metric, $R$ is the Ricci scalar (in string frame), $\Phi$ is the dilaton, $H=dB$ is the 3-form flux and $\Lambda$ is the cosmological constant.

	As an example, the above construction can be realized as follows. Let us consider a stack of $k$ NS5 branes in flat space wrapping a four dimensional compact manifold (\eg\ $T^4$ or $K_3$). The near horizon geometry of the stack of $k$ NS5 branes is given by $\mathbb{R}^{1,1}\times\mathbb{R}_\phi$ with a dilaton that is linear in the radial coordinate $\phi$ (where $\phi=\log(\sqrt{k} U)$). The string coupling goes to zero near the boundary (\ie\ $U\to \infty$) whereas it grows unboundedly as one goes deep in the bulk (\ie\ $U\to 0$).  Next, let's add $p$ (with $p\gg1$) F1 strings stretched along $\mathbb{R}^{1,1}$. This stabilizes the dilaton and the string coupling saturates as $g_s\sim 1/\sqrt{p}$. Thus for large $p$ the string coupling is weak and one can trust string perturbation theory. The F1 strings modifies the IR geometry (\ie\ $U\ll 1/\sqrt{k\lambda}$) to $AdS_3$. The smooth interpolation between   $\mathbb{R}^{1,1}\times\mathbb{R}_\phi$ in the UV to $AdS_3$ in the IR corresponds to interpolation between near horizon geometry of the NS5 brane system to that of the F1 strings \cite{Chakraborty:2020swe,Chakraborty:2020yka}. The spacetime theory interpolates between a CFT$_2$ with central charge $6kp$ in the IR to two-dimensional LST in the UV. The theory is nonlocal in the sense that the short distance physics is not governed by a fixed point.
	
	LST can be realized as the decoupled theory on the NS5 branes.  
	It has properties that are somewhat intermediate between a local quantum field theory and a full fledged critical string theory. Unlike a local field theory, at high energy $E$, LST has a Hagedorn density of states $\rho\sim e^{\beta_HE}$ where $\beta_H=2\pi l_s\sqrt{k\lambda }$. On the other hand, LST has well defined off-shell amplitudes \cite{Aharony:2004xn} and upon quantization it doesn't give rise to massless spin 2 excitation. Both these properties are very similar to  local quantum field theories. For a detailed review of LST see \cite{Aharony:1998ub,Kutasov:2001uf}.
	
One can generalize this scenario further by turning on holomorphic and antiholomorphic currents in the spacetime theory $J(x), \overline{J}(\overline{x}))$
\cite{Chakraborty:2019mdf, Kutasov:1999xu}. In that case, parallel to the construction of $D(x, \overline{x})$, one can also construct \emph{single trace} operators, namely, $A(x, \overline{x})$ and $\overline{A}(x, \overline{x})$ of dimension $(1, 2)$ and $(2, 1)$ respectively \cite{Kutasov:1999xu}. $A(x, \overline{x})$ has the same conformal dimension and OPE’s with the currents as the irrelevant double trace $J(x)\overline{T}(\overline{x})$ operator. Analogously, the single trace marginal $\overline{A}(x, \overline{x})$  is related to the irrelevant double trace $T(x)\overline{J}(\overline{x})$ operator in the spacetime CFT. In the symmetric product CFT, one can think of the operator of $A,\overline{A}$ as
\begin{equation}
A(x, \overline{x}) = \sum_{j=1}^{p} J_j (x)\,\overline{T}_j (\overline{x}); \,\overline{A}(x,\overline{x}) = \sum_{j=1}^{p} T_j (x)\,\overline{J}_j(\overline{x}).
\end{equation}
Turning on $A, \overline{A} (x,\overline{x})$, in addition to the $D(x,\overline{x})$ operators, in the spacetime corresponds to the perturbing the worldsheet llagrangian by the following marginal operators,
\begin{equation}
\delta\mathcal{L}_{WS} =\lambda\, J^{-}_{SL}(z)\overline{J}_{SL}^{-} (\overline{z}) + \epsilon_{+}\, K (z) \overline{J}_{SL}^{-} (\overline{z}) +\epsilon_{-}\, J_{SL}^{-} (z) \overline{K}(\overline{z}).
\end{equation}
One has to strictly consider the positive sign of the coupling $\lambda$ because only for that sign of the coupling the spectrum of the deformed theory is real and the theory is unitary.

The worldsheet $U(1)$ currents $K(z)$ and $\overline{K} (\overline{z})$ are associated with left and right-moving momenta on a $S^1$ in the bulk spacetime labelled by the coordinate $y$. Such a deformation will lead to the sigma model action \cite{Chakraborty:2019mdf},
\begin{equation}
S(\lambda, \epsilon_+,\epsilon_-)=\frac{k}{2\pi}\int d^2z \left( \partial \phi \overline{\partial}\phi +h \partial \overline{\gamma} \overline{\partial} \gamma+\frac{2\epsilon_{+} h}{\sqrt{k}} \partial y \overline{\partial} \gamma +\frac{2\epsilon_{-} h}{\sqrt{k}} \overline{\partial} y \partial \overline{\gamma} +\frac{f^{-1} h}{k} \partial y\overline{\partial} y \right)
\end{equation}
where $f^{-1}=\lambda+e^{-2\phi}$, $h^{-1}=\lambda-4\epsilon_+\epsilon_- +e^{-2\phi}$. This corresponds to the $4d$ background \cite{Chakraborty:2019mdf},
\begin{equation}
\frac{ds^2}{{l_{s}}^2}=k h \left(d\gamma+ \frac{2\epsilon_-}{\sqrt{k}} dy\right)  \left(d\overline{\gamma}+ \frac{2\epsilon_+}{\sqrt{k}} dy\right) + k d\phi^2 +dy^2
\end{equation}
with a dilaton
\begin{equation}
e^{2\Phi} = g_s^2 e^{-2\phi} h,
\end{equation}
and a NS-NS B-field,
\begin{equation}
B_{\gamma\bar{\gamma}}=-\frac{hk}{2},\qquad B_{y\gamma}=-B_{\gamma y}=\epsilon_{+}h\sqrt{k},\qquad B_{y\overline{\gamma}}=-B_{\overline{\gamma}y}=-\epsilon_{-}h\sqrt{k}
\end{equation}
See Appendix \ref{sigma model to 4d} for some of the details omitted here.
\subsection{The Holographic $2+1$-d background} \label{holbg}

Upon performing a KK reduction along the $y$-circle \cite{Chakraborty:2019mdf},  target space NS-NS sector background described by the $3$d metric
\begin{equation}
ds^2=kl_s^2\frac{  h(\phi )}{f(\phi )}\text{d$\phi $}^2+kl_s^2\frac{  h(\phi )^2}{f(\phi )}d\gamma d\bar{\gamma}-kl_s^2 h(\phi )^2 (\epsilon_+ \text{d$\gamma $}+\epsilon_- \text{d$\bar{\gamma} $})^2,
\end{equation}
and the dilaton, $\Phi$ and a 2-form gauge field $H$ background\footnote{In addition there are $U(1)$ gauge fields originating from the KK reduction of the $4$d metric and $4$d B-field, refer to Appendix \ref{KK redone} for the full list.},
\begin{align}
	e^{2\Phi}&=g_s^2e^{-2\phi}\sqrt{f(\phi)h(\phi)}, & B_{\gamma\bar{\gamma}}&=\frac{k h(U)l_s^2}{2}.
\end{align}
The functions $h,f$ are defined by
\begin{align}
	h(\phi)&=\frac{1	}{\lambda-4\epsilon_+\epsilon_-+e^{-2\phi}}, & f(\phi)&=\frac{1}{\lambda+e^{-2\phi}},
\end{align}
where $\lambda, \,\epsilon_{\pm}$ are the irrelevant dimensionless couplings for $ T\bar{T}, J\bar{T}, \,\&\, \bar{J}T$ deformations respectively.  Here $\phi$ is the radial coordinate while the $\gamma, \overline{\gamma}$ are lightlike coordinates parallel to the boundary. In this work we will work instead with the following coordinates,
	\begin{align*}
	e^{\phi}&=\sqrt{k}U, &	x&=\frac{\sqrt{k}l_s}{2}(\gamma+\bar{\gamma}), & t&=\frac{\sqrt{k}l_s}{2}(\gamma-\bar{\gamma})
\end{align*}
Thus, $U$ is the radial coordinate (RG scale) while $t,x$ are boundary time and space coordinate. In terms of these new coordinates metric reads,
\begin{multline}
				ds^2=\frac{  h(U )}{f(U )}\Bigg[k l_s^2\frac{dU^2}{U^2}-h(U)\,\left(1+f(U)(\epsilon_+-\epsilon_-)^2\right)dt^2\\-2\,h(U)\,f(U)\,(\epsilon_+^2-\epsilon_-^2))\,dt dx+h(U)\,\left(1-f(U)(\epsilon_++\epsilon_-)^2\right)dx^2\Bigg]. \label{stationary coordinate metric}
\end{multline}
while  the dilaton and the Kalb-Ramond field are given by,
		\begin{align}
				e^{2\Phi}
				&=\frac{g_s^2}{\sqrt{\left(k \lambda  U^2+1\right) \left(k U^2 \left(\lambda -4 \epsilon _- \epsilon _+\right)+1\right)}},
				& dB&=\frac{2 h(U)}{k^{3/2}l_s U^2\sqrt{1-4 \epsilon_+ \epsilon_- f(U)}{}}\epsilon_3. \label{stationary coordinate dilaton and H}
		\end{align}
Here we have,
		\begin{align*}
			h(U)&=\frac{kU^2}{1+(\lambda-4\epsilon_+\epsilon_-)kU^2}, & \,\, f(U)&=\frac{kU^2}{1+\lambda k U^2},
		\end{align*}
(We notice that if we replace $\lambda \rightarrow \lambda'= \lambda -4\epsilon_+ \epsilon_-$, then $f(U)\rightarrow h(U)$. This fact will be put to use in the calculations to follow in the coming sections). This background interpolates between $AdS_3$ in the IR to linear dilaton flat spacetime in the UV. In the dual sense this geometry represents an integrable RG flow connecting a Lorentz invariant local CFT (fixed point) in the IR to a Lorentz violating nonlocal theory in the UV, namely a deformed little string theory (LST).\\

\section{Holographic Volume Complexity}   \label{0T complexity}

In this section we employ holography to compute the computational complexity of the LST deformed by irrelevant single trace $J\bar{T}$ and $ T\bar{J}$ deformation following the Complexity-Volume (CV) \cite{Susskind:2014rva} prescription. Computational complexity, just like entanglement entropy, is a manifestly UV-divergent quantity, and for ordinary quantum field theories the UV divergence structure of computational complexity is rigidly constrained \cite{ Carmi:2016wjl, Reynolds:2016rvl}. In this section we reveal the UV-divergences which might arise in a nonlocal and lorentz violating field theory, such as two-dimensional CFT deformed by single trace  $J\bar{T}$ and $ T\bar{J}$ and compare and contrast them with those arising in a lorentz invariant local quantum field theory (\eg\ a CFT$_2$). The volume complexity prescription computes the complexity of the dual boundary theory in terms of the volume of a maximal volume spacelike slice, $\Sigma$,
\begin{equation}
	C_V = \frac{V_{\Sigma}}{G_N\,L}~,\ \  \text{ with } \ \   V_\Sigma = \int_\Sigma d^{D-1}x\, \sqrt{\gamma_\Sigma}~,\label{eq:cv}
\end{equation}
where $\gamma_{\mu\nu}$ is the pullback metric on the maximal volume slice. As mentioned before, $L$ represents a suitable characteristic scale of the geometry. Here, we are working in the string frame with a non-trivial dilaton background and the volume complexity proposal needs to be generalized. The appropriate generalization is given by \cite{Klebanov:2007ws},
\begin{equation}
	C_V  = \frac{ \tilde{V}_{\Sigma}}{\kappa_0^2\, L}~, \ \ \ \text{with } \ \  \tilde{V}_{\Sigma} = \int_\Sigma d^{D-1}x\,  e^{-2\left(\Phi-\Phi_\infty\right)} \sqrt{\gamma_\Sigma}~. \label{CV_str}
\end{equation}

One can check that this generalization furnishes the correct powers of $G_N$\footnote{See \cite{Klebanov:2007ws} for a similar prescription for the Ryu-Takayanagi formula for the entanglement entropy} in the denominator using the string theory convention, $\kappa_0^2 e^{-2(\Phi_\infty-\Phi_0)} = 8 \pi G_N$ where $e^{\Phi_\infty}$ is the flat space string coupling and $e^{\Phi_0}$ is the string coupling of $AdS_3$. In anticipation of the fact that the dual boundary field theory is Lorentz violating, we compute the volume complexity in two different Lorentz frames and the comparison is drawn between the results.

\subsection{Volume Complexity in stationary coordinates ($x,\, t$)}

We specify the a spacelike hypersurface by the condition, $t=t(U),\,\forall x$. The pullback of the ambient metric in the so called stationary coordinates (\ref{stationary coordinate metric}) on the hypersurface becomes:
\begin{multline}
	ds^2_{\Sigma}=\left(\frac{k l_s^2}{U^2}-h(U)(1+f(U)(\epsilon_+-\epsilon_-)^2)t'(U)^2\right)dU^2-2h(U)f(U)(\epsilon_+^2-\epsilon_-^2))t'(U)dU dx\\+h(U)(1-f(U)(\epsilon_++\epsilon_-)^2)dx^2.
\end{multline}
The general form of the volume of any hypersurface in string frame with appropriate inclusion of the dilaton factors in the integral measure is,
\begin{equation}
	\begin{split}
		V_{\Sigma}(t_*) &= e^{2(\Phi_\infty-\Phi_0)} \int dx\, dU \,  e^{-2\left(\Phi-\Phi_0 \right)} \sqrt{\gamma_{\Sigma}},\nonumber \\
		&= \frac{ k l_s L_x}{e^{-2(\Phi_\infty-\Phi_0)}} \int_0^{\infty} dU\, \sqrt{1+k  \mu\, U^2} \sqrt{1-\frac{t'(U)^2 U^4}{l_s^2 \left(1+k \mu \,U^2\right) }}.
	\end{split}
\end{equation} 
where, $L_x$ is the IR cutoff of the boundary LST and we have defined 
\begin{equation}
\mu:=\lambda-\left(\epsilon _-+\epsilon _+\right){}^2
\end{equation}
for later convenience. To find the maximal volume one needs to extremize this volume functional. The corresponding Euler-Lagrange equation is,
\begin{align}
	-U l_s^2 t''(U) \left(1+k \mu U^2\right)+l_s^2 t'(U) \left(3 k \mu  U^2+4\right)-2 U^4 t'(U)^3=0.
\end{align}
To solve this nonlinear differential equation perturbatively, we employ the near boundary power series expansion of the form:
\begin{equation}
	t(U)=T+\frac{a}{U}+\frac{b}{U^2}+\frac{c}{U^3}+\frac{d}{U^4}+....
\end{equation}
Plugging this ``large $U$" expansion in the Euler Lagrange equation and solving iteratively in powers of $U^{-1}$ we get all coefficients to vanish, $a=b=c=d=\ldots =0$.  With this knowledge, the volume $V_{\Sigma}$ of the maximal slice turns out to be:
\begin{equation}
	\begin{split}
		V_{\Sigma}(t_*)&= \frac{ k l_s L_x}{e^{-2(\Phi_\infty-\Phi_0)}}\int dU  \sqrt{1+k U^2 \mu}.
	\end{split}
\end{equation}
Therefore by \eqref{eq:cv}, the volume complexity turns out to be
\begin{eqnarray}
	C_{V} \equiv  \frac{V_\Sigma(t_*)}{\kappa_0^2\,L}  & = & \frac{k\,l_s\,L_x}{G_N\,L}\left[\frac{l_s}{2\epsilon} \sqrt{1+\frac{k\,\mu\,l_s^2}{\epsilon^2}}  +\frac{\sinh^{-1}\left(\frac{\sqrt{k\,\mu}\,l_s}{\epsilon}\right)}{2\sqrt{k\,\mu}} \right]~.
\end{eqnarray}
Note that by convention  the length scale $L$ appearing here is the characteristic length scale associated with the geometry. Comparison with results from action complexity  helps us resolve this ambiguity $L=\ell=\sqrt{k}\,l_s$, the AdS radius, and the volume complexity is after evaluating the integral is
\begin{equation}
	\begin{split}
		C_{V}(T)
		&=\frac{c L_x }{3 \epsilon }\sqrt{\frac{3 \rho _H^2}{4 \pi ^2 \epsilon ^2}+1}+\frac{2 \pi  c L_x }{3 \sqrt{3} \gamma _H}\sinh ^{-1}\left(\frac{\sqrt{3} \rho _H}{2 \pi  \epsilon }\right),\\
		&=\frac{c L_x }{3 \rho_H }\left(\frac{\rho_H}{\epsilon}\sqrt{1+\frac{3 \rho _H^2}{4 \pi ^2 \epsilon ^2}}+\frac{2 \pi  }{\sqrt{3}}\sinh ^{-1}\left(\frac{\sqrt{3} \rho _H}{2 \pi  \epsilon }\right)\right).\label{CV2}
	\end{split}
\end{equation}
Here as before, $\epsilon$ is the UV cutoff required to regularize the divergent integral by placing the boundary at $U=\frac{l_s}{\epsilon}$ and  $c=\frac{3\sqrt{k}l_s}{2G_N}$ is the Brown-Hanneaux central charge of the IR $CFT_2$. The expression in the first line is rewritten in terms of the Hagedorn density of states, $\rho_H$  \cite{Chakraborty:2020xyz} in the second line:
\begin{equation}
	\rho_H=\frac{2\pi}{\sqrt{3}}\sqrt{k \mu}l_s.
\end{equation} 
We immediately notice that the leading divergence is quadratic followed by a logarithmic divergence. The quadratic (and logarithmic) divergence now depend on both the parameters controlling non-locality \emph{and} lorentz violation. However it appears that the Lorentz violation effects and nonlocality effects are combined into a single parameter, namely $\mu = \lambda -  \left( \epsilon_+ + \epsilon_{-}\right)^2$ and there is no way to cleanly separate the effects of one from the other. This is perhaps mildly disappointing since our hope was to be able to see the effects of nonlocality and Lorentz violation in separate UV divergence structures. Also, we see that in order for the notion of complexity to make sense, we have to restrict $\mu \geq  0$. This condition is important in ensuring the existence of a smooth dual gravity background geometry as mentioned in earlier works \cite{Chakraborty:2020udr,Chakraborty:2020cgo}. As a consistency check we note that the complexity expression (\ref{CV2}) smoothly reduces to the previously known $\mathcal{M}_3$ expression as the lorentz violating couplings $\epsilon_{\pm}$ \cite{Chakraborty:2020fpt} are turned off.\\

Let's now examine the behavior of the theory in the two opposite extreme limits. 
Thinking of $\rho_H$ as the distance scale below which non-local and lorentz violating effects kicks in, one of the interesting limit to study would be the UV limit $\epsilon/\rho_H<<1$ where the short distance physics is that of the non-local lorentz violating field theory:
\begin{equation}
	\lim_{\epsilon/\rho_H\to0}C_V=\frac{cL_x}{\sqrt{3}\rho_H}\left(\frac{ \rho_H^2 }{2 \pi  \epsilon ^2}+\frac{\pi  }{3 } \ln \left(\frac{3 \rho_H^2}{\pi ^2 \epsilon ^2}\right)+O\left(\epsilon/\rho_H\right)\right). \label{CV1div}
\end{equation}
The divergence structure as is evident from this expression, does not match with that of the lorentz covariant local field theory. For the latter case, the complexity being an extensive quantity counting the degrees of freedom in the field theory is expected to diverge linearly i.e. $L_x/\epsilon$.\\

Another interesting regime to study is the IR behavior where, $\rho_H/\epsilon<<1$.
\begin{equation}
	\lim_{\rho_H/\epsilon\to0}C_V=\frac{2 c L_x}{3\, \epsilon }.\label{CV1vanish}
\end{equation}
This expression reproduces the expected result for a local field theory \cite{Reynolds:2016rvl} by correctly counting the total number of degrees of freedom.\\

\subsection{Volume complexity in static ($X,\, T$) coordinates}

As alluded to in the introduction, due to the presence of additional irrelevant \{$\epsilon_{\pm}$\} couplings, the field theory is lorentz violating. As a result, the bulk geometry also inherits this character. Therefore we feel it is instructive to repeat the CV calculation in a different Lorentz frame, namely the ``static coordinate system" obtained after performing the following lorentz boosts on the stationary coordinate system of the previous section,
\begin{align}
	X&=\frac{1}{2\sqrt{\epsilon_{+}\epsilon_{-}}}((\epsilon_{+}+\epsilon_{-})x+(\epsilon_{+}-\epsilon_{-})t),\nonumber\\
	T&=\frac{1}{2\sqrt{\epsilon_{+}\epsilon_{-}}}((\epsilon_{+}-\epsilon_{-})x+(\epsilon_{+}+\epsilon_{-})t), \label{boost}
\end{align}
the resulting metric is:
\begin{align}
	ds^2&= k l_s^2\frac{dU^2}{U^2}-h(U)\, dT^2+ f(U)\,dX^2. \label{static frame metric}
\end{align}
Using CV prescription, the maximal codim-1 surface $\Sigma$ is required to be given by the equation $T=T(U)$ with appropriate functional form which extremizes the volume element. Since there are no crossterms of form $dt dX$, it is appropriate to refer this as a static coordinate system. \\

The induced metric  is
\begin{align}
	ds_\Sigma^2& \equiv \gamma_{ab} dx^a dx^b,\nonumber\\&= \left(\frac{k l_s^2}{U^2}-h(U) T'(U)^2\right)dU^2+ f(U)\,dX^2.
\end{align}
In the string frame, the volume of such a spacelike slice anchored at a time $T_*$ on the boundary is,
\small{\begin{align}
		\tilde{V}(T) &= e^{2(\Phi_\infty-\Phi_0)} \int dx\, dU \,  e^{-2\left(\Phi-\Phi_0 \right)} \sqrt{\gamma_{\Sigma}},\nonumber \\
		&= \frac{ k^{3/2} l_s L_x}{e^{-2(\Phi_\infty-\Phi_0)}} \int_0^{\infty} dU\,\frac{k U^2}{\sqrt{f(U) h(U)}}\,\sqrt{f(U) \left(\frac{k l_s^2}{U^2}-h(U) T'(U)^2\right)}.
\end{align} }
Here $L_x=\int dx$ is the spatial extent (IR cutoff) of the boundary theory target space and $\lambda'$ is defined to be
\begin{equation}
\lambda'\equiv\lambda-4\epsilon_{+}\epsilon_{-}.
\end{equation}
Extremizing this volume leads to the following Euler-Lagrange equation:
\begin{align}
	l_s^2 \left(U T''(U) \left(1+k \lambda'  U^2\right)+T'(U) \left(3 k \lambda'  U^2+4\right)\right)-2 U^4 T'(U)^3=0. \label{EL_0T}
\end{align}
The solution is found by employing series expansion method, lets assume the near boundary expansion of $T(U)$ of the form:
\begin{eqnarray}\label{maxvslice}
	T(U ) & = & T_*+\frac{a_1}{U }+ \frac{a_2}{U^2}+\frac{a_3}{U ^3}+\dots.
\end{eqnarray}
And plugging back in \eqref{EL_0T} and solving them order by  order in $\frac{1}{U}$, we obtain the result that all the coefficients vanish. Thus the maximal volume slice is $T(U)=T_*$, a result that can be anticipated from the time reflection symmetry: $T\rightarrow-T$, of the background. Thus, the volume of the maximal volume slice is,
\begin{align}
	\tilde{V}_\Sigma(T)&=\frac{ k^{3/2}\, l_s\, L_x}{e^{-2(\Phi_\infty-\Phi_0)}}\int_0^\infty dU \frac{ U }{\sqrt{h(U)}},\\
	&=\frac{ k\, l_s\, L_x}{e^{-2(\Phi_\infty-\Phi_0)}} \int_0^\infty dU \sqrt{1+k U^2\lambda'}, 
\end{align}
which diverges as $U \rightarrow \infty$. So we impose a UV cutoff at $U=l_s/\epsilon$ to regulate it. Also, we have defined $\lambda'$ to be $\lambda -4 \epsilon _- \epsilon _+$. The regulated volume is then,
\begin{eqnarray}
	\tilde{V}_\Sigma(T)  = \frac{k\,l_s\,L_x}{e^{-2(\Phi_\infty-\Phi_0)}}\left[\frac{l_s}{2\epsilon} \sqrt{1+\frac{k\,\lambda'\,l_s^2}{\epsilon^2}}  +\frac{\sinh^{-1}\left(\frac{\sqrt{k\,\lambda'}\,l_s}{\epsilon}\right)}{2\sqrt{k\,\lambda'}} \right]~. \label{stationary frame volume}
\end{eqnarray}
As expected, due to time translation symmetry the expression is independent of $T_*$. Therefore from \eqref{CV_str} volume complexity turns out to be
\begin{eqnarray}
	\mathcal{C}_{V} \equiv  \frac{\tilde{V}_\Sigma}{\kappa_0^2\,L}  & = & \frac{k\,l_s\,L_x}{G_N\,L}\left[\frac{l_s}{2\epsilon} \sqrt{1+\frac{k\,\lambda'\,l_s^2}{\epsilon^2}}  +\frac{\sinh^{-1}\left(\frac{\sqrt{k\,\lambda'}\,l_s}{\epsilon}\right)}{2\sqrt{k\,\lambda'}} \right]~.\label{stationary frame complexity}
\end{eqnarray}
Again following the remarks of the preceding section, $L=\ell=\sqrt{k}\,l_s$, the AdS radius and the volume complexity is thus,

\begin{eqnarray}
	\mathcal{C}_V=\frac{cL_x}{3\beta'_H}\left[\frac{\beta'_H}{2\epsilon}\sqrt{4+\frac{\beta_H'^2}{\pi^2\epsilon^2}}+2\pi~\sinh^{-1}\left(\frac{\beta'_H}{2\pi\epsilon}\right)\right]~,\label{CVl}
\end{eqnarray}
where, $\beta'_{H}$ is the inverse Hagedorn temperature
\begin{eqnarray}
	\beta'_H=2\pi l_s\sqrt{k\lambda'}~.
\end{eqnarray}

We would like to draw the reader's attention to the important fact that the holographic volume complexity expression in the static frame \eqref{CVl} does not match with that in the stationary frame \eqref{CV2}. This is the artifact of the dual field theory being Lorentz violating in nature i.e. the complexity measured in different frames related by a Lorentz boost transformation do not agree. Similar observation had also been made in regard to entanglement entropy in \cite{Chakraborty:2020udr}. This is indeed gratifying to note.

\subsubsection{A comment on the nonlocality and Lorentz violation}\label{lstlim}

Let us recall that $\beta'_H$  can be thought of the length scale below which nonlocality and Lorentz violation effects kicks in. Thus, an  interesting limits to study would be $\epsilon/\beta'_H\ll1$ where the short distance physics is that of a nonlocal and Lorentz violating theory. In this limit the volume complexity takes the form
\begin{eqnarray}
	\lim_{\epsilon/\beta'_H\to 0} \mathcal{C}_{V} &=&\frac{cL_x}{3\beta'_H}\left[\frac{\beta_H'^2}{2\pi \epsilon^2}+2\pi \log\left(\frac{\beta'_H}{\pi\epsilon}\right)+ \pi +O\left(\frac{\epsilon}{\beta'_H}\right)\right]~.\label{CV@0T}
\end{eqnarray}
Evidently the divergence structure of the volume complexity \eqref{CV@0T} does not appear like that of a local quantum field theory. 

For the case of a local quantum field theory, complexity being an extensive quantity should be proportional to the degrees of freedom given by the number of lattice sites $\propto L_x/\epsilon$ \ie\ scales inversely with the cutoff $\epsilon$ (lattice spacing). 
The quadratic and logarithmic divergences in \eqref{CV@0T} are a reflection of the fact that the boundary theory, being a LST, is a nonlocal, Lorentz violating field theory and fittingly a special combination of nonlocality and Lorentz violation parameters, namely $\beta'_H$, features in the coefficient of this quadratic as well as the logarithmic divergences. One can check, that by making the nonlocality and Lorentz violation vanish in the limit $\epsilon/\beta'_H\gg1$, the volume complexity expression \eqref{CVl} indeed reduces to that of a local field theory,
\begin{equation}\label{CVir}
	\lim_{\epsilon/\beta'_H\gg 1 } \mathcal{C}_V =\frac{2c}{3\beta'_H}~\frac{L_x}{(\epsilon/\beta'_H)}= \frac{2c}{3}~\frac{L_x}{\epsilon}~. 
\end{equation}
This expression of complexity (being proportional to the product of $c$, the central charge \ie\ the number of degrees of freedom per lattice site, and $L_x /\epsilon$, which gives the total number of lattice sites) counts the total number of degrees of freedom  in a local field theory.

This quadratic UV divergence of the LST in 1-space dimensions,
i.e. a ``hypervolume'' divergence is a fascinating observation. Let compare and contrast it with the divergence structure arising in (holographic) entanglement entropy (EE). The EE for nonlocal field theories one encounters a similar phenomenon, the RT prescription
yields a volume law instead of a perimeter (area) law for a subregion
EE, e.g. see \cite{Barbon:2008ut, Karczmarek:2013xxa, Shiba:2013jja, Pang:2014tpa}
in addition to the LST EE \cite{Chakraborty:2018kpr}.
However, physical understanding of the volume dependence (divergence)
of EE for nonlocal field theories is perhaps intuitively obvious.
Given a subregion, for a local field theory the entangling degrees
of freedom are localized on the boundary. Once the theory becomes
nonlocal, the entangling degrees of freedom are not only localized
on the boundary of the subregion but also along direction orthogonal
to the boundary, i.e. throughout the volume of the subregion. Hence
the appearance of the volume divergence for EE. However, for the case
of complexity is qualitatively different, it is \emph{already} a
volume law for local field theories, so it is not obvious intuitively
why the ``hypervolume'' law and in particular why the power of divergence
is ``$\text{volume}+1$'' instead of ``$\text{volume}+n$'' for
arbitrary positive integral $n$. At this point we can only speculate
which specific aspect of nonlocality of LST is captured by the hypervolume
divergence: in the strong dilaton region (UV), the LST effectively
behaves like it has grown an extra spatial dimension, much alike IIA
string theory which grows an extra dimension when the dilaton turns
strong (strong coupling). This appearance of an effective extra (noncompact) spatial
dimension could potentially explain the ``$\mathbf{volume}+1$''
divergence structure. Although this analogy is not exact or direct
since the LST studied in our work is obtained from NS5 branes in $IIB$
frame/theory, while the $10$ dimensional string to $11$ dimensional
$M$-theory is realized in the $IIA$ frame, and the emergent dimension is a compact (circle). Similar observations/
suggestions i.e. LST behaving like it develops an extra spatial dimension
at strong coupling, have been made in early works in LST in $IIA$ frame, e.g. in \cite{Minwalla:1999xi}. Perhaps a more definitive statement can only be made when the holographic complexity of nonlocal field theories which are not necessarily LST (or derived from string theory) are computed. These theories will not share the stringy property of developing extra spatial dimensions at strong coupling and might have a different kind of divergence structure. 

Next consider the coefficient of the log term (which is universal) in the expression of volume complexity \eqref{CV@0T} in the deep UV (\ie\ $\epsilon\ll \beta'_H$), which is 
\begin{equation}
	\tilde{N}= c\frac{L_x}{\beta'_H}~. \label{total dof}
\end{equation}
Evidently, this coefficient counts the total number of  ``regularized/effective" degrees of freedom in the theory if we regard the lattice spacing of LST to be the Hagedorn scale, $\beta'_H$ instead of the UV cutoff $\epsilon$ of the original IR CFT, namely, $ N\sim c\, \frac{L_x}{\epsilon}$. Regarding the universality of the log divergence piece in volume complexity
and action complexity: We regret that the language in the draft led
the referee to infer that we are claiming that the log divergence
is universal across different holographic proposals of complexity
(e.g. volume and action). There are now numerous proposals of holographic
complexity (complexity$=$volume, complexity$=$action, complexity$=$spacetime
volume 2.0, etc., finally culminating in the claim by Belin et. al. \cite{Belin:2021bga} that there might be infinite number of such spatial codimension-one bulk geometric prescriptions of holographic complexity which are as good candidates as the ones suggested originally. It has been generally observed that although the leading divergence
pieces across different prescriptions match, the coefficients of the
subleading divergences do not match, either in sign or in magnitude.
It could be that various prescriptions of holographic complexity correspond
to field theory duals which are distinct but are in the same universality
class in the sense of RG (although the study of field theory complexity
is at a very premature stage to make such classifications of universality
classes under RG). However, \emph{once we pick a proposal},
the coefficient of the log divergence is universal in the usual sense -
if we rescale the UV energy scale, the coefficient of the log
divergence does not get rescaled. Such coefficients which do not get
rescaled usually capture some universal physics e.g. in the RT proposal it gives the $c$-function.\\

Another interesting fact emerges if we rewrite the quadratic UV divergence term \eqref{CV@0T}, in a manner which mimics a local field theory:
\begin{equation}
	\mathcal{C}_V = \, \frac{c L_x \beta'_H}{6\pi ^2 \epsilon^2} + \ldots =\,  \frac{2\tilde{c}(\epsilon)}{3} \frac{L_x}{\epsilon} + \dots, \ \text{ where }\ \ \tilde{c}(\epsilon)= c \frac{\beta'_H}{4 \pi^2 \epsilon}~,
\end{equation}
If we pretend this is a local theory with a linear divergence structure, then the coefficient of the linear divergence if identified as an ``effective central charge'' is now a UV scale dependent parameter $\tilde{c}(\epsilon)$, in fact it is a monotonically increasing function of UV scale (energy), $\frac{1}{\epsilon}$. So this ``effective central charge" diverges as the UV cutoff is withdrawn.  Similar observations have been made about LST, namely a divergent central charge, elsewhere in the literature. \\

\begin{figure}[H]
	\centering
	\includegraphics[width=.5\textwidth]{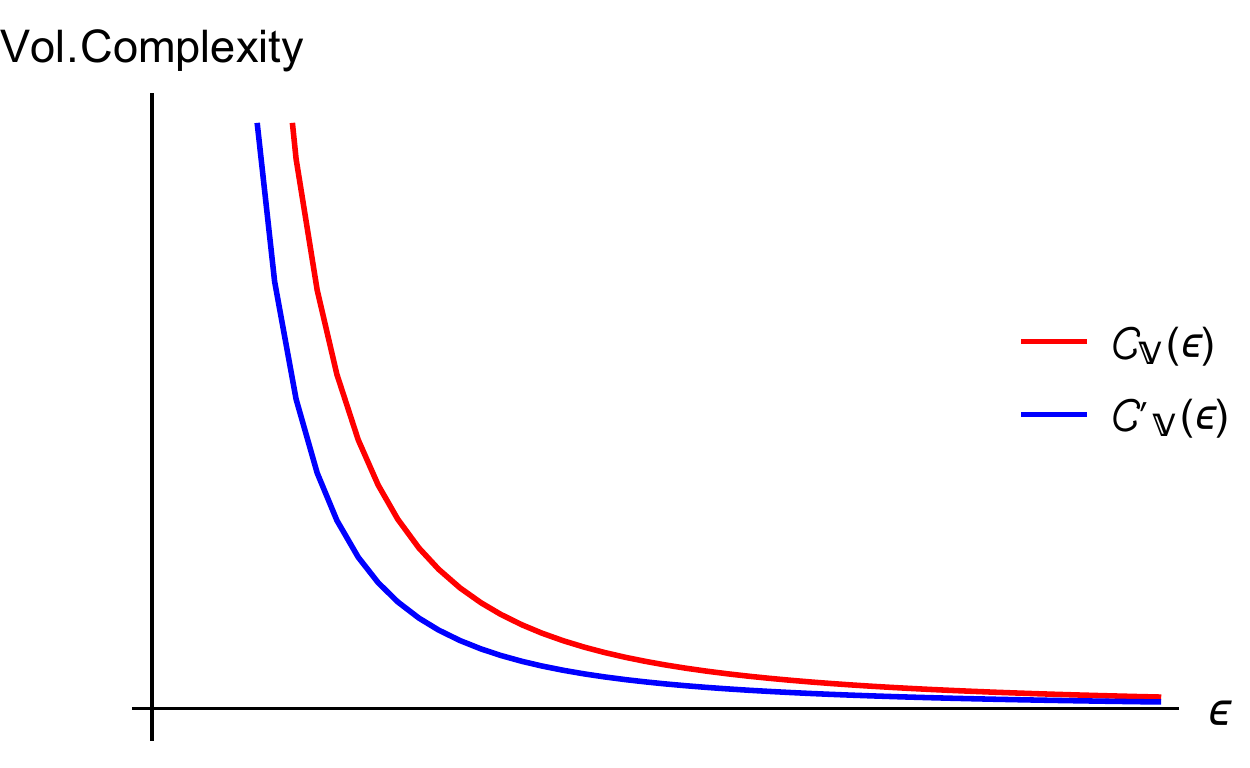}
	\caption{ Static frame complexity, $\mathcal{C}_V$ and stationary frame complexity, $\mathcal{C}'_V$ as a function the UV cutoff scale $\epsilon$.}
	\label{fig1}
\end{figure}

Now we compare the complexity results for the static frame and the stationary frame.  They share several common features:
\begin{enumerate}
	\item{The static frame volume complexity,  $\mathcal{C}_V$ \eqref{CVl} as a function of $\epsilon/\beta'_H$ and the stationary frame complexity, $\mathcal{C}'_V$ \eqref{CV2} as function of $\epsilon/\rho_H$ are always positive and monotonically decreases from UV to IR.}
	
	\item{In the extreme UV regime (\ie\ when $\epsilon \ll \beta'_H, \rho_H$), $\mathcal{C}_V$ diverges as \eqref{CV@0T}}, and $C'_V$ too diverges as \eqref{CV1div}.
	
	\item{In the extreme IR regime, (\ie when $\epsilon \gg \beta'_H, \rho_H$), $\mathcal{C}_V$ decreases to 0 as \eqref{CVir} and so does $\mathcal{C}'_V$ according to \eqref{CV1vanish}.}
	
\end{enumerate}
Finally we plot the static frame complexity, $\mathcal{C}_V$ and the stationary frame complexity, $\mathcal{C}'_V$ as a function of the cutoff, $\epsilon$ in figure \ref{fig1}.

In conclusion, we notice that the volume complexity for the LST deformed with Lorentz violating and nonlocal couplings leads to the exact same kind of divergences which nonlocality by itself would have produced (quadratic and logarithmic divergences). The distinctive signature of Lorentz boost violation is that the coefficients of the quadratic and logarithmic divergences as well as the finite piece in complexity differ in the two frames related by a Lorentz boost. 

\section{Subregion volume complexity } \label{subregionC}
The volume complexity computed in the last section was unable to capture  the distinguishing signatures of lorentz violation form nonlocality as far as the type of UV divergences appearing was concerned. In the hope that the subregion complexity might have something more to tell us about the differences between signatures of nonlocality and Lorentz violation, in this section we explore the features of subregion volume complexity for a boundary subregion of size $L$. To this end we have to focus our attention to the portion of the maximal volume slice which is contained within the Ryu-Takayanagi (RT) surface (curve) homologous to the aforementioned boundary spacelike interval (of size $L$) following the prescription of \cite{Alishahiha:2015rta}.\\

First we briefly review the construction of the the RT surface in the dimensionally reduced $2+1$-d bulk gravity background, which is a codimension 2 surface in the bulk, homologous to the boundary subregion. The volume functional on a codimension-2 slice (in this case a curve) is obtained by looking at the constant time section of the $U=U(x)$ surface.
\begin{equation}
	\begin{split}
		ds^2&=\left(k l_s^2\frac{U'^2}{U^2}+h(U)(1-f(U)(\epsilon_++\epsilon_-)^2)\right)dx^2,
	\end{split}
\end{equation}
where, the prime denotes the derivative with respect to the parameter $x$ parameterizing the RT curve. In the string frame, this co-dimension-2 surface has the following volume functional, which in the present case turns out to be the length of the curve
{\footnotesize{\begin{equation}
			\begin{split}
				\int dx \, \mathcal{L}(U(x),U'(x),x)
				&=\frac{1}{U}\sqrt{k U'^2 l_s^2 \left(k \lambda  U^2+1\right) \left(k U^2 \lambda'+1\right)+k U^4 \left(k U^2\mu +1\right)}.
			\end{split}
\end{equation}}}
where, the primes over the quantities denotes their derivative with respect to $x$. Next, one has to minimize this string frame length functional to obtain the RT curve. However, instead follow the procedure of \cite{Chakraborty:2020udr} and start by analyzing the integrals of motion. 
The condition that lagrangian is independent of time, gives us the first integral of motion 
\begin{equation}
	\begin{split}
		c_1&=\frac{\partial \mathcal{L}}{\partial t'}\,~(=0),
	\end{split}
\end{equation}
Since the lagrangian is cyclic in parameter $x$, the corresponding ``hamiltonian" should be conserved:
\begin{equation}
	\begin{split}
		c_2&=U'\frac{\partial \mathcal{L}}{\partial U'}-\mathcal{L},\\
		&=\frac{-k U^3 \left(k U^2 \mu+1\right)}{\sqrt{k U'^2 l_s^2 \left(k \lambda  U^2+1\right) \left(k U^2 \lambda'+1\right)+k U^4 \left(k U^2 \mu+1\right)}},\\
		&=-\sqrt{k} U_0\sqrt{k   U_0^2\mu+1},
	\end{split}
\end{equation}
$c_2$ being a constant, we have used the boundary conditions at $x=0$ to evaluate $c_2$,  i.e. $U(0)=U_{0}$ and $U'(0)=0$ to evaluate it.\\

Solving for $U'(x)$ and choosing the positive root,
\begin{equation}
	\begin{split}
		U'(x)&=\frac{U^2 \sqrt{k \mu  U^2+1} \sqrt{k \mu  \left(U^4-U_0^4\right)+U^2-U_0^2}}{ U_0 l_s \sqrt{\left(k \lambda  U^2+1\right) \left(k \lambda'  U^2+1\right)} \sqrt{k \mu  U_0^2+1}}.
	\end{split}
\end{equation}
To obtain the subregion size we integrate the above equation by specifying the appropriate limits of integration.
\begin{align}
	\int_{0}^{x} dx'&=\int_{U_0}^{U}
	\frac{d\tilde{U}}{\tilde{U}'(x)}.
\end{align}
After the integration limits had been specified, the subregion size is given as the function of the turning point $U_0$ by:
\begin{align}
	L  &=\int_{U_0}^{\infty}
	dU\,	\frac{dU}{U'(x)} \label{L(U_0)} 
\end{align}
If we choose to look
at deep  linear dilatonic region, ($k\lambda U^2>>1$) we obtain a simplification:
\begin{align}
	L&=\frac{\pi\sqrt{ k \lambda\lambda' } l_s  }{2\sqrt{\mu}}+O\left(\frac{1}{U_0^2}\right).
\end{align}
We can analytically solve for $L$  only perturbatively, but the characteristic features of subregion length in the linear dilaton region are immediately obvious. 
As we move $U_0$ closer to the boundary, $L$ asymptotes to a constant value:
\begin{align}
	L_{c}&=\frac{\pi\sqrt{ k \lambda\lambda' } l_s  }{2\sqrt{\mu}}. \label{Lc}
\end{align} 
This behaviour is typical of the theory having a Hagedorn phase transition as had already been alluded to before in the literature \cite{Chakraborty:2020udr} in the context of the study of entanglement entropy using the $3+1$-d dual bulk background. (The critical length turns out to be the same).\\

We now perform some sanity checks by reproducing established results for different special cases from the general form equation (\ref{L(U_0)}) : 
\begin{itemize}
	\item\textbf{AdS (Case $\lambda=\epsilon_{\pm}=0)$:} The simplest of the all is the pure AdS geometry which has been the subject of an extensive study for which, the relation between the subregion length and the turning point is well known:  
	\begin{equation}
		\begin{split}
			L&=\frac{2l_s}{U_0}.	
		\end{split}
	\end{equation}
\item \textbf{ WAdS (Case $\lambda=\epsilon_+=0)$:}
The next case is when only the $J\overline{T}$ coupling is turned on. This case had also been thoroughly investigated and the gravity dual is warped AdS (WAdS) spacetime \cite{Azeyanagi:2012zd}.
\begin{align}
	\int_{0}^{L/2}dx^{'}&=\int_{U_{0}}^{\frac{l_s}{\epsilon}}dU\frac{U_{0}l_s\sqrt{1-kU_{0}^2\epsilon_-^2}}{U^2\sqrt{1-kU^2\epsilon_-^2}\sqrt{U^2-U_{0}^2+k\epsilon_-^2(-U^4+U_{0}^4)}}\label{4.11}\\
	\Rightarrow L&=\frac{2l_s}{U_{0}}+2kU_{0}l_s\epsilon_-^2\ln\left({\frac{2l_s}{U_{0}\epsilon}}\right)+O(\epsilon_-^4).
\end{align}
Upon turning off the coupling ($\epsilon_-\to 0$), one simply recovers the pure AdS result.
	\item\textbf{$\mathcal{M}_3$\,(Case $\epsilon_+=\epsilon_-=0)$:}
	When only $T\overline{T}$ coupling is turned on,
	\begin{equation}
		\begin{split}
			\int_{0}^{L/2} dx'&=\int_{U_0}^{l_s/\epsilon}
			dU\frac{U_0 l_s \sqrt{k \lambda  U^2+1} \sqrt{k \lambda  U_0^2+1}}{U^2 \sqrt{\left(U^2-U_0^2\right) \left(k \lambda  \left(U^2+U_0^2\right)+1\right)}},\\
			L	&=\frac{1}{2} \pi  \sqrt{k \lambda } l_s+O\left(\frac{1}{U_0^2}\right),
		\end{split}
	\end{equation}
	we recover the result already encountered earlier in \cite{Chakraborty:2018kpr}.
	
	Alternatively, treating  coupling as the perturbation parameter,
	\begin{equation}
		\begin{split}
			L	&= \frac{2l_s}{U_0}+ k^2 \lambda ^2 U_0^3 l_s \ln \left(\frac{2 l_s}{U_0 \epsilon }\right)+O(\lambda^3).
		\end{split}
	\end{equation}
	With $\lambda\to0$, we again recover the AdS result.
\end{itemize}

Thus we have successfully reproduced the features of the RT curves for the special cases of the 	pure AdS, warped AdS and the $\mathcal{M}_3$ from our general formula relating the RT curve turning point and the subregion length (\ref{L(U_0)}). We will use this relation to obtain the expression for the subregion volume complexity next. This will be done numerically, since the analytic expressions can only be obtained perturbatively. Since we are not interested in perturbative answers, we will do this exactly but numerically. Just as we have done for the RT curve, before presenting the final results for the general case, we first perform sanity checks by studying various special cases where the effects of locality and Lorentz violation are removed and comparing those expressions to existing results in the literature obtained in the contexts where the boundary dual is a local CFT$_2$, instead of an LST$_2$. 
\subsection{Subregion volume (complexity) for $\lambda=\epsilon_{\pm}=0$: Poincare patch of AdS$_3$ } \label{AdS2 subregion complexity}
The first check is the maximal volume corresponding to the subregion size  ($\mathcal{V}(L)$) for the simplest of the cases - pure $AdS_{2+1}$. This can be evaluated analytically exactly. The induced metric on a codimension-$1$ submanifold of a constant time  slice is
\begin{equation}\begin{split}
		ds^2&=\frac{kl_s^2}{z^2}(dz^2+dx^2).
	\end{split}
\end{equation}
Then subregion volume is, \begin{equation}
	\begin{split}
		\mathcal{V}(L)&=\int^{l_s/\epsilon}_{0} dz		\int_{0}^{x(z)} dx' e^{-2(\Phi(z)-\Phi_{\infty})}\sqrt{\gamma},\\
		&=kl_s^2\left(\frac{L}{2 \epsilon }-\frac{\pi }{2}\right).
	\end{split}
\end{equation}
The linear UV divergence is expected of any lorentz covariant local $CFT$ in one spatial dimension. This is a well known result \cite{Alishahiha:2015rta}.

\subsection{Subregion volume complexity $\epsilon_{\pm}=0$: $T\overline{T}$ deformation or $\mathcal{M}_3$}
The next case  we treat is a new result, although appropriately it belongs to the subject matter of the preceding work \cite{Chakraborty:2020fpt} on pure $\mathcal{M}_3$ complexity. Since the subregion complexity calculation was omitted there, for the sake of completeness, we reproduce here the corresponding result for subregion complexity.
In this case, the pullback of the ambient metric on codimension-$1$ surface bounded by the RT curve for the pure $\mathcal{M}_3$ case is:
\begin{equation}
	\begin{split}
		ds^2&=k l_s^2\frac{dU^2}{U^2}+h(U)dx^2.
	\end{split}
\end{equation}
The volume corresponding to this subregion of the (maximal volume slice) as the function of the turning point $U_0$ is given by
\begin{equation}
	\begin{split}
		\mathcal{V}&=	\int_{U_0}^{l_s/\epsilon} dU		\int_{0}^{x(U)} dx \,e^{-2(\Phi(U)-\Phi_{0})}\sqrt{\gamma},\\
		&=kl_s\int_{U_0}^{l_s/\epsilon} dU	\sqrt{1+k\lambda U^2}\int_{U_0}^{U}\frac{d\tilde{U}}{\tilde{U}'(x)}.
	\end{split}
\end{equation}
We have to eliminate $U_0$ in favor of $L$ to express the maximal volume in terms of subregion length. However if we insist on analytical expression, then the inversion can only be achieved iteratively or perturbatively. In order to not compromise on precision, we instead perform the calculations numerically to illustrate the quantitative features of the subregion complexity.\\
\begin{figure}
     \centering
     \begin{subfigure}[b]{0.3\textwidth}
         \centering
         \includegraphics[width=\textwidth]{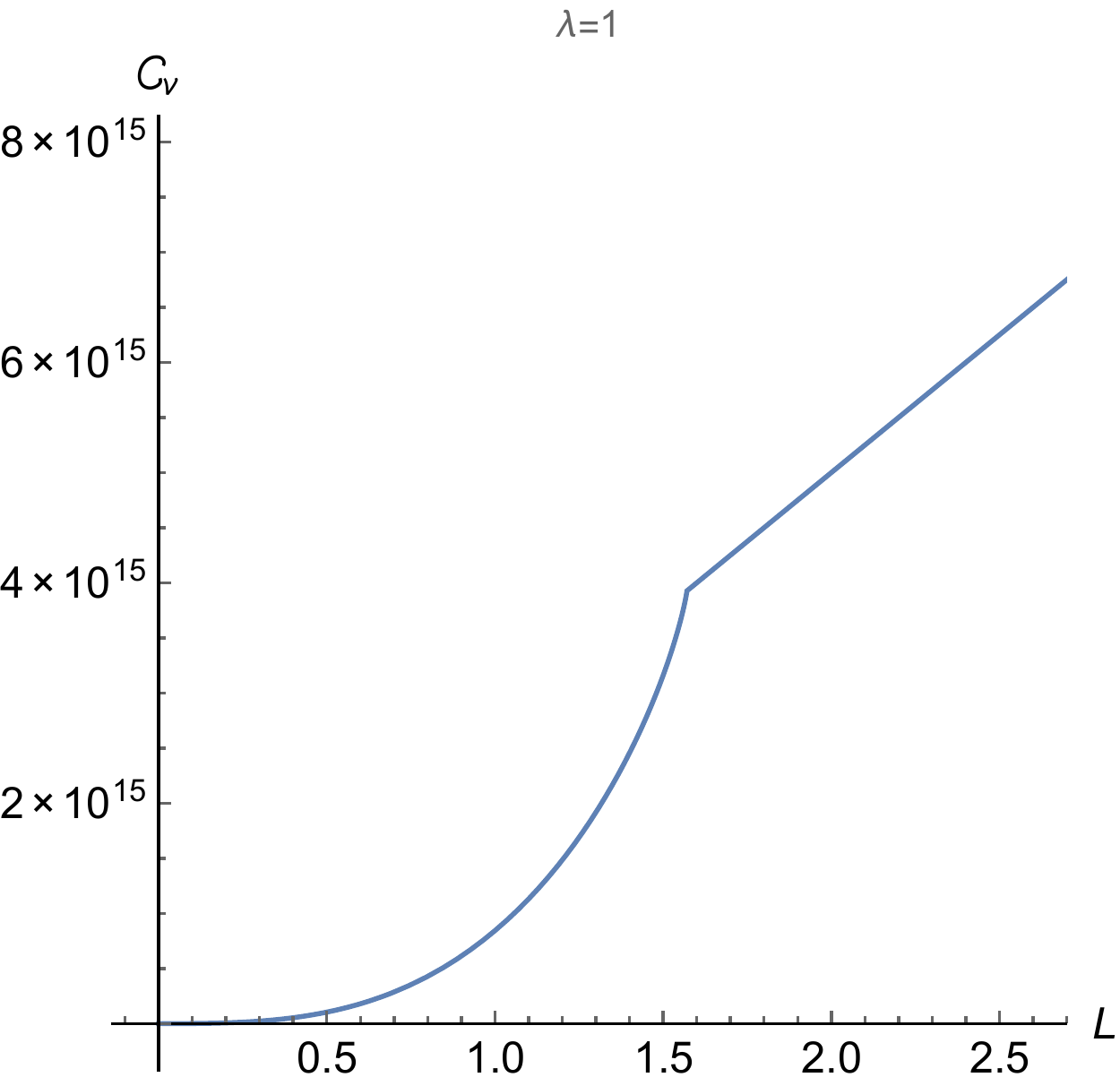}
         \caption{$\mathcal{C}_V\,vs\, L$ \,for \,$\lambda=1$}
         \label{fig:M3 case}
     \end{subfigure}
     \hfill
     \begin{subfigure}[b]{0.3\textwidth}
         \centering
         \includegraphics[width=\textwidth]{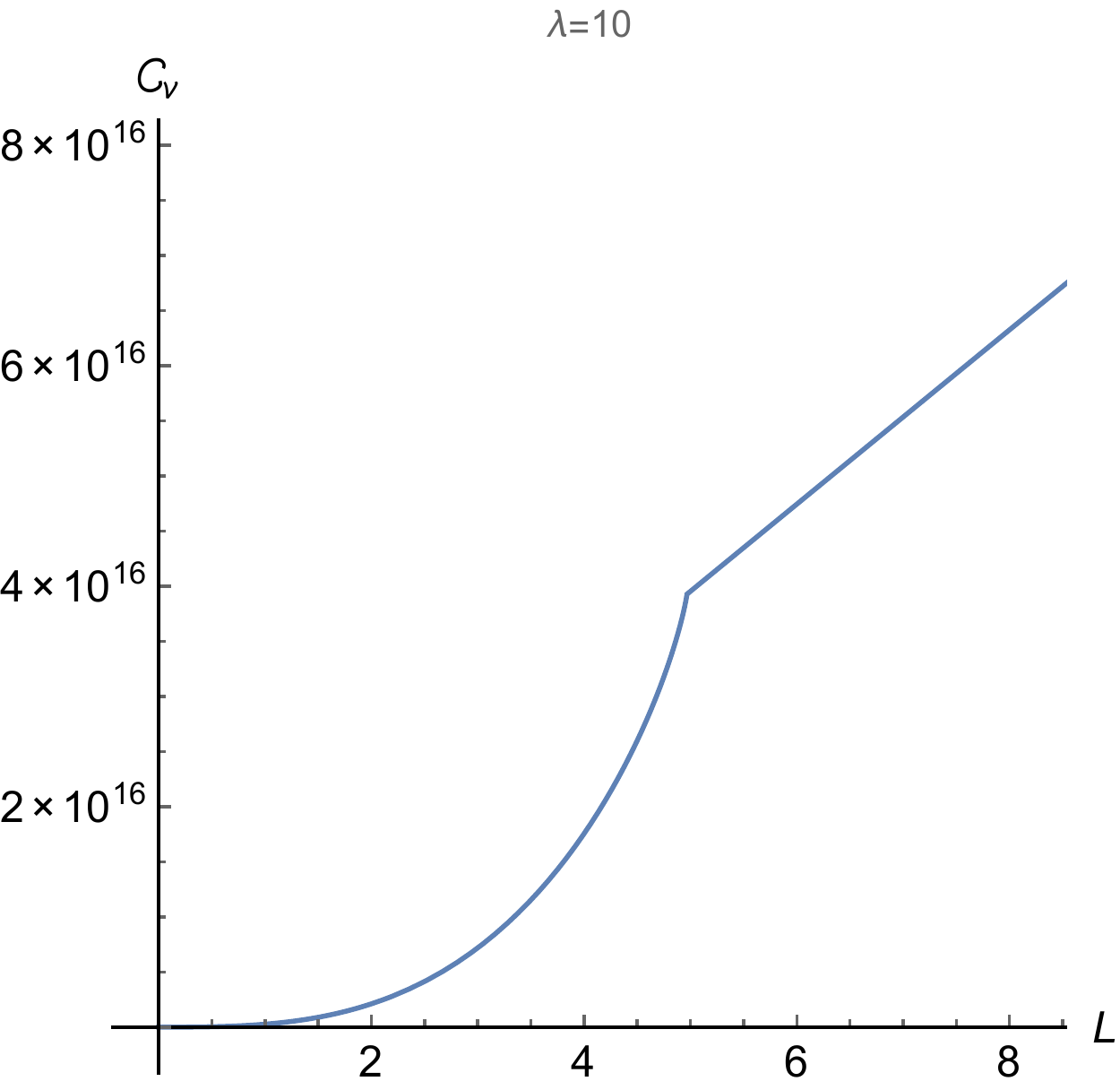}
         \caption{$\mathcal{C}_V\,vs\, L$ \,for \,$\lambda=10$}
         \label{fig:three sin x}
     \end{subfigure}
     \hfill
     \begin{subfigure}[b]{0.3\textwidth}
         \centering
         \includegraphics[width=\textwidth]{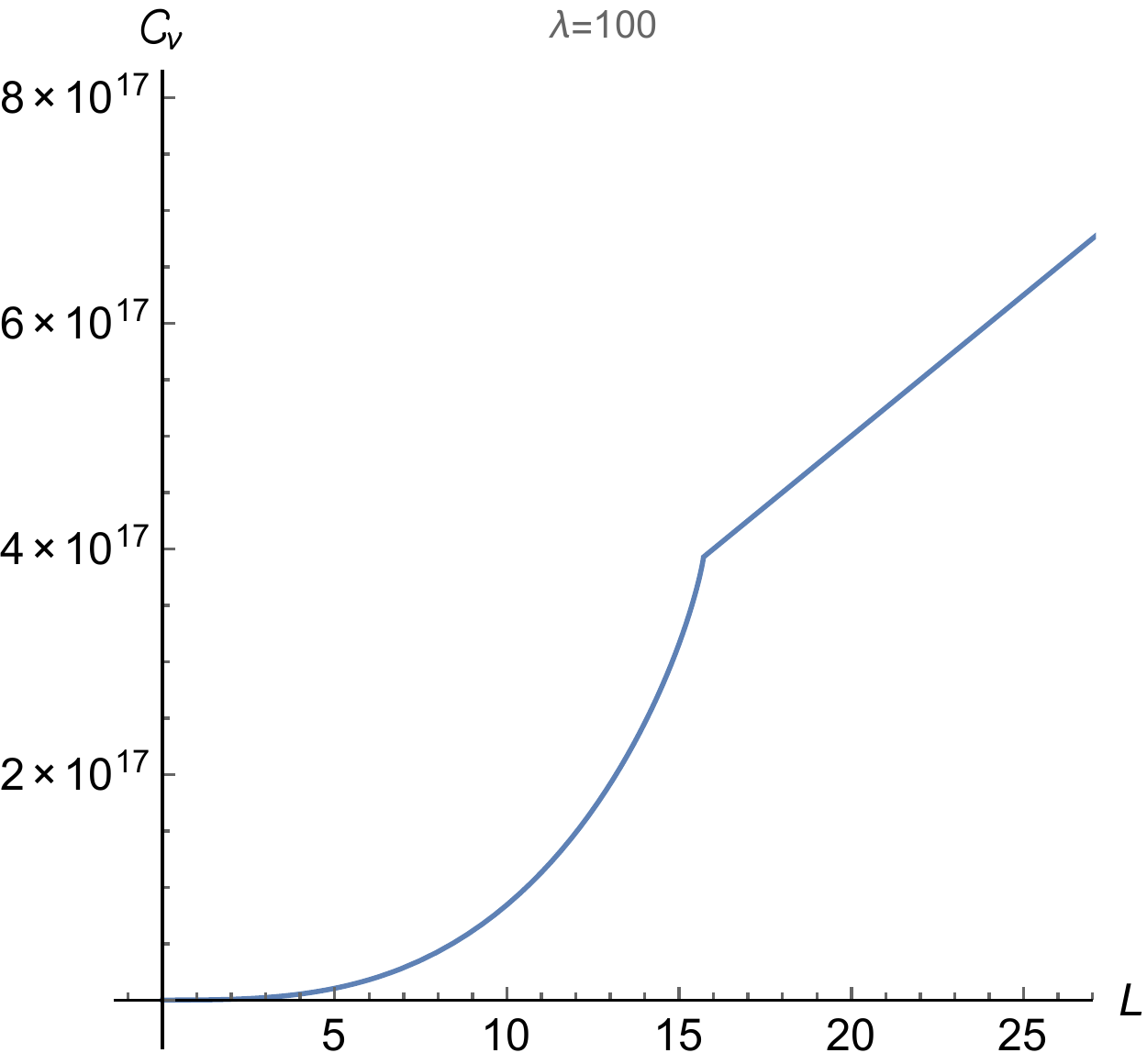}
         \caption{$\mathcal{C}_V\,vs\, L$ \,for \,$\lambda=100$}
         \label{fig:five over x}
     \end{subfigure}
        \caption{Subregion volume complexity ($\mathcal{C}_V$) vs. subregion size ($L$) graphs for $T\overline{T}$ deformed CFT$_2$ (LST) for different values of the deformation parameter ($T\overline{T}$ coupling) $\lambda$ exhibiting Hagedorn phase transition. The critical suregion size at the transition point increases monotonically with $\lambda$.}
        \label{fig:three graphs}
\end{figure}

In Fig. \ref{fig:three graphs} we present the numerical plots demonstrating the dependence of complexity (modulo the factor of $8\pi G_N \sqrt{k}l_s$) on the subregion length $L$ for three different values (differing orders of magnitude) of $\lambda$, the $T\overline{T}$ deformation coupling parameter.  All plots display the following universal traits 
\begin{itemize}
       \begin{item}
       The subregion volume complexity is a monotonically increasing function of the subregion size.
       \end{item}
     
       \begin{item}
       The subregion volume complexity undergoes a sharpe (phase) transition as the subregion size is increased beyond a certain critical size as depicted by the presence of a kink in each of the plots.
       \end{item}       
       
       \begin{item}
       Once the subregion size is larger than the critical subregion (kink), the subregion volume complexity grows linearly with subregion size. This is characteristic of the AdS geometry as pointed out in the previous subsection \ref{AdS2 subregion complexity}. The RT curve extends deep into bulk where the geometry is close to AdS$_3$. 
       \end{item}
       
       \begin{item}
       The parabolic portion of the curve, for subregion size (length) is less than the critical length, pertains to the linear dilaton region because that is where the subregion size slowly approaches to a constant value $L_c$ regardless of the position of the turning point $U_0$ of the RT curve. The RT curve here remains stuck mostly in the deep UV region i.e. near the boundary.
       \end{item}       
\end{itemize}
The linear growth of the complexity with the subregion size when the subregion size is larger then the Hagedorn scale (see next section for more details) is plausible because in this situation the LST behaves more like a local CFT and the number of degrees of freedom in the Lorentz covariant local theory can be thought of as uniformly distributed along the boundary subregion. The kink signifies the termination of the linear dilaton geometry and bulk being subsequently taken over by the AdS geometry. It will turn out that same universal features will emerge when we introduce Lorentz violating effects in the system i.e. when the couplings $\epsilon_\pm$ are nonzero. In order to avoid repetition, we will leave further quantitative discussion to the next section where we tackle the case when Lorentz violating effects are turned on. \\

\subsection{Subregion volume complexity for nonzero $\lambda, \epsilon_{\pm}$: $ T\bar{T}$, $ J\bar{T}$ and $ \bar{J}T$ LST }\label{subregionC in stationary}
Armed with the hints and insights from the previous sections for the various subcases (i.e.  pure $AdS$ and $\mathcal{M}_3$), we are now ready to tackle the most general case with both the locality violating, and Lorentz violating couplings turned on and obtain the general characteristics for the subregion volume complexity. As was done in the previous section, we first record the pullback of the ambient metric on codimension two maximal volume surface, namely the constant $t$ surface, bounded by  RT curve for the general case of the nonlocal as well as both of the Lorentz violating couplings turned on:
\begin{equation}
	ds^2=k l_s^2\frac{dU^2}{U^2}+h(U)(1-f(U)(\epsilon_++\epsilon_-)^2)dx^2.
\end{equation}
The maximal volume arising from the above geometry is given by:
\begin{equation}
	\begin{split}
		\mathcal{V}&=	\int_{U_0}^{l_s/\epsilon} dU		\int_{0}^{x(U)} dx e^{-2(\Phi(U)-\Phi_{0})}\sqrt{\gamma},\\	&=kl_s\int_{U_0}^{l_s/\epsilon} dU\sqrt{  1+k\mu U^2}\int_{U_0}^{U}\frac{d\tilde{U}}{\tilde{U'}(x)}.
	\end{split}
\end{equation}
For the reasons alluded to earlier in the previous subsection, we again opt for a numerical approach to uncover the features of subregion complexity. The corresponding plots for $\mathcal{V}\,\, vs\,\, L$ (note that modulo the factor of $8\pi G_N \sqrt{k}l_s$, the complexity $\mathcal{C}_V$ and maximal slice volume $\mathcal{V}$ are the same) for various values of the Lorentz violating couplings against the fixed value of $\lambda=170$, are appended below in Fig. \ref{fig:six stationary graphs}. In all the plots we notice some features which are in common with the $\mathcal{M}_3$ ($T\overline{T}$ deformation) set up, namely

\begin{itemize}
          \begin{item}
          The subregion volume complexity is a monotonically increasing function of the subregion size 
          \end{item}
          \begin{item}
          The subregion complexity undergoes a phase transition as the subregion size is varied. For small subregions, we encounter a parabolic growth up to a kink at some critical subregion length, followed by the linear growth which is characteristic of dual AdS geometry in the deep IR. The parabolic region of the curve corresponds to the UV region i.e. linear dilaton geometry because that is where, the subregion length slowly approaches zero.
          \end{item}
          \begin{item}
          For fixed $\lambda$ (i.e. nonlocality scale held fixed), the critical subregion size at the phase transition point in the plots increases (shifts rightwards) as the Lorentz-violating coupling $\epsilon_+$ ($\epsilon_{-}$) is increased. Interestingly the critical subregion size changes (increases) even if \emph{just one} of the couplings $\epsilon_+$ is made nonzero. We will keep this in mind when we are looking at the static frame complexity where it will turn out that the critical subregion size is a function of the product $\epsilon_{+} \epsilon_{-}$.
          \end{item}
\end{itemize} 
In order to facilitate comparisons, its convenient to scale all the diagrams in a single plot by plotting the logarithms of complexity (modulo the factor of $8\pi G_N \sqrt{k}l_s$) and the subregion size $L$. From the graphs, one can directly appreciate the appearance of the transition point $L_c$ where the complexity characteristics transitions sharply from parabolic to the AdS like linear dependence. A phase transition in the holographic entanglement entropy as a function of subregion size for this same system has been shown in \cite{Chakraborty:2020udr}. However, what is interesting is that, complexity not only undergoes an analogous phase transition, but that the subregion complexity phase transition \textbf{\emph{occurs at the exact same critical subregion length as that of the entanglement entropy phase transition}}, as evident from table \ref{table for EE and CV points} displaying the numerical value of the critical length extracted from the plots and the theoretical expression for the critical subregion length for entanglement entropy (refer to Eq. (\ref{Lc})).\\
\begin{figure}[H]
  \begin{subfigure}{7cm}
    \centering\includegraphics[width=7cm]{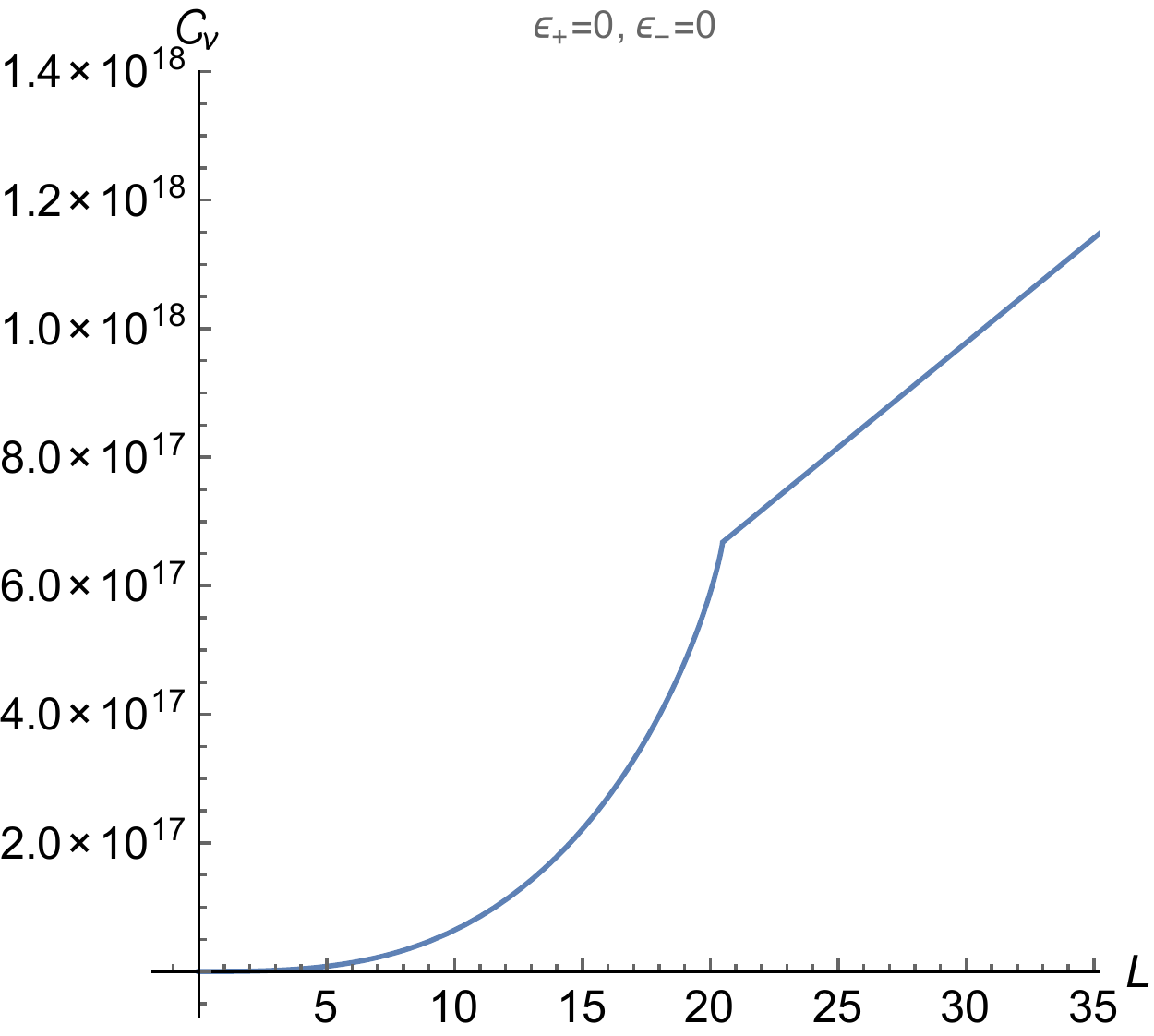}
    \caption{$\lambda = 170, \epsilon_{\pm}=0$}
  \end{subfigure}
  \begin{subfigure}{7cm}
    \centering\includegraphics[width=7cm]{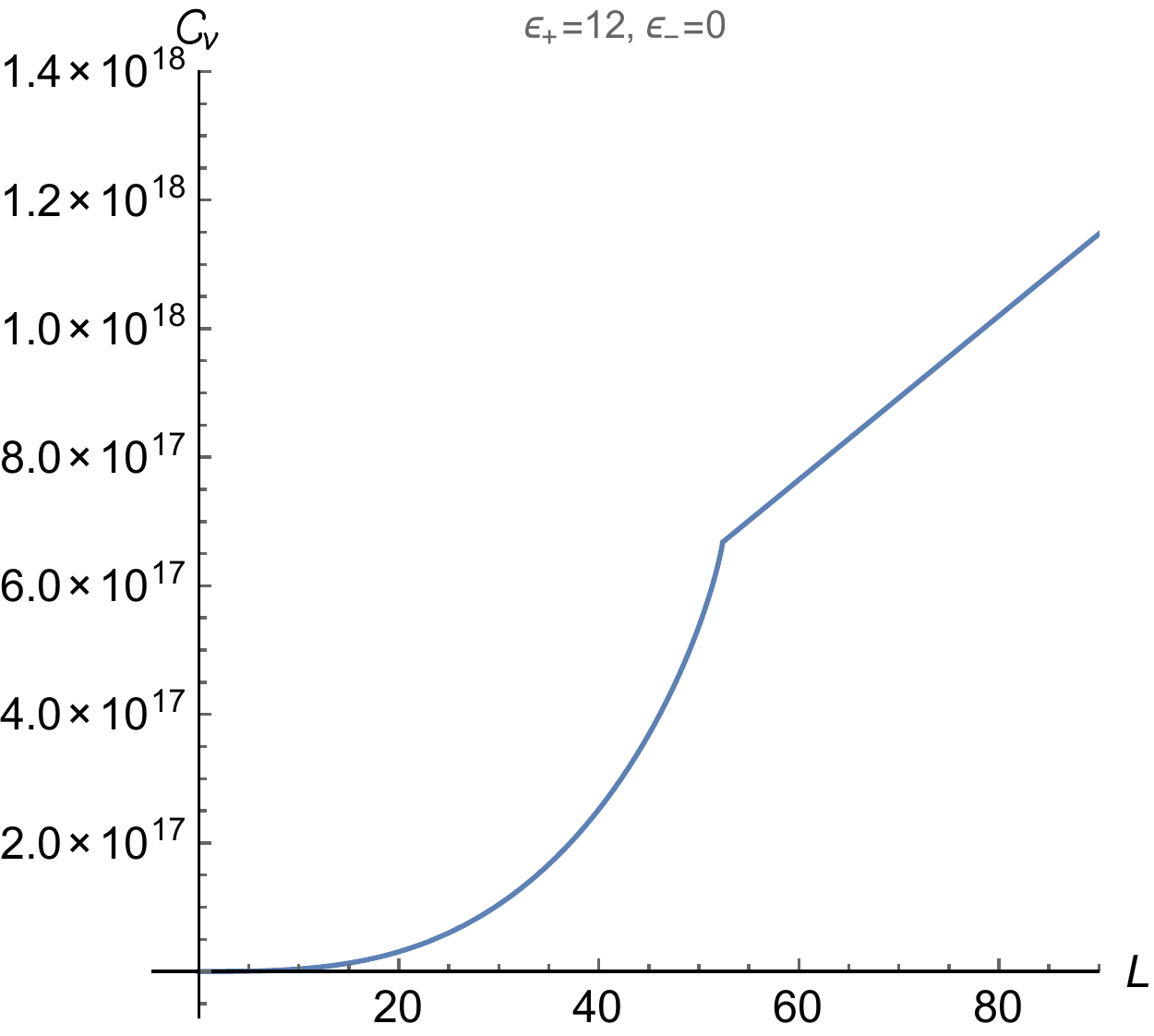}
    \caption{$\lambda = 170$, $\epsilon_+=12, \epsilon_{-}=0$}
  \end{subfigure}
 \vspace{0.5cm}
 \\
  \begin{subfigure}{7cm}
    \centering\includegraphics[width=7cm]{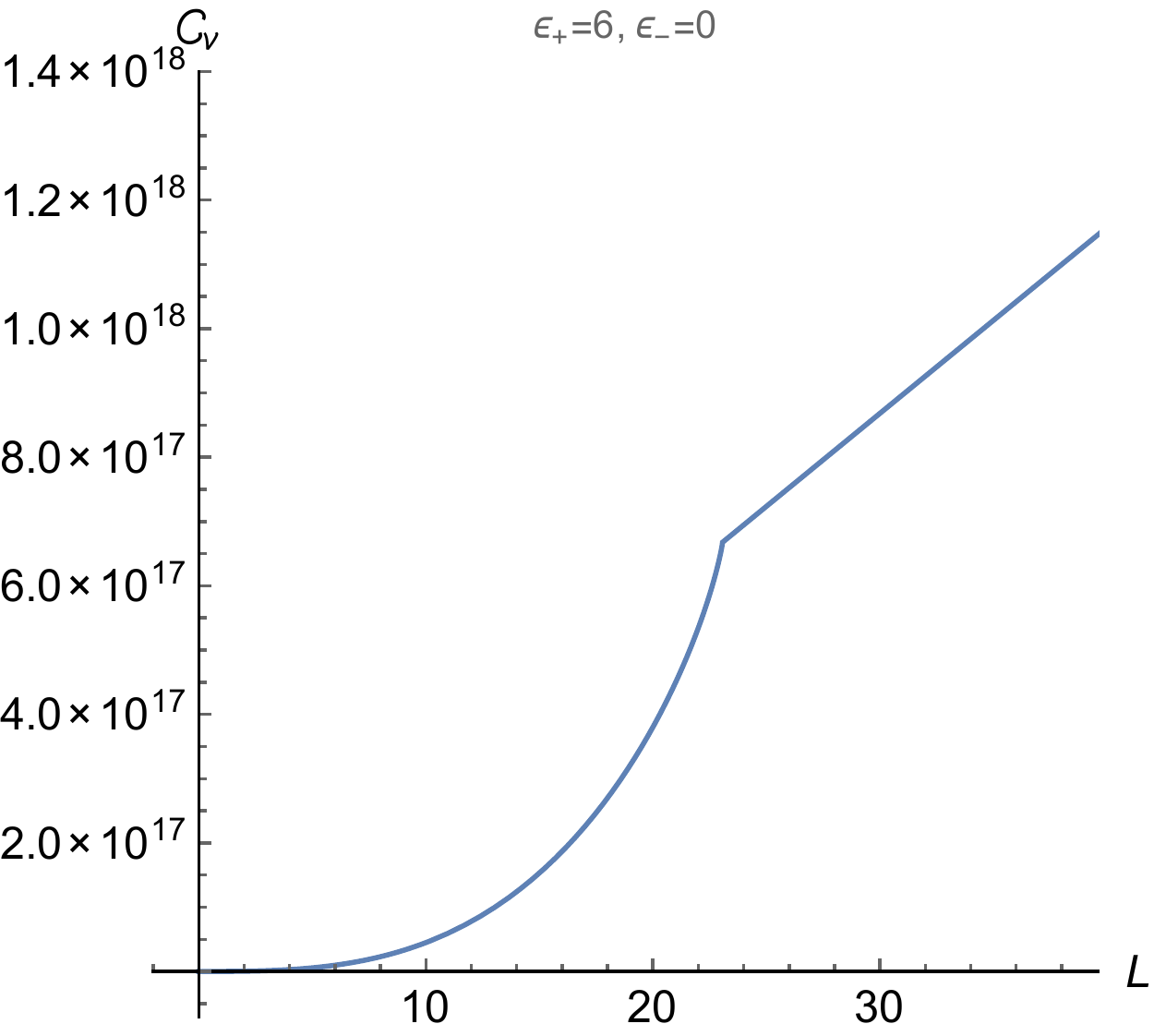}
  \caption{$\lambda = 170$, $\epsilon_+=6, \epsilon_{-}=0$}  \end{subfigure}
  \begin{subfigure}{7cm}
    \centering\includegraphics[width=7cm]{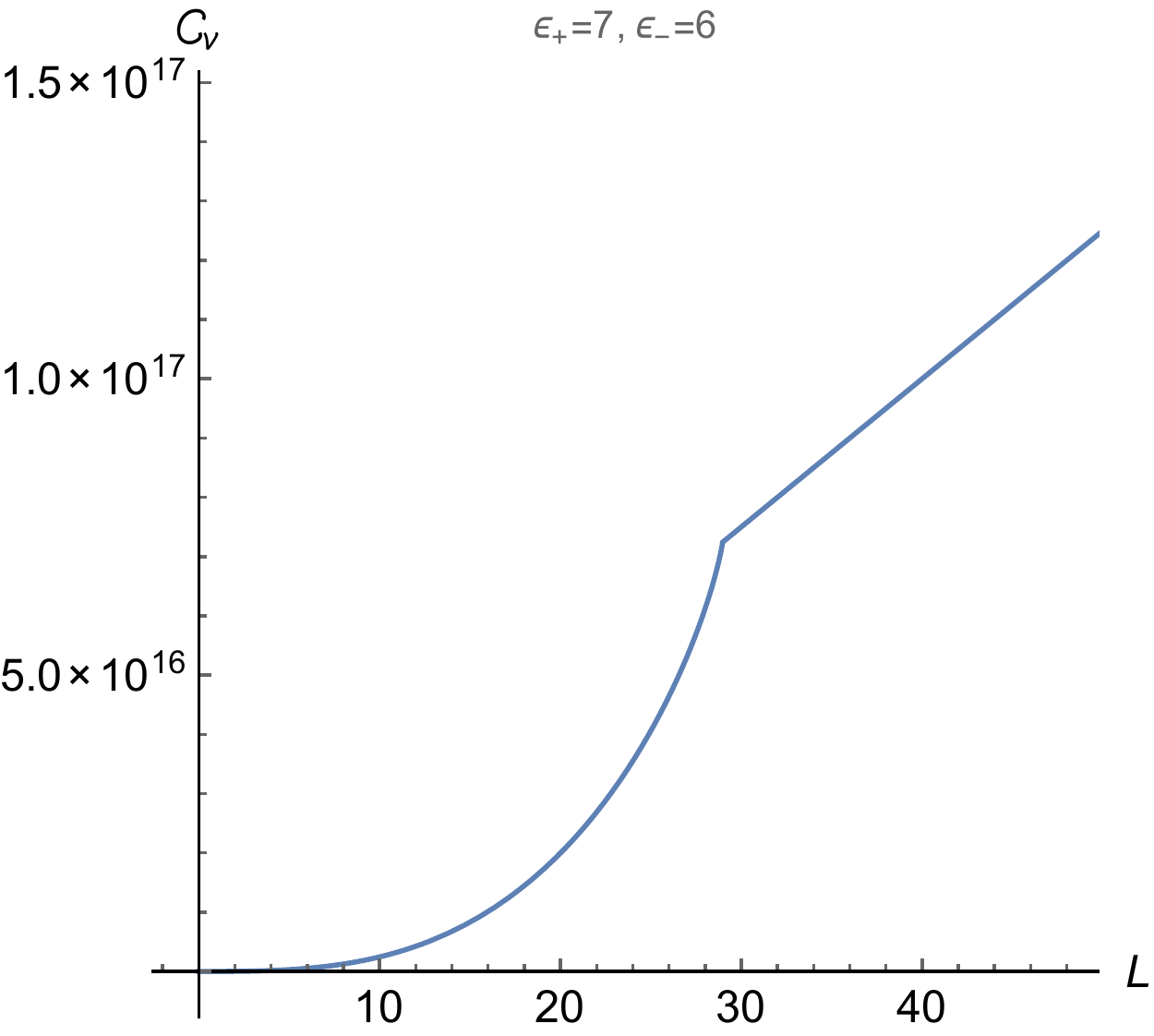}
  \caption{$\lambda = 170$, $\epsilon_+=7, \epsilon_{-}=6$}  \end{subfigure}
 \vspace{0.5cm}
 \\
    \begin{subfigure}{7cm}
    \centering\includegraphics[width=7cm]{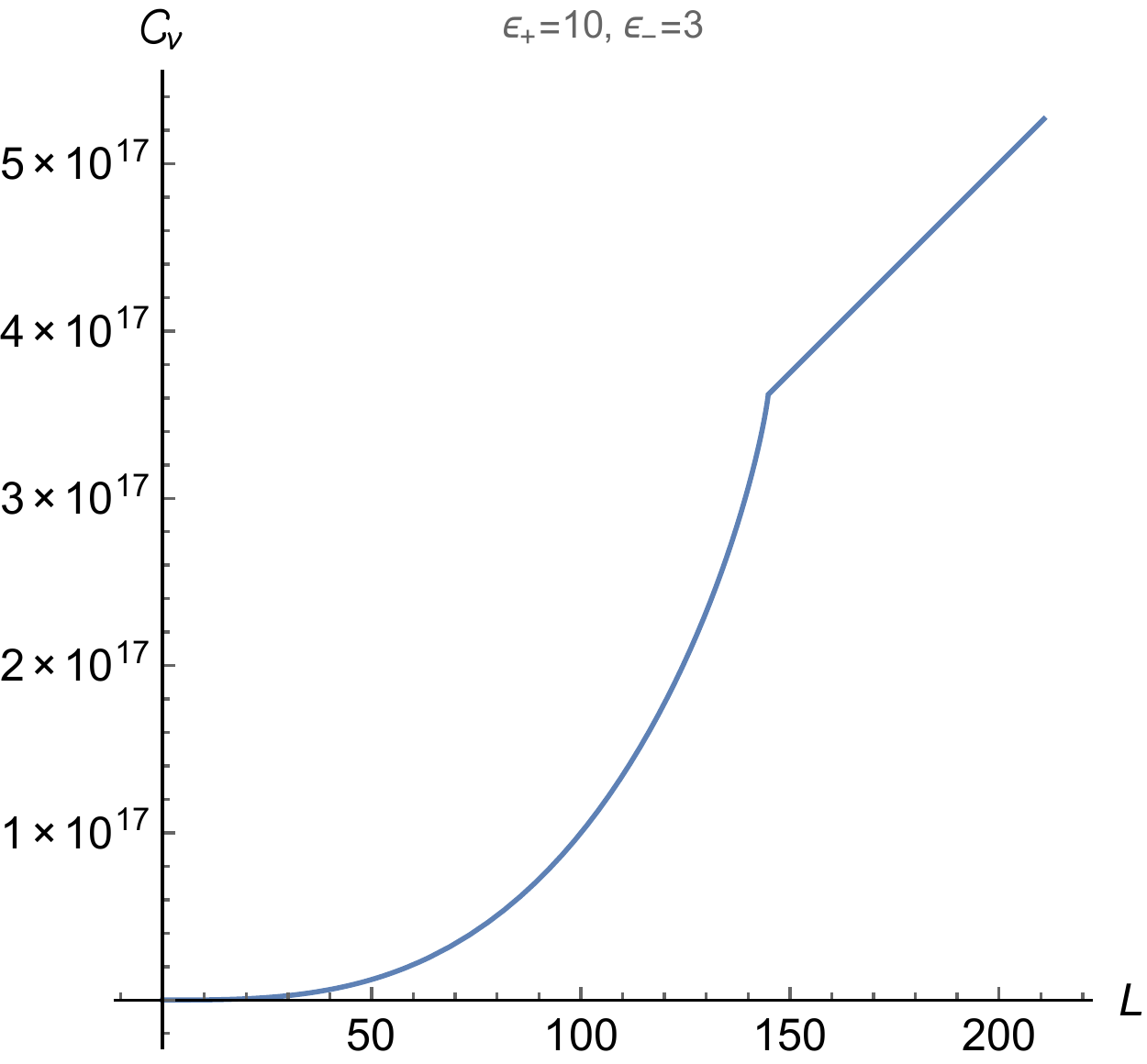}
  \caption{$\lambda = 170$, $\epsilon_+=10, \epsilon_{-}=3$}  \end{subfigure}
  \begin{subfigure}{7cm}
    \centering\includegraphics[width=7cm]{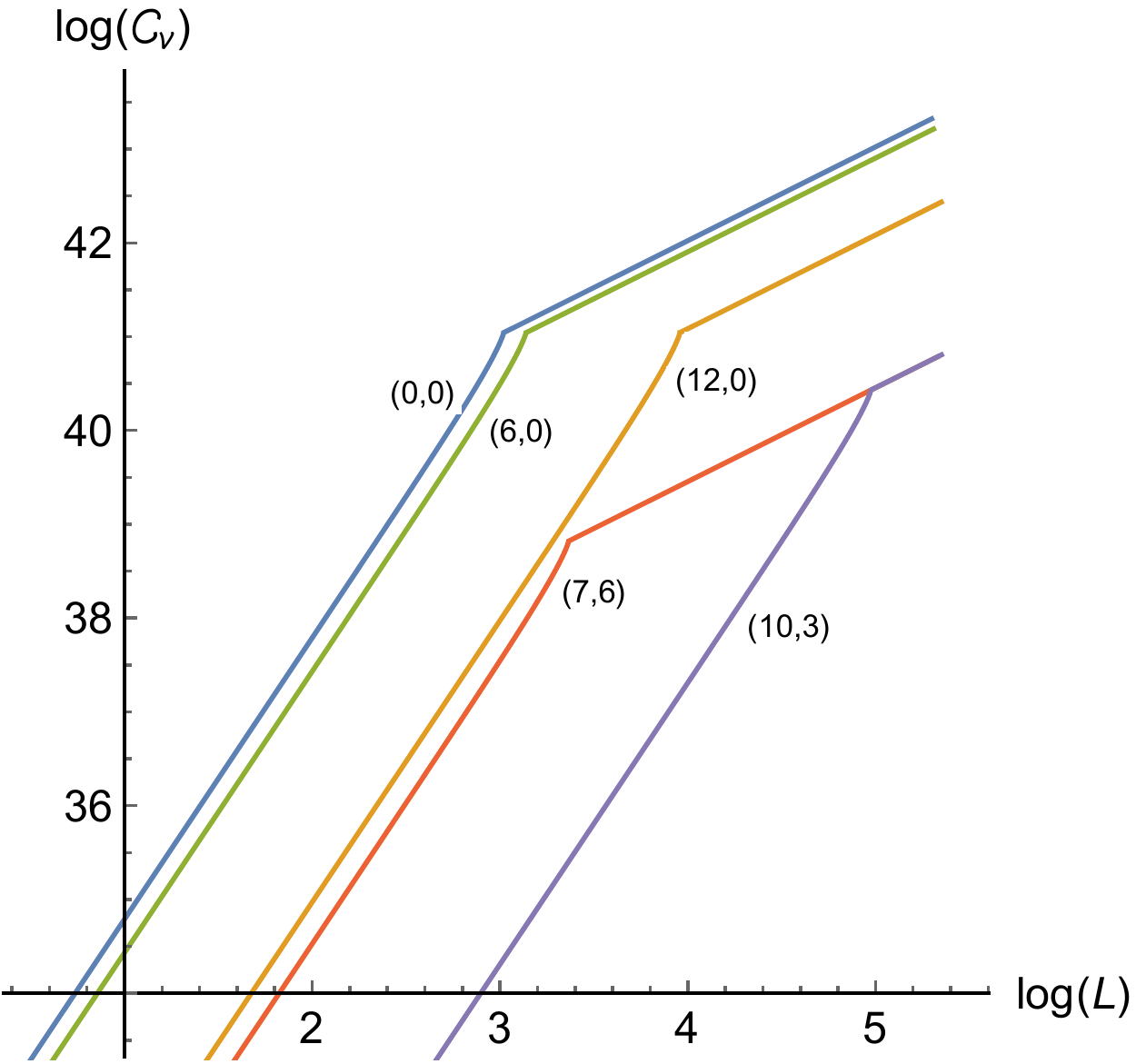}
  \caption{$\ln \mathcal{C}_V$ vs. $\ln L$ \,for $\lambda = 170$}   \end{subfigure}
  \caption{Subregion volume complexity ($\mathcal{C}_V$) vs. subregion size ($L$) graphs for $T\overline{T}$, $J\overline{T}$ \& $\overline{J}T$ deformed CFT$_2$ (LST) for fixed values of the deformation parameter ($T\overline{T}$ coupling) $\lambda =170$ exhibiting Hagedorn phase transition. The last plot is a log-log graph clearly displaying the scaling exponents (slopes).} 
        \label{fig:six stationary graphs}
\end{figure}
\begin{table}[h]
\centering
\begin{tabular}{ |p{4cm}|p{4cm}|p{4cm}|  }
	\hline
	\multicolumn{3}{|c|}{Critical subregion size for complexity for $\lambda=170$} \\
	\hline
	Lorentz violating couplings	($\epsilon_+$,$\epsilon_-$)& Entanglement entropy transition point length $L_c$ (in units of AdS radius) & Subregion complexity transition point from graphs (in units of AdS radius)\\
	\hline
	(0,0)& 20.48    &20.47\\
	(6,0)&   23.06  & 23.05 \\
	(12,0) &52.37 & 52.33\\
	(7,6)   &28.96 &28.94\\
	(10,3)&  144.82  & 144.9\\
	(13,0)& 267.03  & 266.84  \\
	\hline
\end{tabular}
\caption{Table comparing the critical subregion size for phase transition in Entanglement entropy from theory and the critical subregion size for subregion volume complexity extracted from the plots} \label{table for EE and CV points}
\end{table}
This fact that the subregion complexity undergoes a phase transition for the exact same the transition point (critical region size) as entanglement entropy, lends credence to the claim that complexity is a very effective physical observable (perhaps more useful that entanglement entropy) capable of detecting phase transitions (in the present case the Hagedorn phase transition) which perhaps cannot always be captured by usual field theory probes such as correlation functions of local operators.\\
\subsection{Subregion volume complexity in static frame} \label{subregionC in static}
\begin{figure}[H]
  \begin{subfigure}{7cm}
    \centering\includegraphics[width=7cm]{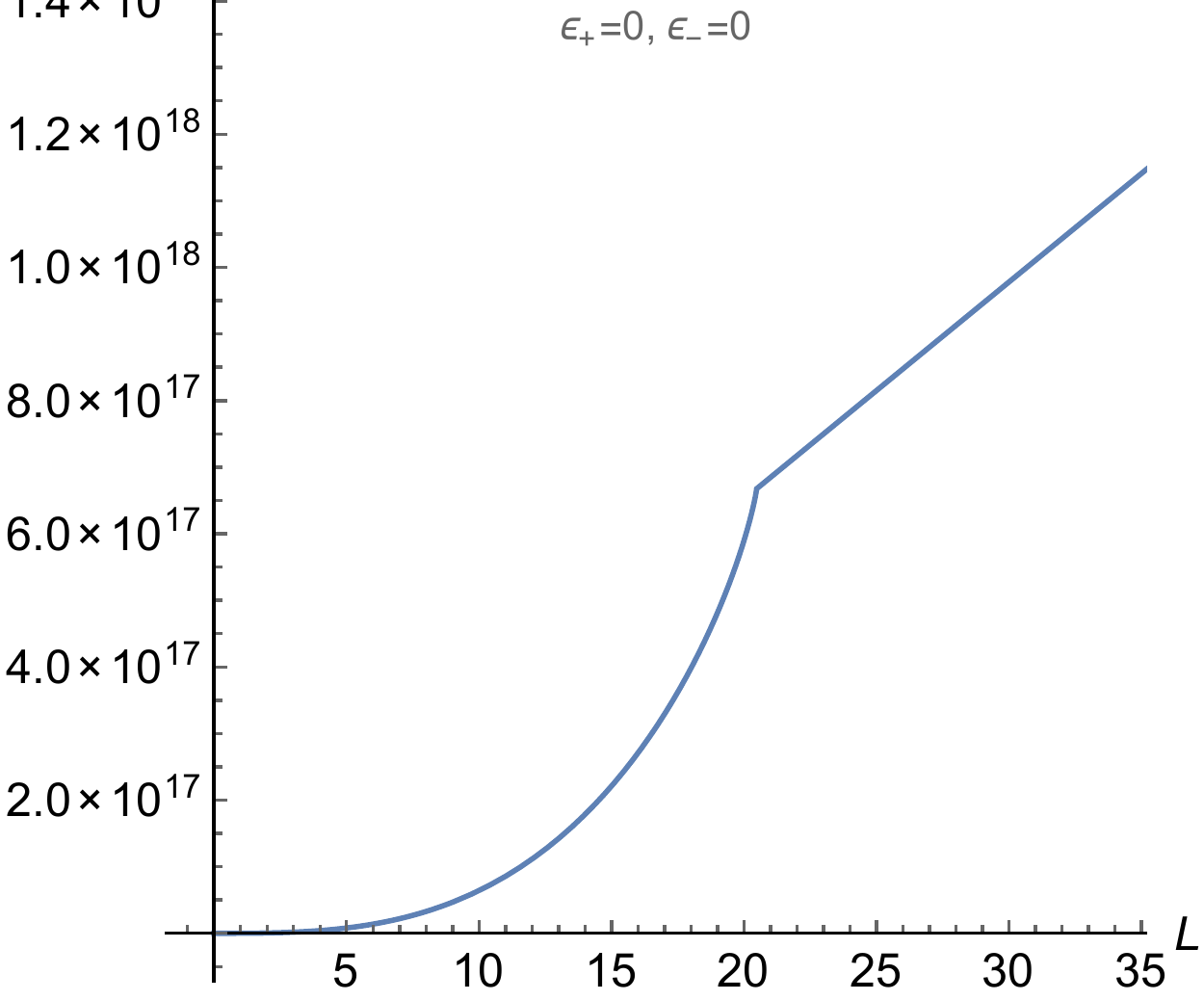}
    \caption{$\lambda = 170, \epsilon_{\pm}=0$}
  \end{subfigure}
  \begin{subfigure}{7cm}
    \centering\includegraphics[width=7cm]{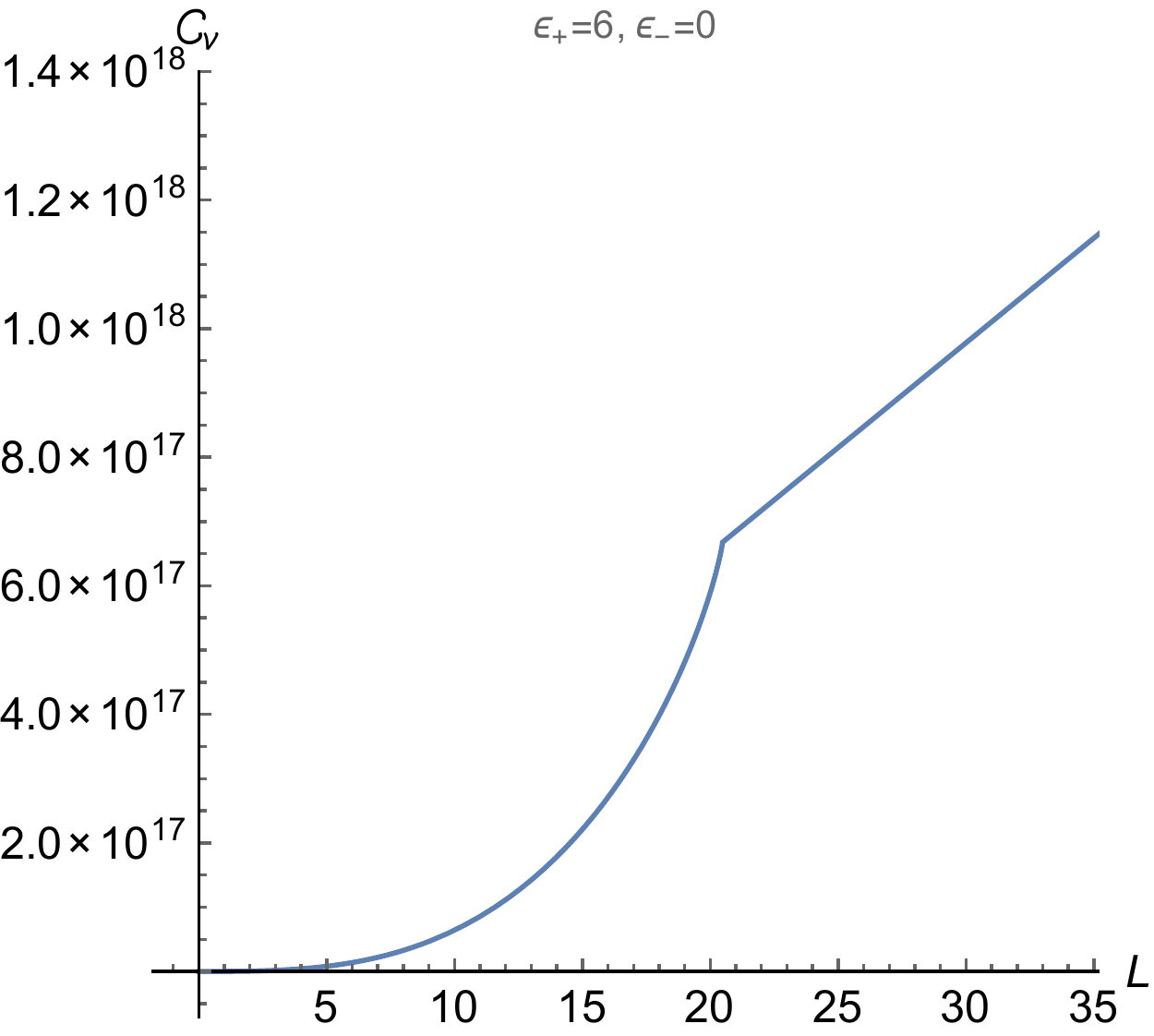}
    \caption{$\lambda = 170$, $\epsilon_+=6, \epsilon_{-}=0$}
  \end{subfigure}
 \vspace{0.5cm}
 \\
  \begin{subfigure}{7cm}
    \centering\includegraphics[width=7cm]{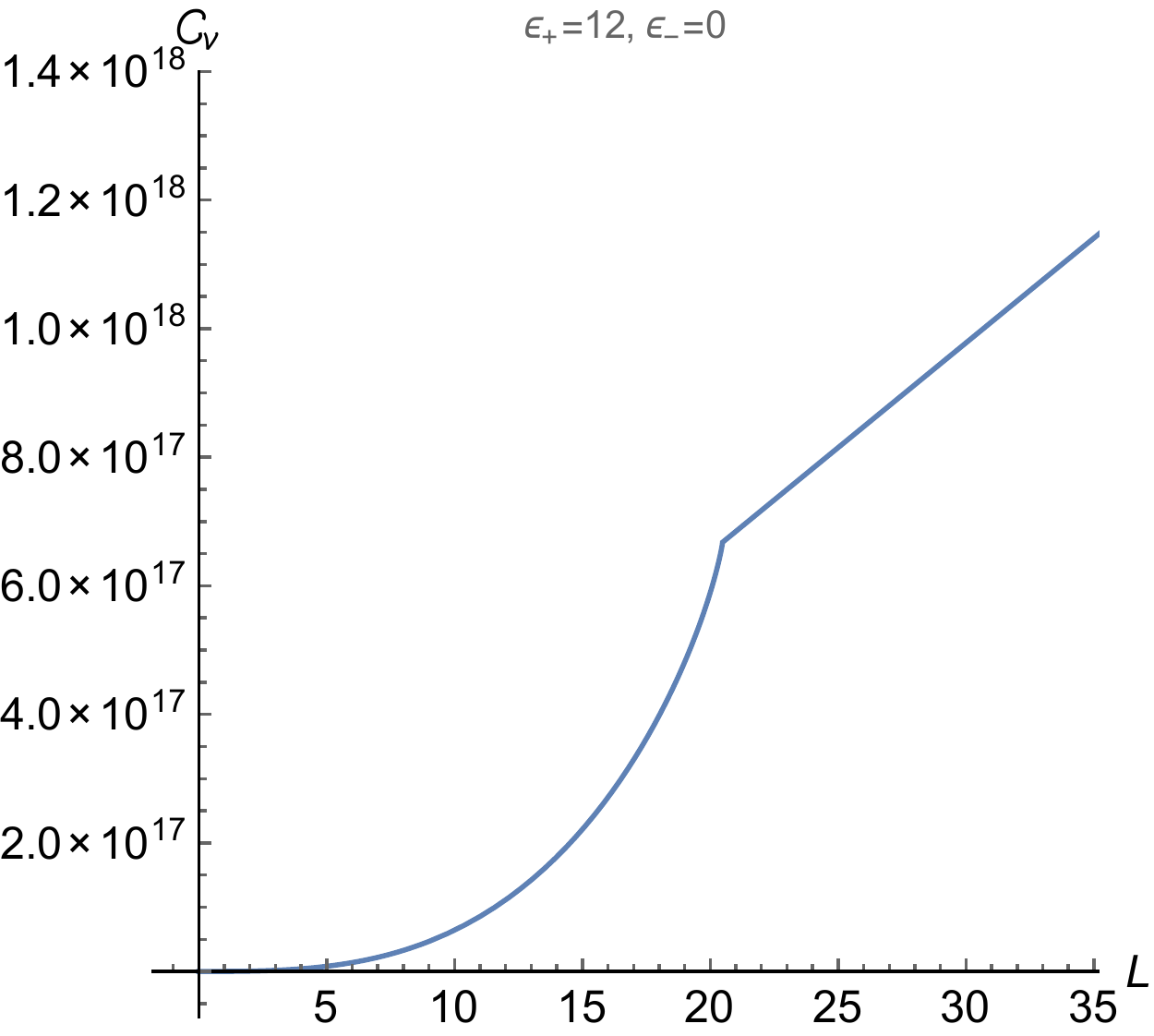}
  \caption{$\lambda = 170$, $\epsilon_+=12, \epsilon_{-}=0$}  \end{subfigure}
  \begin{subfigure}{7cm}
    \centering\includegraphics[width=7cm]{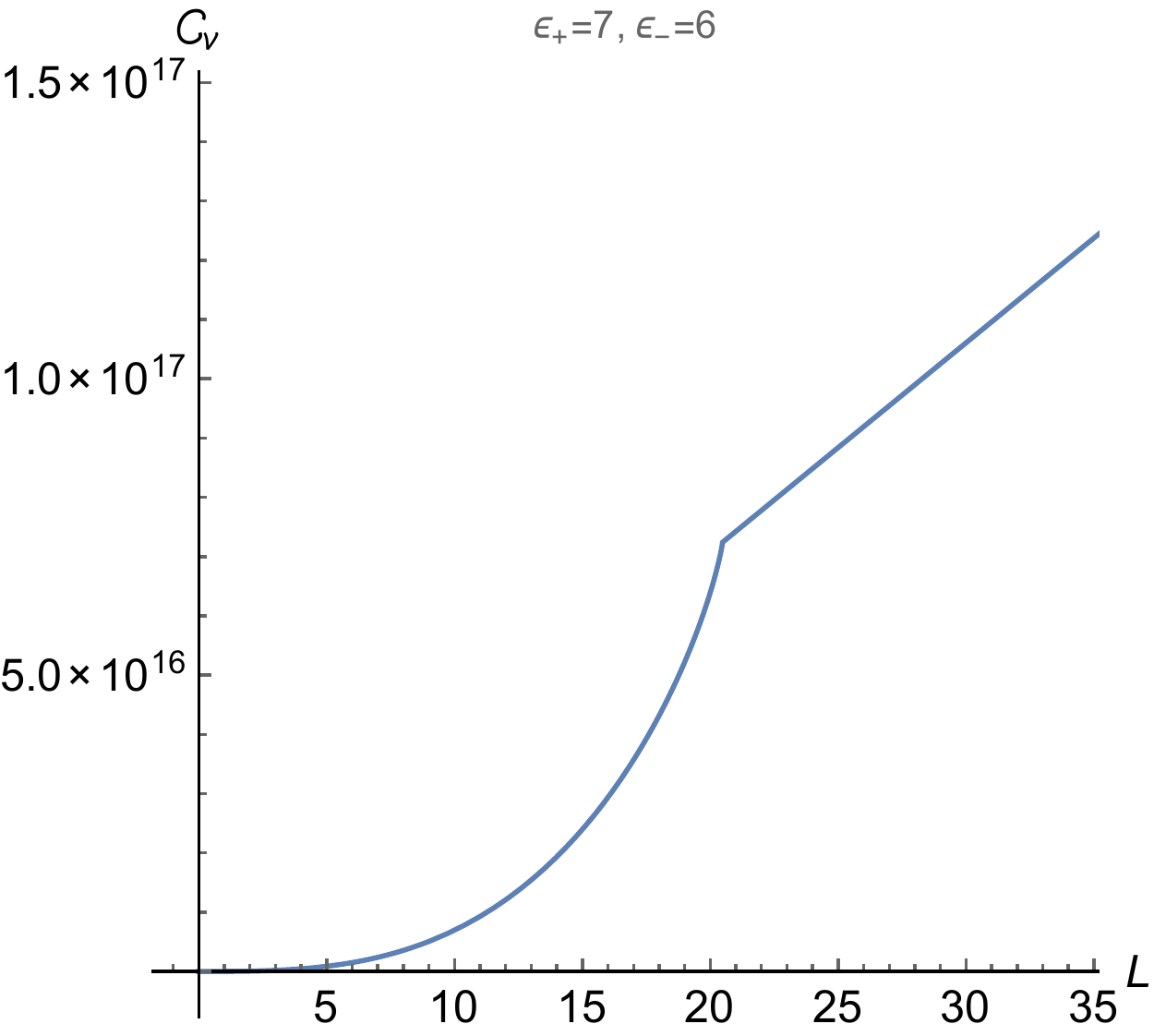}
  \caption{$\lambda = 170$, $\epsilon_+=7, \epsilon_{-}=6$}  \end{subfigure}
 \vspace{0.5cm}
 \\
    \begin{subfigure}{7cm}
    \centering\includegraphics[width=7cm]{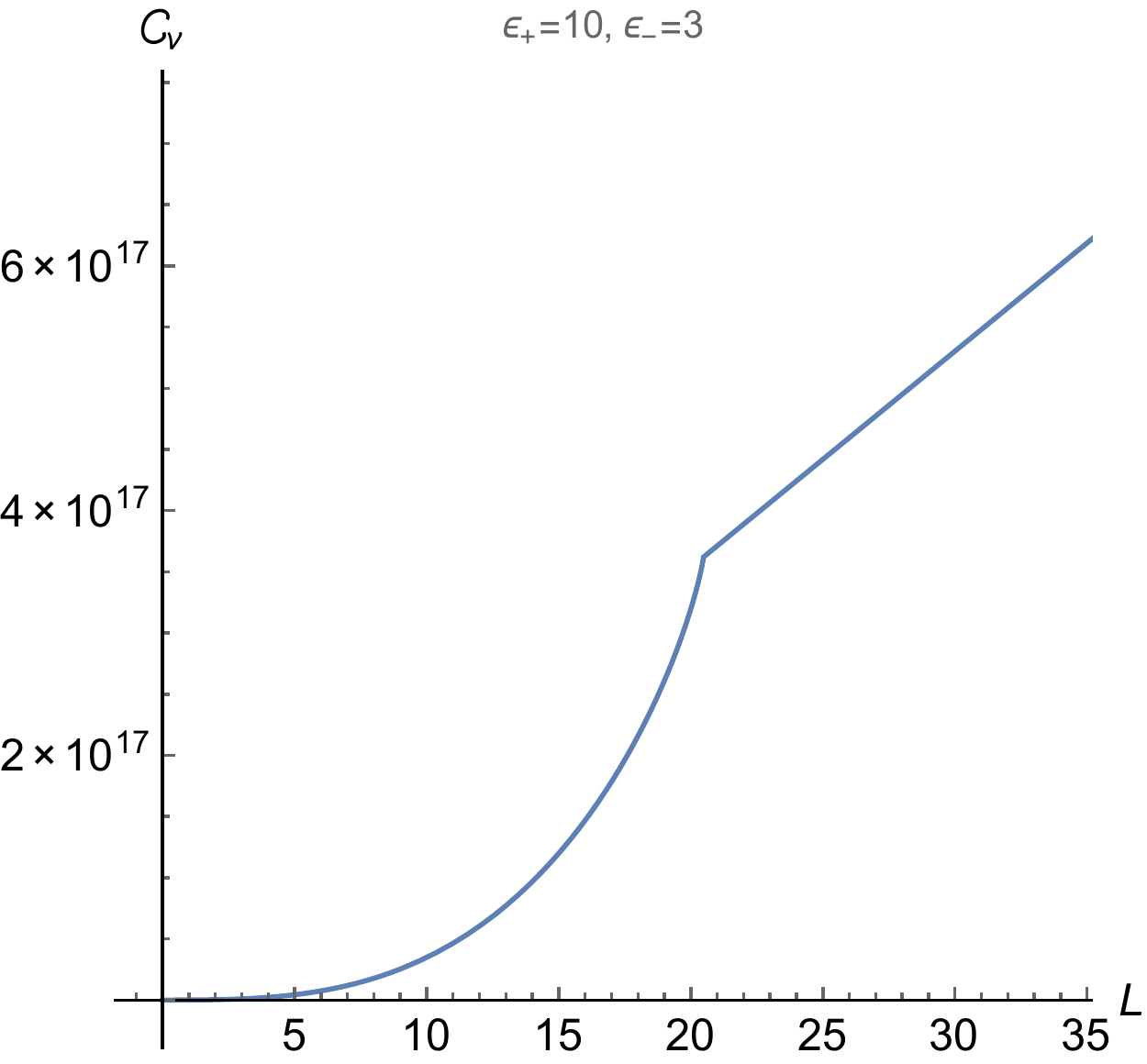}
  \caption{$\lambda = 170$, $\epsilon_+=10, \epsilon_{-}=3$}  \end{subfigure}
  \begin{subfigure}{7cm}
    \centering\includegraphics[width=7cm]{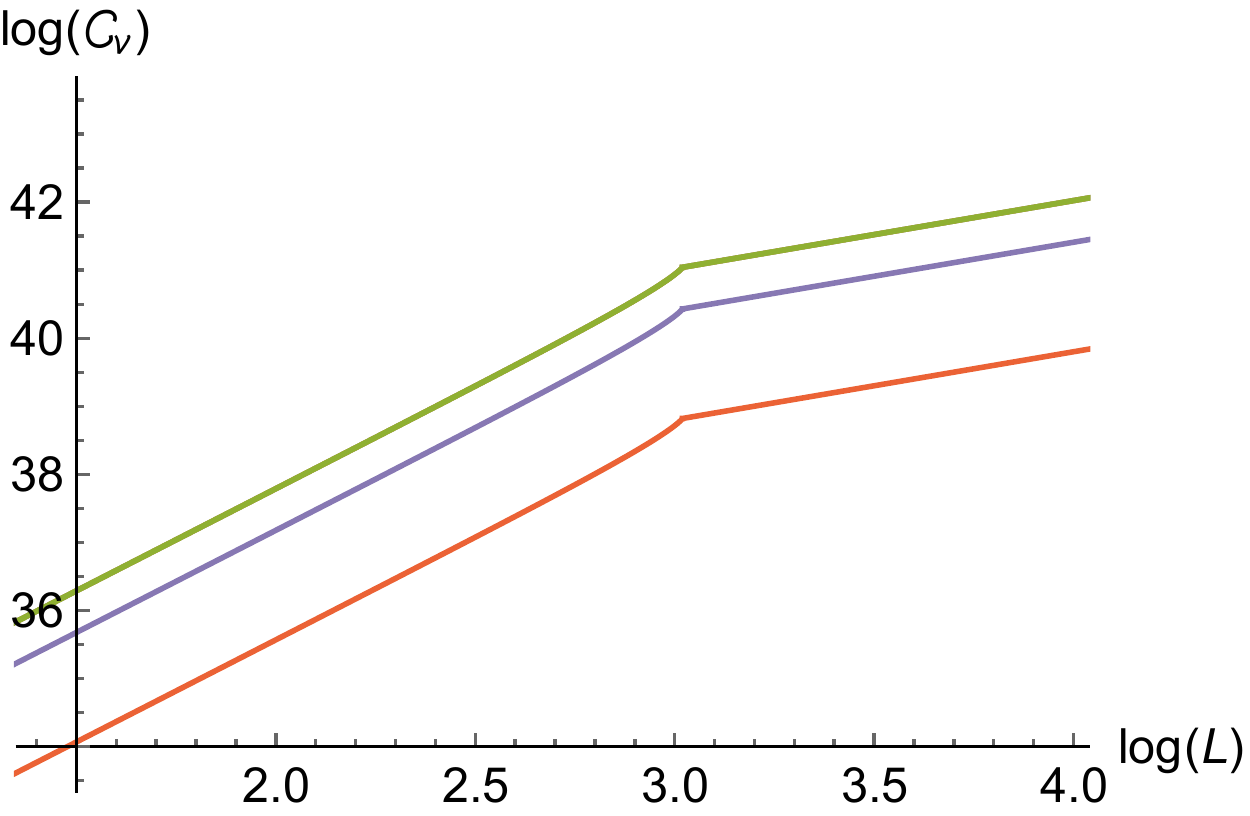}
  \caption{$\ln \mathcal{C}_V$ vs. $\ln L$ \,for $\lambda = 170$}   \end{subfigure}
  \caption{Static frame subregion volume complexity ($\mathcal{C}_V$) vs. subregion size ($L$) graphs for $T\overline{T}$, $J\overline{T}$ \& $\overline{J}T$ deformed CFT$_2$ (LST) for fixed $\lambda =170$ exhibiting Hagedorn phase transition. The last plot is a log-log graph clearly displaying the scaling exponents (slopes).} 
        \label{fig:six static graphs}
\end{figure}

\begin{figure}[H]
	\centering
	\includegraphics[width=0.6\textwidth]{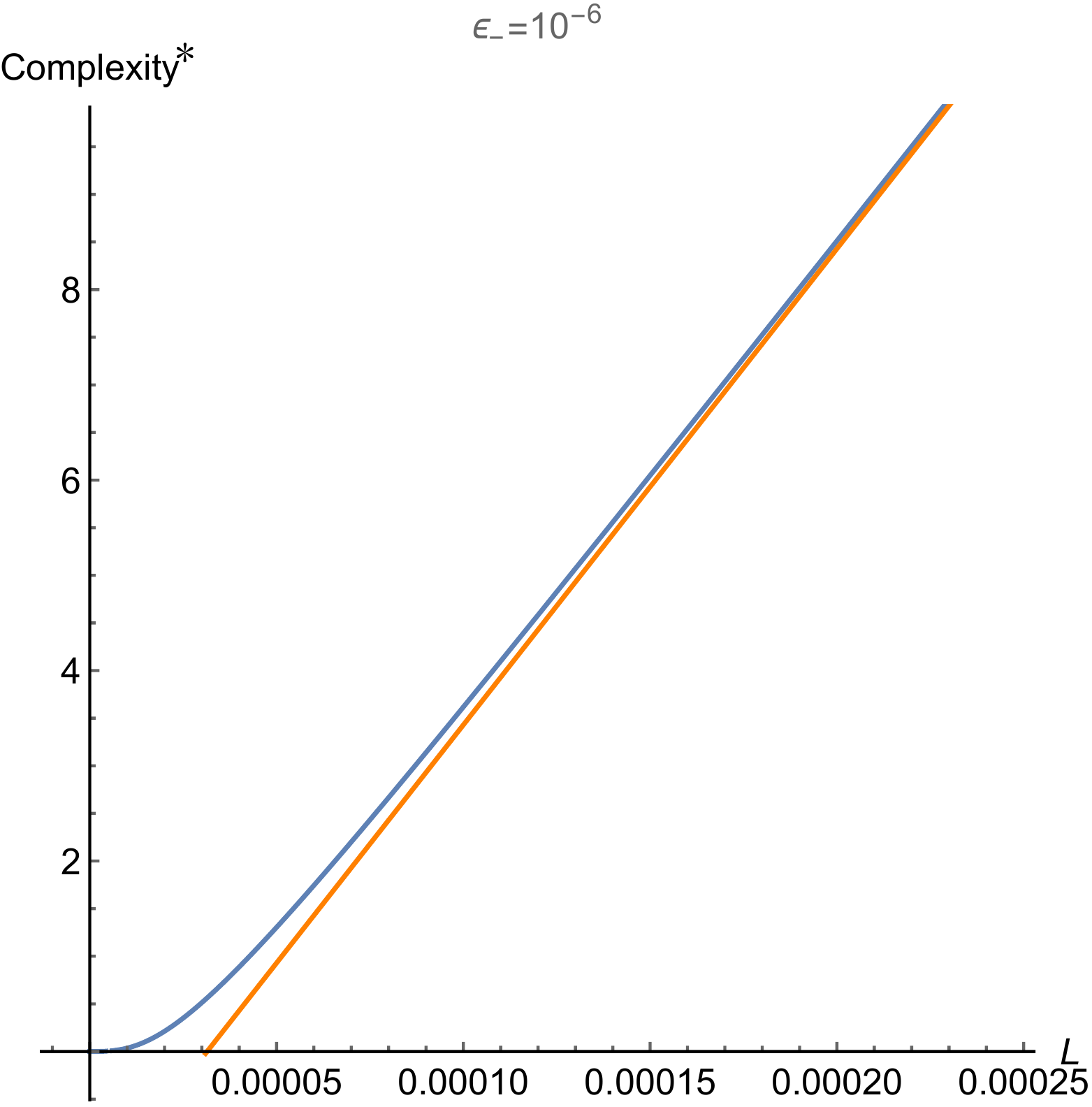}
	\caption{Subregion Volume Complexity vs $L$ plot for null warped AdS$_3$. For this plot we have set $k=10^4$, $l_s=10^{-2}$ and $\epsilon=10^{-5}$. Here, the $y$-axis represents $Complexity^{*}=Complexity\times 8\pi G_N$ while the $x$-axis is the subregion size $L$ in units of the AdS radius ($\sqrt{k} l_s =1.0$). The orange curve is the plot for pure AdS$_3$ while the blue curve is the plot for null WAdS$_3$.} \label{subregionCV for WAdS3}
\end{figure}

Lorentz violating effects are our principal object of interest in this paper and in particular for the system under study i.e. LST our aim is to disentangle the effects of Lorentz violating from nonlocality. One way to perhaps isolate the characteristics of complexity corresponding to Lorentz violating effects in field theories is to examine complexity in different inequivalent Lorentz (boosted) frames. With such hope in this section we compute subregion complexity in a boosted frame (static frame) Eq. (\ref{static frame metric}). To determine the RT curve we will first need the pullback ($\gamma_{ab}$) of the static metric on the one dimensional prospective RT curve is
\begin{align}
	ds_{\gamma}^2=\left(kl_s^2\frac{U'^2}{U^2}+f(U)\right)dX^2.
\end{align}
The length functional for this curve, parameterized as $U=U(X)$, in the string frame is given by
\begin{align}
	\int dX\,\, \mathcal{L}\left(U(X),U'(X),X \right)&=\int dX e^{-2(\Phi-\Phi_{0})}\sqrt{\gamma}=\int dX 
	\frac{kU^2}{\sqrt{f(U)h(U)}}\sqrt{kl_s^2\frac{U'^2}{U^2}+f(U)}.
\end{align}
Employing the same set of steps employed in the previous section we first compute the integrals of motion.
\begin{align}
	C_2&=U'\frac{\partial \mathcal{L}}{\partial U'}-\mathcal{L}=-\frac{k U^3 \sqrt{f(U)}}{\sqrt{h(U)} \sqrt{U^2 f(U)+k U'^2 l_s^2}}=-\frac{k U^3 \sqrt{k U^2 \lambda'+1}}{\sqrt{k U'^2 l_s^2 \left(k \lambda  U^2+1\right)+k U^4}}
\end{align}
Integrals of motion after applying boundary conditions $U(0))=U_0 ~\text{and}~ U'(0)=0$ at the turning point to the above equation is
\begin{align}
C_2	&=-\sqrt{k} U_0 \sqrt{k U_0^2 \lambda'+1}.
\end{align}
Equating and solving for $U'(X)$ gives
\begin{align}
	U'&=\frac{U \sqrt{f(U)} \sqrt{U^4 h\left(U_0\right)-U_0^4 h(U)}}{\sqrt{k} U_0^2 \sqrt{h(U)} l_s}\label{U prime}
\end{align}
Inverting this, the subregion size can be expressed in terms of the turning point $U_0$ as
\begin{equation}
	L= 2	\int_{0}^{L/2} dX=2\int_{U_0}^{\infty}
	\frac{dU}{U'}=2\sqrt{k} U_0^2 l_s\int_{U_0}^{\infty}\frac{ \sqrt{h(U)} }{U \sqrt{f(U)} \sqrt{U^4 h\left(U_0\right)-U_0^4 h(U)}} \label{L}
\end{equation}
In the linear dilaton region, $(k U^2>> 1)$, one can perturbatively solve the above equation to obtain
\begin{align}
	L
	&=\frac{1}{2} \pi  \sqrt{k \lambda} l_s+O\left(1/U_0^2\right)
\end{align}
This limiting value of $L$  in the static frame, named $L'_c$ gives critical length of the subregion at the point of Hagedorn phase transition
\begin{align}
	L'_c=\frac{1}{2} \pi  \sqrt{k \lambda} l_s \label{L'_c}
\end{align}
This is the critical subregion size where entanglement entropy (RT curve) undergoes the Hagedorn phase transition in the static frame. An important thing to note here that despite the Lorentz violating couplings, $\epsilon_\pm$, being turned on this leading order expression is independent of $\epsilon_\pm$, and instead depends just on the nonlocality parameter $\lambda$. The issue of whether this is true to all orders will be settled After determining the RT curve, now we determine the subregion complexity as the string frame area of the co-dimension one maximal area spacelike surface bound from the inside by the RT curve. The (pullback) metric on the maximal area (spacelike) hypersurface is:
\begin{equation}
	\begin{split}
		ds^2&=k l_s^2\frac{dU^2}{U^2}+f(U)dX^2.
	\end{split}
\end{equation}
Thus the maximal volume arising from the above hypersurface bounded between the RT curve and the boundary is:
\begin{align}
			\mathcal{V}&=	\int_{U_0}^{l_s/\epsilon} dU		\int_{0}^{X(U)} d\tilde{X} e^{-2(\Phi(U)-\Phi_{\infty})}\sqrt{\gamma}=k l_s \int_{U_0}^{l_s/\epsilon} dU\sqrt{1+k U^2 \lambda'}\int_{U_0}^{U}
			\frac{d\widetilde{U}}{\widetilde{U}'}
		\end{align}
	We again opt for exact but numerical means in computing the subregion volume (complexity) instead of an analytic but perturbative (approximate) approach. The plots for subregion complexity vs. subregion size for various representative values of the Lorentz violating couplings $\epsilon_{\pm}$ and a fixed value the $T\overline{T}$ coupling $\lambda=170$ are displayed in figure (\ref{fig:six static graphs}).\\	
Here we list the salient features of these plots:
\begin{itemize}
           \begin{item}
           Subregion volume complexity is a monotonically increasing function of the subregion size and it undergoes a phase transition as the subregion size is varied (just like subregion complexity in the stationary frame)  beyond a certain critical length, which turns out to be $L'_c$ of Eq. (\ref{L'_c}), i.e. the same critical subregion size at which entanglement entropy undergoes a phase transition  (refer to table \ref{table for EE and CV points in static} ).
           \end{item}
           \begin{item}
           For subregion size less than the critical size, $L'_c$, complexity grows quadratically with subregion size while for subregion sizes greater than $L'_c$, complexity grows linearly as evident from the log-log plot. The physics of this is the same as in that of the stationary frame - for small subregion sizes the RT curve is confined to the near boundary linear dilaton region, i.e. the deep UV regime of the boundary theor which is a nonlocal theory (LST on scales comparable to the string length scale), while for large subregion sizes the RT curve is well inside the bulk where the geometry is AdS, i.e. the deep IR regime of the boundary theory - LST on length scales far larger than the string scale and hence can be regarded as local CFT. 
           \end{item}
           \begin{item}
           Unlike in the stationary frame, in the static the critical subregion size at the transition point, extracted from the location of the kinks in the plots, does not change as the Lorentz violating couplings $\epsilon_{\pm}$ are varied while keeping the nonlocality scale $\lambda$ fixed. Refer to the table This strongly hints that perhaps the static frame complexity is a probe which is better suited to isolate or extract the effects of nonlocality while the complexity in the stationary frame manifests a mixed characteristic of both nonlocality and Lorentz violation. 
           \end{item}
           \begin{item}
           When either one or both of Lorentz violating couplings $\epsilon_{\pm}$ vanish, their graphs overlap to overlap. This is potentially due to the fact that the static frame subregion complexity becomes effectively the function of $\lambda' = \lambda - 4\epsilon_{+}\epsilon_{-}$, so that it is insensitive to distinguish between the various values of $\lambda'$ for vanishing value of the product $\epsilon_{+} \epsilon_{-}$. So the characteristic signatures of Lorentz violation in the divergence structure is the one which is accompanied by the coefficient $\epsilon_{+} \epsilon_{-}$.
           \end{item}
\end{itemize}
\begin{table}[h]
\centering
\begin{tabular}{ |p{4cm}|p{4cm}|p{4cm}|  }
	\hline
	\multicolumn{3}{|c|}{Location of the critical length $L'_c$ for $\lambda=170$} \\
	\hline
	Lorentz violating couplings	($\epsilon_+$,$\epsilon_-$)& Critical suregion size for entanglement entropy (EE) $L'_c$ computed from Eq. (\ref{L'_c}) & Critical subregion size for subregion complexity in static frame extracted from $\mathcal{C}_V$-$L$ graphs Fig. \ref{fig:six static graphs}\\
	\hline
	(0,0)& 20.48    &20.47\\
	(6,0)&   20.48  & 20.48 \\
	(12,0) &20.48 & 20.49\\
	(7,6)   &20.48 &20.48\\
	(10,3)&  20.48  & 20.48\\
	(13,0)& 20.48  & 20.48  \\
	\hline
\end{tabular}\\
\caption{Table comparing the critical subregion size for phase transition in Entanglement entropy from theory and the critical subregion size for subregion volume complexity extracted from the plots in the static Lorentz frame} \label{table for EE and CV points in static}
\end{table}
As before, in the case of stationary frame subregion complexity, here we find it instructive to supply the table listing the critical subregion size from the plots for various cases of  Loerntz violating couplings at a fixed $\lambda$ and compare it with the perturbatively calculated analytical estimate Eq. (\ref{L'_c}).
It is evident form table \ref{table for EE and CV points in static} subregion volume complexity displays a phase transition \emph{at the exact same critical subregion size} as that of entanglement entropy in the static frame. Thus, we can echo the same message from the end of the previous section regarding the utility of complexity as a physical probe for detecting phase transitions (perhaps even in those circumstances where other probes such as correlators of local operators might fail). However, unfortunately as long as $\lambda$ remains nonzero it appears one cannot isolate the effects of Lorentz violation from nonlocality in this system (LST) in this static frame. In fact, to the contrary, what we have seen in this exercise is that even if $\epsilon_\pm$ are not both zero, but if the product $\epsilon_+ \epsilon_-$ vanishes, then the subregion complexity phase transition point is a pure function of the the nonlocality scale $\lambda$ i.e. in this boosted frame, the phase transition is independent of the Lorentz violating effects. 

\section{Holographic volume complexity of null WAdS$_3$} \label{null WAdS3 CV}
In this section we consider a special limit in the parameter space of the irrelevant couplings of the LST, for which the bulk dual is the null Warped AdS geometry, which is smoothly realised by sending $\lambda \rightarrow 0$ and one of the Lorentz violating coupling (say $\epsilon_-$) to zero. In this limit, the (stationary frame) bulk metric Eq. (\ref{stationary coordinate metric}) becomes, 
\begin{equation}
	ds^2=kl_s^2\frac{dU^2}{U^2}-h(U)\left(1+f(U)\epsilon_-^2\right)\,dt^2+2h(U)\,f(U)\,\epsilon_-^2\,dt\,dx+h(U)\left(1-f(U)\,\epsilon_-^2\right)dx^2\label{WADS}
\end{equation}
The Lorentz parameter $\epsilon_-$ can be identified with the warping parameter $\epsilon_{-}$.  The boundary theory in this case is a \emph{warped} CFT  \cite{Detournay:2012pc, Hofman:2014loa}, a highly nonlocal Lorentz violating field theory with the CFT symmetry algebra now reduced to a semidirect product of Virasoro (left) and a $U(1)$ Kac-Moody algebra (right). In particular, for the null warped WAdS$_3$, the dual warped CFT is not UV complete, beyond a certain critical energy (deep UV) the theory is nonunitary since the energy spectrum is complex \cite{Chakraborty:2020xyz}. Although correlation functions are hard compute in this warped CFT, we demonstrate that this feature (UV incompleteness) easily captured by complexity.
\subsection{Volume Complexity }
The maximal volume spacelike slice does not need to be worked out afresh as it can treated as a special case of the stationary frame metric of the generic $T\overline{T}$, $J\overline{T}$-$\overline{J}T$ deformed bulk geometry. However this limit could be singular so instead of indirectly evaluating the volume (complexity) by taking the naive $\lambda, \epsilon_{+}\rightarrow 0$ limit of the maximal volume expression of the string frame (\ref{stationary frame volume}) (or complexity (\ref{stationary frame complexity})) we compute the integral directly,
\begin{equation}
	\begin{split}
		V_{\Sigma}(T)&=L_xk l_s \int dU  \sqrt{1- \epsilon _-^2k U^2} =\frac{L_x k l_s^2}{2 \epsilon } \sqrt{1-\frac{k \epsilon _-^2 l_s^2}{\epsilon ^2}}+\frac{\sqrt{k} l_s L_x }{2 \epsilon _-}\sin ^{-1}\left(\frac{\sqrt{k} \epsilon _+ l_s}{\epsilon }\right)\\
		\Rightarrow \mathcal{C}_{V}&=\frac{L_x \sqrt{k} l_s}{2 G_N\epsilon } \sqrt{1-\frac{k \epsilon _-^2 l_s^2}{\epsilon ^2}}+\frac{ L_x }{2 \epsilon _-G_N}\sin ^{-1}\left(\frac{\sqrt{k} \epsilon _- l_s}{\epsilon }\right)\label{5.2}
	\end{split}
\end{equation}
Here, we see that the resultant complexity, unlike for the generic case, (\ref{CV2}), fails to remain real unless $\epsilon>\sqrt{k}l_s\, \epsilon_{-}$. Thus the UV cut off cannot be made arbitrarily small. This validates our faith that complexity successfully captures the UV incompleteness of the warped CFT dual to null warped AdS$_3$. In the limit of a small warping parameter $\epsilon_{-}$ (to be precise expanding in $\epsilon_{-}\sqrt{k}l_s/ \epsilon$), the leading term is linearly divergent,
\begin{equation}
\mathcal{C}_V \sim \frac{L_x \,\sqrt{k}\,l_s}{G_N \,\epsilon} - \frac{L_x\,\left(\sqrt{k} l_s\right)^3\epsilon_{-}^2}{6 G_N\, \epsilon^3}
\end{equation}
i.e. like a local field theory in one space dimensions (or pure AdS bulk)! Again this is a reflection that the UV regime (near boundary region) where the nonlocality effects kicks in is excluded from consideration. The complexity characteristics of Warped CFT in general from both the holographic and field theory methods has been taken up in greater detail in a separate paper \cite{Bhattacharyya:2022ren}.
\subsection{Subregion volume complexity for null WAdS$_3$}
\label{s6}
Finally we work out the subregion volume complexity for the interesting special case of null warped AdS$_3$. In this case first the result will be obtained analytically in the approximation of small warping $\epsilon_{-}$ to underscore the fact that subregion complexity is a better probe of nonlocality in this example compared to other probes such as subregion entanglement entropy. Then the exact result will be presented by evaluating the subregion complexity integral numerically without any approximations. First, recall that In this case, the turning point of the RT surface (curve) in terms of the subregion size is already worked out by inverting \eqref{4.11},
\begin{equation}
	U_{0} = \frac{2l_s}{L}+\frac{8kl_s^3\epsilon_-^2}{L^3} \ln\left(\frac{L}{\epsilon}\right) +O[\epsilon_-^4] \label{6.4}
\end{equation}
In terms of the turning point, the subregion volume complexity calculation of null WAdS$_3$ is given by the nested integral,
\begin{align}
	C_{V}&=\frac{\sqrt{k}l_s}{8\pi G}\int_{U_{0}}^{\frac{l_s}{\epsilon}}dU\sqrt{1-kU^2\epsilon_-^2}\int_{U_{0}}^{U}d\tilde{U}\frac{U_{0}\sqrt{1-kU_{0}^2\epsilon_-^2}}{\tilde{U}^2\sqrt{1-k\tilde{U}^2\epsilon_-^2}\sqrt{\tilde{U}^2-U_{0}^2+k\epsilon_-^2(-\tilde{U}^4+U_{0}^4)}}\label{6.6}\\
	&=\frac{\sqrt{k}l_s}{8\pi G}\int_{U_{0}}^{\frac{l_s}{\epsilon}}\bigg[\frac{\sqrt{U^2-V^2}}{UV}-\bigg(\frac{kU\sqrt{U^2-U_{0}^2}}{2U_{0}}-kU_{0}\ln{\bigg(\frac{U+\sqrt{U^2-U_{0}^2}}{U_{0}}\bigg)}\bigg)\epsilon_-^2+O\left(\epsilon_-^4\right)\bigg]dU\label{6.9}\\
	&=\frac{\sqrt{k}l_s}{8\pi G}\bigg[\bigg(1+\frac{4 k l_s^2}{L^2}\epsilon_-^2\bigg)\frac{L}{2\epsilon}-\frac{kl_s^2L}{12\epsilon^3}\epsilon_-^2-\frac{\pi}{2} O\left(\epsilon_-^4\right) \bigg]\label{6.10}
\end{align}
In order to get eqn \eqref{6.10}, we have expanded the integrands in equation \eqref{6.6} in a Taylor series with respect to $\epsilon_-$.  In the above expression of $C_V$, we can clearly see that if we take the warping factor to be zero, equation \eqref{6.10} reproduces the subregion complexity for the pure $AdS_3$.  For, pure $AdS_3$, we recover the expression for subregion volume complexity \cite{Alishahiha:2015rta},
\begin{equation}
	C_{V}=\frac{\sqrt{k}l_s}{8 \pi G}\bigg[\frac{L}{2\epsilon}-\frac{\pi}{2}\bigg]\label{6.5}
\end{equation}
Another very important feature of this result \eqref{6.10} is that the subregion $C_V$ (the divergence structure) reflects the nonlocal nature of the dual warped conformal field theory unlike the holographic entanglement entropy in the appendix \ref{HEE}. While we are using a string background with all NS-NS sector fields turned on in the bulk, this local theory like divergence structure (linear divergence) of EE for Warped AdS$_3$ has been reported using other holographic backgrounds where the bulk theory is either topologically massive gravity (TMG) \cite{Anninos:2013nja, Castro:2015csg} or new massive gravity (NMG) \cite{Basanisi:2016hsh}.  Thus in this example, we see that subregion complexity is a more sensitive or refined probe of nonlocality and Lorentz violation compared to entanglement entropy.
\\
Next we present the plot\footnote{We have used parametric plot function in mathematica here and used Eq. \eqref{4.11} for the expression of $L$ as,
\begin{equation}
	L=\int_{U_{0}}^{\frac{l_s}{\epsilon}}dU\frac{2U_{0}l_s\sqrt{1-kU_{0}^2\epsilon_-^2}}{U^2\sqrt{1-kU^2\epsilon_-^2}\sqrt{U^2-U_{0}^2+k\epsilon_-^2(-U^4+U_{0}^4)}}\label{6.14}
\end{equation}} of Subregion $C_V$ as a function of $L$ in the figure below obtained by direct numerical evaluation of \eqref{6.6}.  
\\
The value of $\epsilon_{-}=10^{-6}$ is used for this plot because from Eq. \eqref{5.2}, it is clear that the value of $\epsilon_-$ has to be smaller than the value of $\epsilon$. Also note that here, the $y$-axis is actually complexity scaled by the universal constant $8 \pi G_N$. We summarize the salient features of this plot
\begin{itemize}

  \item Subregion complexity monotonically increases as a function of the subregion size.
  
  \item Unlike in the case of generic nonvanishing $\lambda, \epsilon_{+}$ the subregion complexity does not undergo any phase transition.

	\item The effect of nonlocality or Lorentz violation is very small in general and only prominent when the subregion is orders of magnitude smaller than the AdS radius. For larger generic subregion sizes the WAdS subregion $\mathcal{C}_V$ coincides with that for pure AdS$_3$ (this part needs to be discussed later.)
	\item Sensible plots are only obtained when the cut off $\epsilon >\epsilon_{-}$\footnote{To be precise one must keep $\epsilon>\epsilon_{-} \sqrt{k}l_s$ but here we have set $\sqrt{k}l_s=1$ so effectively we must keep $\epsilon_{-}<\epsilon$}. This is again a reflection of the fact that the dual boundary theory is not a UV complete theory, it is best thought of as an effective theory with the spectrum truncated at high energies.
\end{itemize}

One might ask whether one can switch to the static frame and evaluate the complexity of LST dual to null warped AdS$_3$ in the that frame just like it was done in the case of generic $J\overline{T}, \overline{J}T$ couplings to demonstrate/check for boost symmetry violation. However in this regard we would like to point out that boost transformation (\eqref{boost}) is singular in the null warped AdS$_3$ limit, i.e. when $\epsilon_+\rightarrow0$. Thus neither such a boost transformation and by extension, nor does a static frame exist for null warped AdS$_3$. In this special point of the parameter space, we will have to be content with the volume complexity, subregion volume complexity and action complexity in the stationary coordinate system exclusively.\\
\section{Action complexity} \label{CA static}
In this section we compute the action complexity, $C_\mathcal{A}$, for the LST obtained by $T\overline{T}, J\overline{T}, \overline{J}T$ deformation of a CFT$_2$  using the holographic dual metric, dilaton and Kalb-Ramond background. Action complexity is an alternative prescription of holographically evaluating the dual boundary theory complexity which offers distinct advantages over volume complexity in that (A) no arbitrary length scales appearing in its definition, just the fundamental constants $\hbar$ and $G_N$, and (B) one does need to solve a variational problem which can be challenging exercise in general for Lorentzian signature spacetimes. Instead one just performs (action) integrals over the \emph{WdW patch}. The WdW patch is defined as the union of all spacelike hypersurfaces in the bulk anchored at a fixed timeslice on the boundary:
\begin{equation}\label{caa}
	C_\mathcal{A}=\frac{S_{WdW}}{\pi \hbar}~.
\end{equation}
The Penrose diagram of the dual bulk geometry (suppressing the transverse boundary direction) is identical to that of the $\mathcal{M}_3$ spacetime, presented in our previous work \cite{Chakraborty:2020fpt}. For completeness, we reproduce the Penrose diagram here with the WdW patch displayed in pink
\begin{figure}[H]
 \centering
 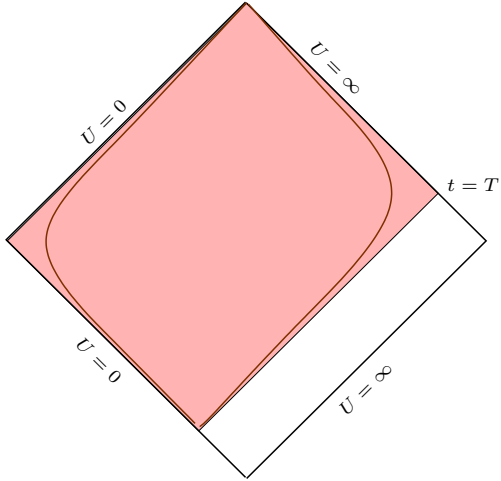
 \caption{Penrose diagram of the dual bulk geometry with the Wheeler-deWitt (WdW) patch shaded in pink for the boundary time $T$. The brown curves are timelike surfaces which can be continuously deformed into the null boundaries of the WdW patch by means of a regulator parameter.}
 \label{fig:0T M3 WdW}
\end{figure}
The bulk action in the string frame is:
\begin{eqnarray}
	\begin{split}\label{fullaction}
		S &= \frac{1}{16\pi G_N}\int_{M} d^{d+1} X \sqrt{-g}e^{-2(\Phi-\Phi_0)}\left( R+4g^{\mu\nu}\partial_\mu\Phi\partial_\nu\Phi-\frac{H^2}{12}-4\Lambda\right) \\
		&   \hspace{1in}+ \frac{1}{8\pi\,G_N} \int_{\sum \partial M} \sqrt{\gamma}\, \left( \cdots\right) + \frac{1}{8\pi\,G_N} \int_{\cap \partial M} \sqrt{h}\, (\cdots)~.
	\end{split}
\end{eqnarray}
The $(\cdots)$'s represent the supplementary surface/boundary ($\cup \partial M$) terms  and joint ($\cap \,\partial M$) terms necessary for the variation of the action to be well defined, as well as reparametrization invariant.  Since (some) boundaries of the WdW patch are null, the usual GHY terms are not the suitable ones. The issue of determining the boundary terms for null boundaries was settled in \cite{Lehner:2016vdi}. However, we will take an alternative prescription spelled out in \cite{Bolognesi:2018ion}\footnote{see also \cite{Parattu:2015gga}.} where the null boundaries of the WdW patch are first deformed into a single smooth timelike surface using a deformation parameter (regulator), and then we are free to use the usual GHY term. After working out the GHY term we remove the regulator and obtain the result for the null WdW boundary. This affords an enormous simplification as it eliminates the necessity to compute the joint terms (\ie\ terms in the action from joints or edges along which two null surfaces intersect) as well as preserving diffeomorphism and reparametrization invariance of the GHY contribution from beginning to end. Our regularization reproduces the same results as the prescription of \cite{Lehner:2016vdi} for the well known cases of pure AdS, AdS-Schwarzschild, AdS-RN etc. but the status of the equivalence of these two prescriptions for arbitrary generic geometries is yet unexplored. In general the issue of different regularization prescriptions is still being investigated e.g. for a comparison of the two regularizations introduced in \cite{Carmi:2016wjl}, see \cite{Akhavan:2018wla, Jafari:2019qns}.\\

Just to remind the reader, that unlike for the $\mathcal{M}_3$ case, here one has to make a choice: one can either work with the full 4 dimensional bulk action, or one might dimensionally reduce (over the $y$-direction fiber) and work with an effective $3$ dimensional bulk action. While computing the entanglement entropy for this system, the authors of \cite{Chakraborty:2020udr} found that the 4d and 3d results disagree and they opted to work with the full 4d bulk. Happily for us, it turns out that both 4d and 3d actions deliver identical results for complexity, \emph{provided one does not drop the total derivative terms in the dimensionally reduced action}. Conventionally in the literature these total derivative terms are dropped since they do not contribute to the classical equations of motion. However while computing complexity these terms do contribute and one cannot discard them if the complexity before and after dimensional reduction has to match. In appendix Sec. \ref{KK redone}, the KK reduction is reviewed and the exact match between the 4D and 3D actions is carried out after retaining all total derivative terms in the dimensionally reduced (3D) action.\\
\begin{align}
	S_{(3D)}&=\frac{1}{16\pi G_N}\int  d^3 x \sqrt{-g} ~ e^{-2\left(\Phi-\Phi_0\right)}\Big(R-\left(\partial\sigma\right)^{2}-2\,\Box\sigma-\frac{1}{4}e^{2\sigma}\mathcal{F}^{2}-4\Lambda \nonumber\\&\hspace{7cm}+ 4g^{\mu \nu } \partial_\mu \Phi \partial_\nu \Phi+ 4g^{\mu \nu } \partial_\mu \Phi \partial_\nu \sigma  \nonumber\\&\hspace{9cm} -\frac{1}{12}\widetilde{H}^{2} -\frac{1}{4}e^{-2\sigma}\widetilde{F}^{2}\Big). \label{3d bulk action volume terms}
\end{align}
It is evident that after dimensionally reducing to $3$ dimensions the KK scalar, $\sigma$ contributes to the action a term which has a second order derivative, namely $\Box \sigma$ (This is a total derivative term which is usually dropped). This will also to lead an extra contribution to the surface terms (GHY terms) in the string frame, apart from the usual GHY surface term arising from the metric. The full surface term contribution for the $2+1$-dimensional bulk in string frame is
\begin{equation}
	S_{GHY}=\frac{1}{8\pi G_N}\int d^2x \sqrt{-\gamma}\, e^{-2(\Phi-\Phi_0)} \left(K+n^{\mu}\partial_{\mu}\sigma\right)~
\end{equation}
For the derivation, the reader may refer to Appendix \ref{section:GHY}. For the action complexity calculation, in the first order of business is to determine the WdW patch, i.e. the light cones emanating from the boundary timeslice. However, solving the WdW patch is very complicated in the stationary frame coordinates \eqref{stationary coordinate metric} where constant $t$ surfaces are not orthgonal to the vector $\partial/\partial t$. Life is much simpler in the static frame coordinates \eqref{static frame metric} as the constant $T$ surfaces are orthogonal to the time direction vector $\partial/\partial T$. So we perform the action complexity calculation exclusively in the static frame \eqref{static frame metric}.\\

\subsection{Volume (EH) pieces of the onshell action}
In this section we present the volume piece contributions (``EH terms") to the action complexity using the dimensionally reduced $2+1$-d background. As mentioned previously, the evaluation of the gravitational action (``EH terms") in the string frame in $4$ dimensional and the dimensionally reduced $3$ dimensional backgrounds give identical results, for this equivalence, the reader is referred to the Sec.~ \ref{DR action match}. \\
First we have to determine the WdW patch. As we have mentioned earlier, the calculation is most straightforward in the static coordinates $(T, X, U)$ because the lightcones are easy to determine. The Wheeler-deWitt patch (WdW) for the boundary time $T=T_*$ is bounded by the null rays
\begin{eqnarray}
	\,dt_{\pm}&=& \mp \sqrt{k}\,l_s\, \frac{1}{U \sqrt{h(U)}}dU ,
\end{eqnarray}
obeying boundary condition, $T(U \rightarrow\infty)=T_*$. The $T$-integrals in the volume terms \eqref{3d bulk action volume terms} (Einstein-Hilbert plus matter type terms) can be readily done:
\begin{equation}
	T_{+}(U)-T_{-}(U) =2\sqrt{k}\,l_s\,\int^{\infty}_{U} d\tilde{U}\frac{1}{\tilde{U} \sqrt{h(\tilde{U})}}~.
\end{equation}
This integral is divergent and hence we will modify our WdW patch to begin at a UV-cutoff surface $U=l_s/\epsilon$ instead of spatial infinity:
\begin{equation}
	T_{+}(U)-T_{-}(U) =2\sqrt{k}\,l_s\,\int^{l_s/\epsilon}_{U} d\tilde{U}\frac{1}{\tilde{U} \sqrt{h(\tilde{U})}}~. \label{WdW patch with cutoff}
\end{equation}
Having determined the WdW patch, we list the various bulk action term contributions \eqref{3d bulk action volume terms} along with their UV and IR limits are listed as follows.\\\\
\noindent \textbf{The Ricci scalar sector term}:
These terms are from the $4$ dimensional Ricci scalar or equivalently in $3$ dimensions from the full KK reduced sector derived from the $4$ dimensional metric. For details the reader is referred to Sec. \ref{Ricci sector match}.
\begin{eqnarray}
	\begin{split}\label{s1}
		S_{R}&\equiv  
		\frac{1}{16\pi G_N}\int_{WdW} d^3 x \sqrt{-g} ~ e^{-2(\Phi-\Phi_0)}\left(R^{(3)}-2\left(\partial\sigma\right)^{2}-2\square\sigma-\frac{1}{4}e^{2\sigma}\mathcal{F}^{2}\right),\label{eq: correct form of 1.13 of Pope in D=00003D4}\\
		&=\frac{L_xk}{8 \pi  G_N}\int_{0}^{l_s/\epsilon}dU\,U\frac{ \left(8 \lambda'  k U^2-6\right)}{\left(\lambda'  k U^2+1\right)^2}\int^{l_s/\epsilon}_{U} dU'\frac{1}{U' \sqrt{h(U')}}~.
	\end{split}
\end{eqnarray}
The above integral can be performed analytically but the full expression is a bit cumbersome. In the deep UV (\ie\ when $\epsilon/\beta_H\ll1$), $S_R$ takes the following form
\begin{eqnarray}
	\begin{split} \label{SR}
		\lim_{\epsilon/\beta_H\ll 1}S_R=&-\frac{c L_x}{6\beta'_H}(7+8\log2)\log\left(\frac{\beta'_{H}}{\pi \epsilon}\right)+\frac{2cL_x }{3\beta'_{H}}\log^2\left(\frac{\beta'_{H}}{\pi\epsilon}\right)\nonumber\\&\hspace{3cm}+\frac{L_x c}{18\beta'_{H}}(\pi^2+24\log2)+O\left(\left(\frac{\epsilon}{\beta'_{H}}\right)^2\right) ~.
	\end{split}
\end{eqnarray}
In the IR (\ie\ when $\epsilon/\beta_H\gg1$), $S_R$ takes the form
\begin{eqnarray}
	\lim_{\epsilon/\beta_H\gg 1}S_R= -\frac{cL_x}{4\pi\beta'_H}\frac{\beta'_H}{\epsilon}+\frac{7cL_x}{288\pi^3\beta'_H}\left(\frac{\beta'_H}{\epsilon}\right)^3+O\left(\left(\frac{\beta'_{H}}{\epsilon}\right)^{4}\right)~.
\end{eqnarray}\\
\noindent \textbf{The dilaton kinetic term in the action:} We refer the reader to Sec. \ref{Dilaton sector match} for the details. 
\begin{eqnarray}
	\begin{split}\label{s2}
		S_{\Phi}&\equiv 
		\frac{1}{16\pi G_N}\int_{WdW} d^3x\, \sqrt{-g}\, e^{-2(\Phi-\Phi_0)}  4g^{\mu \nu } \partial_\mu \Phi \partial_\nu \Phi,\\
		&=\frac{\lambda'^{2}k^{3}L_{X}}{2\pi G_{N}}\int_{0}^{\frac{l_{s}}{\epsilon}}dU\:\frac{U^{5}}{\left(1+\lambda'\:kU^{2}\right)^{2}}\int_{U}^{\frac{l_{s}}{\epsilon}}\frac{dU'}{U'\sqrt{h(U')}}.\label{eq: Dilaton CA contribution as a nested integral}~
	\end{split}
\end{eqnarray}
In the UV regime, $S_\Phi$ takes the following form:
\begin{eqnarray}
	\begin{split}\label{SPhi}
		\lim_{\epsilon\ll\beta'_{H}}S_{\Phi}=\frac{cL_{x}}{24\pi^{2}\beta'_{H}}\left(\frac{\beta'_{H}}{\epsilon}\right)^{2}+\left(3+8\ln2\right)\frac{cL_{x}}{6\beta'_{H}}\ln\left(\frac{\beta'_{H}}{\pi\epsilon}\right)-\frac{2cL_{x}}{3\beta'_{H}}\ln^{2}\left(\frac{\beta'_{H}}{\pi\epsilon}\right)\\+\left(3-2\pi^{2}-48\ln2\right)\frac{cL_{x}}{36\beta'_{H}}+O\left(\frac{\epsilon}{\beta'_{H}}\right).\label{eq: Dilaton contribution UV limit-1}.
	\end{split}
\end{eqnarray}
One might be a bit alarmed at the appearance of the ``log squared" divergences in the expressions \eqref{SR} and \eqref{SPhi}, which did not arise in the volume complexity cases but as it will turns out,  such log squared divergent contributions will cancel out among each other.\\
In the IR, $S_\Phi$ takes the form
\begin{eqnarray}
	\lim_{\epsilon/\beta_H\gg 1}S_\Phi=0+O\left(\beta_H^5/\epsilon^5\right)~.
\end{eqnarray}\\
\noindent  \textbf{The cosmological constant term in the action:} The details are worked out in Sec. \ref{cc term}. Here we present the main results starting from the $3$ dimensional action,
\begin{eqnarray}
	\begin{split}\label{s3}
		S_{\Lambda}&\equiv 
		\frac{1}{16\pi G_{N}}\int d^{3}x\sqrt{-g}e^{-2\left(\Phi-\Phi_{0}\right)}\left(-4\Lambda\right),\\
		&=  \frac{L_x k}{2\pi G_N} \int_0^{l_s/\epsilon} dU\,U\int^{l_s/\epsilon}_{U} dU'\frac{1}{U' \sqrt{h(U')}}~.
	\end{split}
\end{eqnarray}
In the UV, $S_\Lambda$ takes the following form
\begin{eqnarray}\label{Scc}
	\lim_{\epsilon/\beta'_H\ll 1}S_\Lambda=\frac{c}{24\pi^{2}}\frac{L_{x}}{\beta'_{H}}\left(\frac{\beta'_{H}}{\epsilon}\right)^{2}+\frac{c}{6}\frac{L_{X}}{\beta'_{H}}\ln\left(\frac{\text{\ensuremath{\beta'_{H}}}}{\pi\epsilon}\right)+\frac{c}{12}\frac{L_{x}}{\beta'_{H}}+O\left(\frac{\epsilon}{\beta'_{H}}\right)\label{eq: cc term contribution in the UV limit} ~.
\end{eqnarray}
In the IR, $S_\Phi$ takes the form
\begin{eqnarray}
	\lim_{\epsilon\gg\beta_{H}}I_{\Lambda}=\frac{cL_{x}}{6\pi\epsilon}+\frac{cL_{x}}{144\pi^{3}\beta'_{H}}\left(\frac{\beta'_{H}}{\epsilon}\right)^{3}+O\left(\left(\frac{\beta'_{H}}{\epsilon}\right)^{5}\right)\label{eq: cc term contribution in the IR limit}.  
\end{eqnarray}\\
\noindent  \textbf{The Kalb-Ramond term in the action:} Finally the spacetime volume type contribution from the Kalb-Ramond field in $4$ dimensions or the full Kalb-Ramond derived fields in $3$ dimensions after dimensional reduction (a Kalb-Ramond two-form field and a Kalb-Ramond one-form gauge field). The details including the matching before and after dimensional reduction are worked out in the appendix Sec. \ref{Kalb Ramond sector match}. The main results starting with the action piece are presented here,
\begin{eqnarray}
	\begin{split}\label{s4}
		S_{H}&\equiv\frac{1}{16\pi G_{N}}\int\:d^{3}x\;\sqrt{-g}e^{-2(\Phi-\Phi_{0})}\left(-\frac{1}{12}\widetilde{H}^{2}-\frac{1}{4}e^{-2\sigma}\widetilde{F}^{2}\right),\\&=\frac{L_{x}}{4\pi G_{N}k}\int\,dU\,\frac{h^{2}(U)}{U^{3}}\int_{U}^{\frac{l_{s}}{\epsilon}}dU'\frac{1}{U'\sqrt{h(U')}}\label{eq: H-term complexity contribution as a nested integral} ~.
	\end{split}
\end{eqnarray}
In the UV, $S_H$ takes the following form
\begin{eqnarray}\label{SH}
	\lim_{\epsilon/\beta_H\ll 1}S_H=\frac{cL_{x}}{6\beta'_{H}}\ln\left(\frac{\beta'_{H}}{\pi\epsilon}\right)+O\left(\frac{\epsilon}{\beta'_{H}}\right)\label{H-field contribution UV limit}.
\end{eqnarray}
In the IR, $S_H$ takes the form
\begin{eqnarray}
	\lim_{\epsilon/\beta_H\gg 1}S_H=\frac{cL_{x}}{12\pi\beta'_{H}}\left(\frac{\beta'_{H}}{\epsilon}\right)-\frac{cL_{x}}{288\pi^{3}\beta'_{H}}\left(\frac{\beta'_{H}}{\epsilon}\right)^{3}+O\left(\left(\frac{\beta'_{H}}{\epsilon}\right)^{5}\right).\label{eq: H-field contribution IR limit}
\end{eqnarray}\\
\subsubsection{Action Contributions from the null boundaries of the WdW patch.}
The WdW patch action receives surface contributions (GHY  terms) from the boundaries of the $WdW$ patch. The Poincar\' e horizon and the two joint terms (intersection of the null boundaries of the WdW patch with the Poincar\'e horizon) make vanishing contributions. The only non trivial contribution comes from the two null boundaries of the WdW patch.\\
The null boundaries of the WdW patch are defined by
\begin{equation}
	(T-T_*)=\mp \sqrt{k}l_sA(U)~; \quad \text{where }\ \   A(U)=\int_{l_s/\epsilon}^{U}\frac{d\tilde{U}}{\tilde{U}\sqrt{h(\tilde{U})}}~.
\end{equation}
However, we will deform the pair of null surfaces to a single smooth timelike surface  by introducing a dimensionless parameter, $\varepsilon$,\footnote{This is distinct from the UV regulator, $\epsilon$.}
\begin{equation}
	\frac{(T-T_*)^2}{kl_s^2}-(1+\varepsilon)A^2(U)=0~. \label{GHYbdry}
\end{equation}
Taking differentials of both sides leads to,
\begin{eqnarray}
	h(U)dT^2=(1+\varepsilon)kl_s^2\frac{dU^2}{U^2} ~. \label{dts}
\end{eqnarray}
\\
Using \eqref{dts}, the induced metric on this timelike surface can be written as
\begin{eqnarray}
	ds^2=-\varepsilon k l_s^2\frac{dU^2}{U^2}+h(U)(1-4\epsilon_+\epsilon_-f(U))dX^2~.
\end{eqnarray}
The negative sign in the first term clearly indicates that this is a timelike surface. The unit outward normals to the surface \eqref{GHYbdry} are,
\begin{equation}
	n^T=\frac{-(T-T_*)}{\sqrt{(1+\varepsilon)^2A^2(U)-\frac{(T-T_*)^2}{kl_s^2}}}\frac{1}{\sqrt{k}l_s\sqrt{h(U)}},\ \ \ n^U=\frac{-(1+\varepsilon)A(U)}{\sqrt{(1+\varepsilon)^2A^2(U)}-\frac{(T-T_*)^2}{kl_s^2}}\frac{U}{\sqrt{k}l_s},\ \ \ n^X=0.
\end{equation}
The trace of the extrinsic curvature,
\begin{eqnarray}
	\begin{split}
		K\equiv\nabla_Ln^L=\partial_Ln^L+\Gamma^L_{LM}n^M=\partial_Tn^T+\partial_Un^U+\Gamma^L_{LU}n^U~, \label{K}
	\end{split}
\end{eqnarray}
takes the form
\begin{eqnarray}
	K & = & \frac{1}{\sqrt{\varepsilon}\sqrt{k}l_s(1+kU^2\lambda)}+\frac{1}{\sqrt{\varepsilon}\sqrt{k}l_s(1+kU^2\lambda^{'})}~.
\end{eqnarray}
\\
Thus the GHY term for this surface in the null limit ($\varepsilon\rightarrow 0$) is
\begin{eqnarray}
	\begin{split}
		S^{\partial WdW}_{GHY} &=\frac{1}{8\pi G_{N}}\int_{\partial WdW}d^2x\,\sqrt{-\gamma}e^{-2\left(\Phi-\Phi_{0}\right)}\left(K+n^{\mu}\partial_{\mu}\sigma\right),\label{Full GHY term after dimensional reduction}\\
		&=\lim_{\varepsilon\to 0}\frac{L_x\sqrt{k}}{4\pi G_N}\int_{0}^{l_s/\epsilon}\frac{dU}{\sqrt{1+k U^2\lambda^{'}}}~,\nonumber\\
		&=\frac{c L_x}{3 \beta_H^{'} }\ln \left(\sqrt{1+\frac{\beta_H^{'2} }{4 \pi ^2 \epsilon ^2}}+\frac{\beta_H^{'} }{2 \pi  \epsilon }\right) ~.
	\end{split}
	\label{GHYnull0}
\end{eqnarray}
In the UV, $S^{\partial WdW}_{GHY} $ diverges as
\begin{equation}
	\lim_{\epsilon/\beta_H\ll 1}S^{\partial WdW}_{GHY} =\frac{L_x c}{3\beta_{H}^{'}}\ln\left(\frac{\beta_{H}^{'}}{\pi\epsilon}\right)+O\left(\epsilon/\beta_{H}^{'}\right)~ . \label{UVGHY2}
\end{equation}
In the IR one can write
\begin{equation}
	\lim_{\epsilon/\beta_H\gg 1}S^{\partial WdW}_{GHY} = \frac{L_x c}{6\pi\beta_{H}^{'}}\left(\frac{\beta_{H}^{'}}{\epsilon}\right)-\frac{L_x c}{144\pi^3\beta_{H}^{'}}\left(\frac{\beta_{H}^{'}}{\epsilon}\right)^3+O\left(\beta_{H}^{'4}/\epsilon^{4}\right)~. \label{GHY2}
\end{equation}
\subsection{Action Complexity }
\begin{figure}[H]
	\centering
	\includegraphics[width=.8\textwidth]{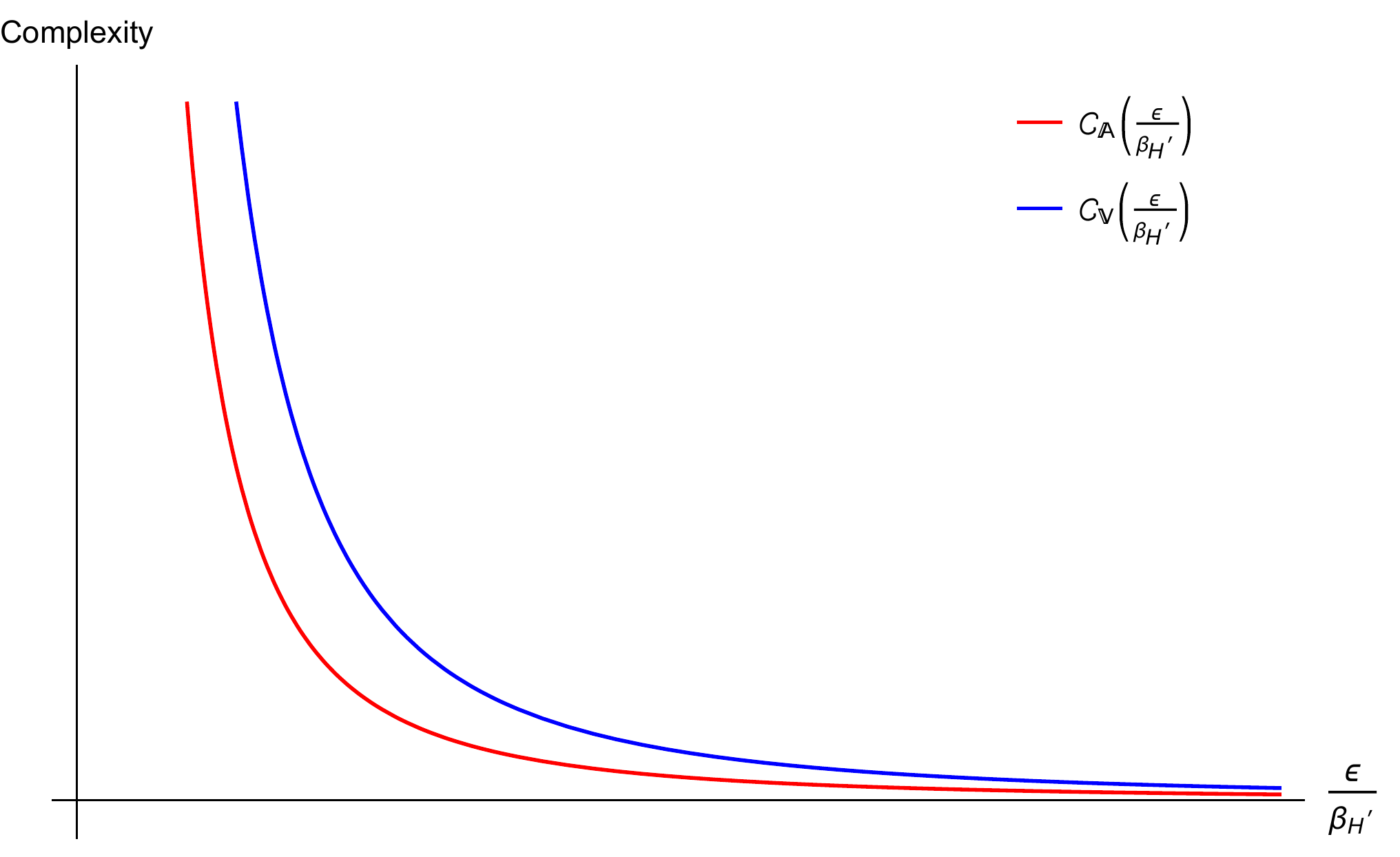}
	\caption{Comparison between $C_V$ and $C_\mathcal{A}$ at zero temperature. For large $\epsilon/\beta_H$, the action complexity decays much faster than volume complexity.}
	\label{img3}
\end{figure}

Gathering all the pieces, the full on-shell action over the WdW patch  is obtained by summing over the contributions \eqref{s1},\eqref{s2},\eqref{s3},\eqref{s4}, and \eqref{Full GHY term after dimensional reduction}.
In the UV regime or the linear dilaton region (\ie\ when $\epsilon/\beta_H\ll1$), the action complexity (obtained by summing over the contributions \eqref{SR},\eqref{SPhi}, \eqref{Scc},\eqref{H-field contribution UV limit},and \eqref{UVGHY2}) diverges as
\begin{equation}\label{CA}
C_\mathcal{A} = \frac{L_x c}{ 3\pi^2 \beta'_{H}} \left[\frac{{\beta'_{H}}^2}{2\pi \epsilon^2}-2\pi \log\left(\frac{\beta'_H}{\pi\epsilon}\right)+ \pi +O\left(\frac{\epsilon}{\beta'_H}\right)\right].
\end{equation}
Comparison this action complexity result with the (static frame) volume complexity expression \eqref{CV@0T} reveals that the leading divergence structure (\ie\ the quadratic divergent term) and the constant term in both cases are identical. The subleading logarithmic divergences differ by a negative sign.
In the IR  (\ie\ when $\epsilon/\beta'_{H}\gg1$) the action complexity takes the form
\begin{eqnarray}
	\lim_{\epsilon/\beta'_{H} \gg 1}C_\mathcal{A}=\frac{cL_x}{18\pi^3\beta'_H}\left(\frac{\beta'_H}{\epsilon}\right)^2+O\left({\beta'_H}^5/\epsilon^5\right)~. \label{CA_IR_0T}
\end{eqnarray}
Thus in pure $AdS_3$ limit, i.e. $\beta'_H=0$, the action complexity vanishes. This is in precise agreement with the analysis performed in \cite{Reynolds:2016rvl}, a dimensional accident (there is a coefficient $C_d=d-2$ in the pure AdS$_{d+1}$ action complexity). Unlike the volume complexity, the action complexity decreases faster. A comparison between volume complexity and action complexity is presented in figure \ref{img3}.
Thus overall, both the volume complexity and the action complexity diverges quadratically in the UV (\ie\ when $\epsilon/\beta_H\to 0$). However, as $\epsilon/\beta_H$ increases, the action complexity decreases (much faster than volume complexity) monotonically eventually going to 0 in the deep IR. This discrepancy can be traced back to the boundary being $1+1$ dimensional, the action complexity has a coefficient $C_d=d-2$. But in general it is understood that complexity is arbitrary or ambiguous up to such numerical factors and in general dimensions the volume and action complexity divergences will match both in the UV and IR.
\\
\section{Action Complexity for null WAdS$_3$} \label{nullWAdS3 CA}
We conclude this work by presenting the results for the action complexity of the null warped AdS$_3$ defined by the limit $\lambda=\epsilon_+=0$. We cannot taking this (singular) limit directly in the action complexity expressions e.g. in \eqref{CA}, because the static frame does not exist in this special case. So instead work it out from scratch in the stationary Lorentz frame in the boundary, for which the dual metric  is also stationary. In this case the dilaton field is simply a constant $	e^{2\Phi}.
=g_s^2$, which points out to the fact the dual boundary theory, a WCFT has a scale (Weyl) symmetry. The null WAdS$_3$ metric in stationary coordinates is
\begin{equation}
	\begin{split}
	ds^{2} & =k\,l_{s}^{2}\frac{dU^{2}}{U^{2}}-kU^{2}\left(1+kU^{2}\epsilon_{-}^{2}\right)dt^{2}+2\left(kU^{2}\right)^{2}\epsilon_{-}^{2}\,dt\,dx+kU^{2}\left(1-kU^{2}\epsilon_{-}^{2}\right)\,dx^{2}\nonumber.
	\end{split}
\end{equation}
To construct the boundaries of the WdW patch for this stationary metric, we first reorganize the null WAdS$_{3}$ metric in the form
\begin{equation}
ds^2=\frac{kU^{2}}{1-kU^{2}\epsilon_{-}^{2}}\left[\frac{l_{s}^{2}\left(1-kU^{2}\epsilon_{-}^{2}\right)}{\left(U^{2}\right)^{2}}dU^{2}-dt^{2}\right]+kU^{2}\left(1-kU^{2}\epsilon_{-}^{2}\right)\left[dx+\frac{kU^{2}\epsilon_{-}^{2}}{1-kU^{2}\epsilon_{-}^{2}}dt\right]^{2}.\label{eq: reorganized metric}
\end{equation}
So the light rays ($ds^{2}=0$) are then given by the equations/conditions,
\begin{align*}
	dt^{2} & =\frac{l_{s}^{2}\left(1-kU^{2}\epsilon_{-}^{2}\right)}{\left(U^{2}\right)^{2}}dU^{2},\\
	dx & =-\frac{kU^{2}\epsilon_{-}^{2}}{1-kU^{2}\epsilon_{-}^{2}}dt.
\end{align*}
These equations can be simultaneously solved by
\begin{equation}
	t_{\pm}(U)=T_{0}\pm l_{s}\int_{U}^{l_{s}/\epsilon}\:dU\,\frac{\sqrt{1-kU^{2}\epsilon_{-}^{2}}}{U^{2}},\label{eq: null ray t}
\end{equation}
\begin{equation}
	x_{\pm}(U)=x_{0}\mp k\,l_{s}\epsilon_{-}^{2}\int_{U}^{l_{s}/\epsilon}\:\frac{dU}{\sqrt{1-kU^{2}\epsilon_{-}^{2}}}.\label{eq: null ray x}
\end{equation}
Here the $\pm$ refer to the future and past directed light rays starting
from the cutoff surface at $x=x_{0},t=T_{0}$. The WdW patch boundary
at time $T_{0}$ is then described by the null surface obtained by
the collection of null rays obtained by varying $U$ and $x_{0}$.
Assuming the range of $x$ is from $\left[-L_{x}/2,+L_{x}/2\right]$,
one sees that for a fixed $U$, 
\[
dx=dx_{0}.
\]
As a result for the spacetime volume terms in the action (Ricci, cc,
Kalb-Ramond piece etc.) the ranges of integration are
\[
\int dx=\int dx_{0}=L_{x}~,
\]
and 
\begin{equation}
	\int_{t_{-}(U)}^{t_{+}(U)}dt=2l_{s}\int_{U}^{l_{s}/\epsilon}\:dU\,\frac{\sqrt{1-kU^{2}\epsilon_{-}^{2}}}{U^{2}}.\label{eq: range of t integration}
\end{equation}
\\

\subsection{Bulk Action terms}
The supergravity action we are going to evaluate in the volume is 
\begin{align*}
S=\frac{1}{16\pi G_N}\int dU dt dx\sqrt{-g}e^{-2(\Phi(U)-\Phi(0))} \left(R-4\Lambda-\frac{1}{12}\tilde{H}^2+4g^{\mu\nu}\partial_{\mu}\Phi\partial_{\nu}\Phi\right)~.
\end{align*}
Integral measure appearing in the string frame metric turns out to be
\begin{align}\nonumber
	e^{-2(\Phi(U)-\Phi_{0})}\sqrt{-g}=k^{3/2}l_sU~.
\end{align}
\begin{itemize}
	\item \textbf{Einstein Hilbert term in the action}
	\begin{align}
		S_{EH}&=\frac{1}{16\pi G_N}\int dU dt dx\sqrt{-g}e^{-2(\Phi(U)-\Phi(0))} \left(R-4\Lambda\right)~,\nonumber\\
		&=\frac{-\sqrt{k}L_x}{4\pi G_N}\int dU\, U \int_{U}^{l/\epsilon}dU'\frac{\sqrt{1-\epsilon_-^2 k U'^2}}{U'^2}~,\nonumber\\
		&=-\frac{c L_x}{24 \pi  \sqrt{k} \epsilon _- l_s}\left(\frac{\sqrt{k} \epsilon _- l_s}{\epsilon }\sqrt{1-\frac{k \epsilon _-^2 l_s^2}{\epsilon ^2}}+\sin ^{-1}\left(\frac{\sqrt{k} \epsilon _- l_s}{\epsilon }\right)\right)~.\label{eq:EH piece}
	\end{align}
	\item \textbf{Kalb Ramond term in the action}
	\begin{align}
		S_{KR}&=\frac{1}{16\pi G_N}\int dU dt dx\sqrt{-g}e^{-2(\Phi(U)-\Phi(0))} \left(-\frac{1}{12}\tilde{H}^2\right)~,\nonumber\\
		&=\frac{-L_x\sqrt{k}}{4\pi G_N }\int dU\,U \int_{U}^{l_s/\epsilon}dU'\frac{\sqrt{1-\epsilon_-^2 k U'^2}}{U'^2}~,\nonumber\\
		&=-\frac{c L_x}{24 \pi  \sqrt{k} \epsilon _- l_s}\left(\frac{\sqrt{k} \epsilon _- l_s}{\epsilon }\sqrt{1-\frac{k \epsilon _-^2 l_s^2}{\epsilon ^2}}+\sin ^{-1}\left(\frac{\sqrt{k} \epsilon _- l_s}{\epsilon }\right)\right)\label{eq:KR piece}
	\end{align}
\item \textbf{Dilaton term in the action}
\begin{align}
S_{KR}&=\frac{1}{16\pi G_N}\int dU dt dx\sqrt{-g}e^{-2(\Phi(U)-\Phi(0))} \left(4g^{\mu\nu}\partial_{\mu}\Phi\partial_{\nu}\Phi\right)~.
\end{align}
On account of dilaton being trivially a constant, the dilaton has vanishing contribution towards the gravitational action. 
	\item \textbf{Gravitational action volume contribution}
	\begin{align}
		S&=\frac{1}{16\pi G_N}\int dU dt dx\sqrt{-g}e^{-2(\Phi(U)-\Phi(0))} \left(R-4\Lambda-\frac{1}{12}\tilde{H}^2+4g^{\mu\nu}\partial_{\mu}\Phi\partial_{\nu}\Phi\right)~,\nonumber	\\
		&=-\frac{c L_x}{12 \pi  \sqrt{k} \epsilon _- l_s}\left(\frac{\sqrt{k} \epsilon _- l_s}{\epsilon }\sqrt{1-\frac{k \epsilon _-^2 l_s^2}{\epsilon ^2}}+\sin ^{-1}\left(\frac{\sqrt{k} \epsilon _- l_s}{\epsilon }\right)\right)~,\label{eq:bulk action}\\
		&=-\frac{c L_x}{6 \pi  \epsilon }+\frac{c k \epsilon _-^2 l_s^2 L_x}{36 \pi  \epsilon ^3}+O(\epsilon_-^4)~.\nonumber
	\end{align}
\end{itemize}
	The choice of the coupling is bounded from above  by the UV cutoff via $\epsilon >\epsilon_- \sqrt{k} l_s$.
\subsection{GHY surface terms from null boundaries of WdW patch}
Now for the GHY calculation let's first check that there is no mixed/cross
term in the induced metric from the $t,x$ part of the metric (\ref{eq: reorganized metric})
 by plugging in (\ref{eq: null ray t},\ref{eq: null ray x}).
For the future boundary of the WdW patch, keeping in mind that $dU<0$,
\[
dt_{+}=-\frac{\sqrt{1-kU^{2}\epsilon_{-}^{2}}}{U^{2}}dU,
\]
and,
\[
dx_{+}=dx_{0}+l_{s}\frac{k\epsilon_{-}^{2}\:dU}{\sqrt{1-kU^{2}\epsilon_{-}^{2}}}.
\]
Then, the $t,x$ part of the metric simplifies to
	\begin{align*}
		kU^{2}\left(1-kU^{2}\epsilon_{-}^{2}\right)\left[dx_{+}+\frac{kU^{2}\epsilon_{-}^{2}}{1-kU^{2}\epsilon_{-}^{2}}dt_{+}\right]^{2} 
		& =kU^{2}\left(1-kU^{2}\epsilon_{-}^{2}\right)dx_{0}^{2}.
	\end{align*}
Similarly, one can also show that for the past boundary of the WdW
patch, the $t,x$ part of the metric (\ref{eq: reorganized metric})
becomes,
\[
kU^{2}\left(1-kU^{2}\epsilon_{-}^{2}\right)\left[dx_{-}+\frac{kU^{2}\epsilon_{-}^{2}}{1-kU^{2}\epsilon_{-}^{2}}dt_{-}\right]^{2}=kU^{2}\left(1-kU^{2}\epsilon_{-}^{2}\right)dx_{0}^{2}.
\]
Now we can turn on the timelike deformation (note that this timelike
deformation is a separate/disjoint deformation of the future and past
null surfaces),
\[
\left(t-T_{0}\right)^{2}=\left(1+\delta\right)l_{s}^{2}\:A^{2}(U),\quad\quad \text{where,}\,\, A(U)\equiv\int_{U}^{l_{s}/\epsilon}\:dU\,\frac{\sqrt{1-kU^{2}\epsilon_{-}^{2}}}{U^{2}}~.
\]
Plugging this in the $t,U$ part of the metric (\ref{eq: reorganized metric}), the timelike (near null) part of the induced
metric on the deformed surface is
\[
\frac{kU^{2}}{1-kU^{2}\epsilon_{-}^{2}}\left[\frac{l_{s}^{2}\left(1-kU^{2}\epsilon_{-}^{2}\right)}{\left(U^{2}\right)^{2}}dU^{2}-dt^{2}\right]=-\delta\,\frac{k\,l_{s}^{2}}{U^{2}}dU^{2}.
\]
Thus the induced metric on the timelike deformed surfaces is,
\begin{equation}
	ds_{\gamma}^{2}=-\delta\,\frac{k\,l_{s}^{2}}{U^{2}}dU^{2}+kU^{2}\left(1-kU^{2}\epsilon_{-}^{2}\right)dx_{0}^{2},\label{eq: metric on the deformed timelike surface}
\end{equation}
and so
\[
\sqrt{-\gamma}=\sqrt{\delta}k\,l_{s}\sqrt{\left(1-kU^{2}\epsilon_{-}^{2}\right)}.
\]
(The range of $x$-integration for this GHY term is $\int dx=\int dx_{0}=L_{x}$.)\\
\\
\\
Now to figure out the normal to the surface we first recall that on
the timelike deformed surfaces the changes in $dt,dx,dU$ are constrained
by the equations
\[
dt=\mp\sqrt{(1+\delta)}l_{s}\frac{\sqrt{1-kU^{2}\epsilon_{-}^{2}}}{U^{2}}dU,\quad dx_{+}=dx_{0}\pm l_{s}\frac{k\epsilon_{-}^{2}\:dU}{\sqrt{1-kU^{2}\epsilon_{-}^{2}}}.
\]
So on the deformed surface the (near null) timelike tangent vector
can be taken to be,
\begin{align*}
	T_{1}^{\mu} & \propto \left(1,\,\mp\sqrt{(1+\delta)}l_{s}\frac{\sqrt{1-kU^{2}\epsilon_{-}^{2}}}{U^{2}},\,\pm l_{s}\frac{k\epsilon_{-}^{2}}{\sqrt{1-kU^{2}\epsilon_{-}^{2}}}\right)
\end{align*}
while the spacelike tangent vector can be taken to be,
\[
T_{2}^{\mu}\propto\left(0,0,dx_{0}\right)=\left(0,0,1\right).
\]
Let the components of the normal vector be,
\[
N_{\mu}=\left(n_{U},n_{t},n_{x}\right)
\]
So the normal vector components satisfy the equation $N_{\mu}T_{1}^{\mu}=N_{\mu}T_{2}^{\mu}=0$
or,
\[
n_{U}\mp\sqrt{(1+\delta)}l_{s}\frac{\sqrt{1-kU^{2}\epsilon_{-}^{2}}}{U^{2}}n_{t}\pm l_{s}\frac{k\epsilon_{-}^{2}}{\sqrt{1-kU^{2}\epsilon_{-}^{2}}}n_{x}=0.
\]
and
\[
n_{x}=0.
\]
Thus only $n_{U}$ and $n_{t}$ are non-zero. Thus the unnormalized
normals can be taken to be,
\begin{align*}
	n_{U}&=1&n_{t}&=\pm \frac{U^2}{\sqrt{\delta +1} l_s \sqrt{1-k U^2 \epsilon _+^2}}&n_{x}&=0.
\end{align*}
Let's first start working with the upper portion of the timelike (near null) surface (regulating surface).
The normalised outward normal to the upper portion of the regulating timelike surface is
\begin{align}
	n^U&=\frac{\sqrt{\delta +1} U}{\sqrt{\delta } \sqrt{k} l_s}~,&n^t&=-\frac{\sqrt{1-k U^2 \epsilon _-^2}}{\sqrt{\delta } \sqrt{k} U}~,&n^x&=\frac{\sqrt{k} U \epsilon _-^2}{\sqrt{\delta } \sqrt{1-k U^2 \epsilon _-^2}}~.\nonumber
\end{align}
The trace of the extrinsic curvature is
\begin{align}
	K&=\nabla_{\mu} n^{\mu}=\partial_U n^U+\partial_t n^t+\partial_xn^x+\underbrace{(\Gamma_{UU}^U+\Gamma_{tU}^t+\Gamma_{Ux}^x)}_{\text{$\frac{1}{U}$}}n^U\nonumber~,\\
	&=\frac{2 \sqrt{\delta +1}}{\sqrt{\delta } \sqrt{k} l_s}~,\nonumber\\
	&=\frac{2}{\sqrt{\delta } \sqrt{k} l_s}+O(\delta^{1/2})~.\label{eq:ext curvature}
\end{align}
Now for the lower portion of the deformed timelike surface, the outward unit normal is
\begin{align}
	n^{U}&=\frac{\sqrt{\delta +1} U}{\sqrt{\delta } \sqrt{k} l_s}~, &n^t&=\frac{\sqrt{1-k U^2 \epsilon _-^2}}{\sqrt{\delta } \sqrt{k} U}~, &n^x&=-\frac{\sqrt{k} U \epsilon _-^2}{\sqrt{\delta } \sqrt{1-k U^2 \epsilon _-^2}}~.\nonumber
\end{align}
The normalization constant and the extrinsic curvature are same as that for the upper portion of the regulating surface.\\

The GHY integral evaluates to be
\begin{align}
		S_{GHY}^{\delta}&=\frac{1}{8\pi G_N}\int_{L/2}^{-L/2} dx_0 \int_{0}^{l_s/\epsilon}dU\, e^{-2(\Phi-\Phi_0)}\sqrt{-\gamma}(K+n^{\mu}\partial_{\mu}\sigma)~,\nonumber\\
		&=\frac{ \sqrt{k}L_x}{4\pi G_N} \int_{0}^{l_s/\epsilon}dU\,\sqrt{1-k U^2 \epsilon _-^2}~,\nonumber\\
		&=\frac{c}{12 \pi  \sqrt{k} \epsilon _- l_s}\left(\frac{\sqrt{k} \epsilon _- l_s}{\epsilon } \sqrt{1-\frac{k \epsilon _-^2 l_s^2}{\epsilon ^2}}+\sin ^{-1}\left(\frac{\sqrt{k} \epsilon _- l_s}{\epsilon }\right)\right)~.\label{eq:GHY boundary}
	\end{align}
Full action complexity obtained by summing (\ref{eq:bulk action}) and (\ref{eq:GHY boundary}) evaluates to be zero.

\section{Discussion \& Outlook} \label{DiscOut}
In this work, we investigated aspects of the little string theory (LST),  which is the holographic (boundary) dual of a string theory in a target space that interpolates between $AdS_3$ in the IR to an anisotropic spacetime with a linear dilaton and NS-NS B-field that violates Lorentz isometry in the UV. This LST can be regarded as a nonlocal Lornetz (boost) noninvariant UV deformation of a local CFT$_2$ by "single trace" analogue of the usual irrelevant $T\overline{T}, J\overline{T}, T\overline{J}$ operators. Our tool of investigation was holographic complexity, specifically the holographic volume complexity (CV) and holographic action complexity (CA) prescriptions. Our aim was to identify and, if possible, isolate the signatures of the Lorentz-violation in holographic complexity. This work extends our previous work \cite{Chakraborty:2020fpt} in two respects. In our previous work we looked at LST with just nonlocality ($T\overline{T}$ deformation) without turning on Lorentz-violating couplings ($J\overline{T}, \overline{J}T$ deformations), and we omitted the interesting or informative case of subregion complexity. Here we summarize of our findings:
\\
\begin{itemize}	
	\item Volume complexity was evaluated for two different frames related by a boost - namely the stationary and static frame, while the action complexity was evaluated only in the static frame. Both the volume complexity and action complexity are UV divergent and hence manifestly regulator dependent. In the regime where the UV cutoff (lattice spacing) is shorter than the Hagedorn scale of the LST, the leading piece diverges \emph{quadratically} with the UV cutoff (cf Eq. \eqref{CV1div}, \eqref{CV@0T}, and Eq. \eqref{CA}). We identify this leading quadratic divergence as the characteristic signature of nonlocal nature of the LST. Modulo an overall factor ambiguity (which is well known in the literature) the leading divergences for both complexities (volume and action) in the static frame agree and have the same sign.
\\
	\item There are subleading logarithmic divergences in both volume complexity \eqref{CV@0T} and action complexity expressions \eqref{CA} which have the same magnitude but differ in sign. The universal coefficient \eqref{total dof} of this log divergent term can be interpreted as the total number of degrees of freedom in the LST  with the Hagedorn scale, $\beta_H$ treated as the lattice spacing.
\\	
   \item The characteristic length scale of nonlocality is different in the stationary and static frames. For the stationary frame, this nonlocality scale is given by $\rho_H=\frac{2 \pi}{\sqrt{3}}l_s\,\sqrt{k\left(\lambda - \left(\epsilon_++\epsilon_-\right)^2\right)}$ while in the static frame it is given by $\beta'_H=2\pi l_s\,\sqrt{k\left(\lambda - 4 \epsilon_+ \epsilon_-\right)}$. This effect of changing the nonlocality scale upon boosting to a different Lorentz frame is the characteristic signature of the fact that the theory is not boost invariant. 
\\	
	\item In the opposite regime, \ie\ when the UV cutoff is much larger than the nonlocality scale, the volume complexity expectedly reduces to that of a local field theory \ie\ having linear divergence (corresponding to a single spatial dimension) \eqref{CVir}  matching that of a CFT with the central charge equal to the Brown-Henneaux expression derived from a pure $AdS_3$ calculation. Similarly, in this limit the action complexity too reproduces the expected pure $AdS_3$ answer \eqref{CA_IR_0T} \cite{Reynolds:2016rvl, Carmi:2016wjl}.
\\		
	\item The subregion volume complexity as a function of the subregion size (length), in both the stationary and static frames, displays a sharp transition as the subregion size is varied across a critical subregion size in both frames. In the stationary frame this critical length is $L_c = \frac{\pi \sqrt{k\,\lambda\,\lambda'}}{2\sqrt{\mu}} l_s$) while in the static frame this critical scale is $L'_c=\frac{\pi \sqrt{k\,\lambda}}{2}\,l_s$ where $\lambda'\equiv\lambda-4\epsilon_+ \epsilon_-$ and $\mu\equiv\lambda -\left(\epsilon_++\epsilon_-\right)^2$. We identify this phase transition of subregion volume complexity with the Hagedorn phase transition which have been previously observed in entanglement entropy as well as the thermodynamics \cite{Chakraborty:2020xyz, Chakraborty:2020udr}.
\\
     \item Upon setting $\lambda=\epsilon_+=0$, one obtains null warped AdS$_3$ metric in the bulk (with nonzero dilaton and B-field turned on). This point in the parameter space is out of the unitarity regime and hence corresponds to a boundary dual WCFT which does not admit a UV completion. Nevertheless, one can still study it as an effective theory, which violates locality and Lorentz boost symmetry. The volume complexity expression confirms that the UV cutoff (lattice spacing) cannot be made arbitrarily small and is bounded by the warping parameter $\epsilon_-$, namely $\epsilon>\sqrt{k} l_s \epsilon_-$. Below this the volume complexity does not make sense (turns imaginary). The action complexity on the other hand vanishes, perhaps due to a dimensional accident akin to the unwarped AdS$_3$ case. The subregion volume complexity is a monotonically increasing function of the subregion size $L$, but there is no Hagedorn like phase transition. Surprisingly, the holographic entanglement entropy of this null warped AdS$_3$ solution dual to the highly nonlocal, Lorentz violating WCFT has a logarithmic divergence, same as that of pure AdS$_3$ dual to local CFT$_2$. The universal coefficient of the log divergence however receives a contribution from the warping parameter.
\\ 
\end{itemize}

The analysis of holographic Wilson loop \cite{Chakraborty:2018aji}, holographic entanglement entropy \cite{Chakraborty:2018kpr,Asrat:2019end,Asrat:2020uib,Chakraborty:2020udr} and thermodynamics \cite{Chakraborty:2020xyz,Chakraborty:2020swe} for the LST naturally reveals the nonlocality scale through some pathologies in the physical observables. For example, the free energy and the entropic c-function diverges as the RG scale approaches the nonlocality scale of LST. The partition function in the thermodynamic limit develops a branch cut singularity as the temperature approaches the Hagedorn temperature of LST. So it is perhaps natural to expect that the subregion complexity would also show such singular/pathological traits when the subregion size approaches the nonlocality scale. This was one of the main reasons to include the subregion complexity in this work. But to the contrary, In our analysis of holographic complexity, we didn't come across such pathologies.\\

      It would be interesting to work out the action complexity in the stationary frame. This will entail solving a more involved technical problem of constructing WdW patches for stationary metrics \cite{AlBalushi:2020ely, Bernamonti:2021jyu}. Although we don't expect any radically different answers for the action complexity in the stationary frame compared to volume complexity (as evidenced by the strong similarities in static frame counterparts), it will still be nice to close this gap. We leave this exercise for a future work as well.\\
      
	  So far everything we have done here corresponds to the zero temperature case. Since the LST is a nonlocal theory for which we do not have much intuition, there might appear novel exotic divergences compared to the zero temperature case - so perhaps it is important that one studies the finite temperature case. In fact such a computation was performed in our previous paper \cite{Chakraborty:2020fpt} for the LST dual to $\mathcal{M}_3$. There we computed the finite temperature the action complexity using the bulk dual black hole geometry. In particular, we consider the thermofield double state of two LST's for which the dual bulk geometry is an eternal $\mathcal{M}_3$ black hole. Qualitatively, the action complexity\footnote{For such a black brane bulk background analytic calculations of the maximal volume slice without any approximations are not possible, and so we abandoned the volume complexity scheme. Instead we numerically computed the action complexity exactly.} at finite temperature showed the same behavior as that of the zero temperature case. More importantly, \emph{no newer divergences} compared to the zero temperature case was found (perturbatively up to second order in finite temperature corrections). For the $J\overline{T}, T\overline{J}$ deformed theory, such a black brane background dual to finite temperature LST with $J\overline{T}, T\overline{J}$ deformations (thermofield double) was recently worked out \cite{Apolo:2021wcn}. It would be interesting to carry out the computation of action complexity for this eternal black blane geometry - although from the insights gathered from our past paper and the patterns established in the current work, we do not expect to see novel UV divergence structures because for this LST Lorentz violating effects seem to be mixed with nonlocality effects and they come in a single joint package. We leave this exercise for future work.\\
	  
Finally, one needs to study the characteristics of holographic complexity for a larger class of nonlocal theories, not necessarily LST as was done for the case of entanglement entropy \cite{Barbon:2008ut, Karczmarek:2013xxa}. This will help us settle the issue of the hypervolume UV divergence structure i.e., whether one should always expect a general "$\text{volume}+1$" scaling for the leading term or something more complicated related to the cause or origin of the nonlocality.\\
	
	A well known issue in the holographic proposals for evaluating circuit complexity of the boundary theory is that there is neither any direct specification of the reference state nor the unitary gates (operators) which constitute the circuits. These issues are still not settled in the holography literature. The only thing one can definitively state is that In the AdS/CFT case the reference state is clearly \emph{not} the CFT vacuum, since the holographic complexity is nonzero for pure AdS geometry (equivalently the CFT vacuum state). We are unable to shed any further light on these issues in our current work. However, at the end of the day, the LST$_2$ we study is obtained by deforming a CFT$_2$ by a set of irrelevant deformations. Hence, we might as well use the \emph{exact same} set of unitary gates and the \emph{exact same} reference state as used for the initial CFT$_2$ which we UV-deformed. This is sensible since the LST complexity obtained here smoothly go over to the CFT (pure AdS) complexity once the UV deformation parameters are set to zero. We \emph{can} comment on the target state though. In the CFT$_2$ case the target state was the CFT vacuum, invariant under the $SL(2,\mathbb{R})\times SL(2,\mathbb{R})$ symmetry. For the LST$_2$ case obtained by a single trace $T\overline{T}$ deformation of the CFT$_2$ which was the subject of our last paper \cite{Chakraborty:2020fpt}, the target state was the ``no string" vacuum state, \ie  the vacuum of the BRST cohomology of the coset $\frac{SL(2,\mathbb{R}) \times U(1)}{U(1)}$ at zero temperature \cite{Chakraborty:2020yka}. For the LST in this paper, obtained after further breaking the Lorentz boost symmetry, it is not yet clear that a coset description can be provided. The states can be labeled by the left over symmetry generators corresponding to time translations, translation in the $x$-direction and the $U(1)$ left and right moving sector charges $J,\overline{J}$. In fact the vacuum here has nontrivial quantum numbers, $U(1)$ charge(s) since after dimensional reduction of the $y$-circle the $3$d bulk has nontrivial KK $U(1)$ gauge fields turned on. \\
	
Since this correspondence between LST and String backgrounds with asymptotically linear dilaton backgrounds is a non-AdS/non-CFT case of holography, perhaps a more direct exercise would be to work out the holographic dictionary in the vein of GKPW and/or as HKLL \cite{Hamilton:2005ju, Hamilton:2006az, Hamilton:2006fh}. As we have already remarked in our previous work \cite{Chakraborty:2020fpt}, one anticipates some surprising twists in the bulk-boundary map/dictionary in this case because such maps will reconstruct \emph{local} supergravity excitations in the bulk, from \emph{nonlocal} excitations of the LST in the boundary. In the traditional AdS/CFT setting such local bulk reconstruction maps are to a great extent determined by the (conformal) symmetry preserved by the boundary state, as well as \emph{locality/microcausality} properties of the boundary CFT correlators, \eg\ in the HKLL recipe locality in the bulk directions parallel to the boundary is a simple and direct consequence of boundary (CFT) locality, and the nontrivial challenge was to understand bulk locality in the emergent radial (holographic) direction from the locality in the boundary (transverse) directions. However in the case of LST, the field theory is \emph{nonlocal} and Lorentz symmetry is broken. It will be interesting to identify which alternative properties of a nonlocal theory such as the LST plays the crucial role in emergence of the quasilocal semiclassical bulk space in both radial as well transverse directions.
\\
\section*{Acknowledgements}
The authors would like to thank Soumangsu Chakraborty for valuable discussions pointing us to numerous woks and results in the $J\overline{T},\overline{J}T$ literature. The work of GK was supported partly by a Senior Research Fellowship (SRF) from the Ministry of Education (MoE)\footnote{formerly the Ministry of Human Resource Development (MHRD)}, Govt. of India and partly from the RDF fund of SR: RDF/IITH/F171/SR. The work of SM is supported by the Department of Science and Technology (DST) of the Ministry of Science and Technology of India by a fellowship under the ``Innovation in Science Pursuit for Inspired Research (INSPIRE)" scheme, DST/INSPIRE Fellowship/2019/IF190561.  The work of SR is supported by the IIT Hyderabad seed grant SG/IITH/F171/2016-17/SG-47.  SR also thanks Arpan Bhattacharyya for valuable discussions, especially drawing attention to the work \cite{Goto:2018iay} and discussions on the WAdS$_3$ case.\\
\appendix
\section{The $\sigma$-model and the $4$ dimensional background} \label{sigma model to 4d}
The euclidean signature worldsheet (bosonic) sigma model action (ignoring the dilaton piece $\alpha'\Phi R$ at leading order in $\alpha'$) is,
\begin{equation}
	I=\frac{1}{4\pi}\int d^{2}\sigma\sqrt{g}\:\left(g^{ab}G_{\mu\nu}+i\epsilon^{ab}B_{\mu\nu}\right)\partial_{a}X^{\mu}\partial_{b}X^{\nu}.\label{eq: worldsheet metric in alpha' =00003D 1}
\end{equation}
where we have set $\alpha'=1$. Working in conformal gauge, i.e. with the worldsheet metric
\[
g_{11}=g^{11}=g_{22}=g^{22}=1,g_{12}=g_{21}=g^{12}=g^{21}=0.
\]
and the Levi-Civita tensor,
\[
\epsilon_{12}=1,\epsilon_{21}=-1,\epsilon_{11}=\epsilon_{22}=0.
\]
\[
\Rightarrow\epsilon^{12}=1,\epsilon^{21}=-1,\epsilon^{11}=\epsilon^{22}=0.
\]
Substituting these in (\ref{eq: worldsheet metric in alpha' =00003D 1}),
we get
\begin{equation}
	I=\frac{1}{4\pi}\int d^{2}\sigma\:\left[G_{\mu\nu}\left(\partial_{1}X^{\mu}\partial_{1}X^{\nu}+\partial_{2}X^{\mu}\partial_{2}X^{\nu}\right)+iB_{\mu\nu}\left(\partial_{1}X^{\mu}\partial_{2}X^{\nu}-\partial_{2}X^{\mu}\partial_{1}X^{\nu}\right)\right].\label{eq: worldsheet metric in alpha' =00003D 1-1}
\end{equation}
Switching to lightcone worldsheet coordinates, $z=\sigma^{1}+i\sigma^{2}$
and $\overline{z}=\sigma^{1}-i\sigma^{2}$,  the sigma model action,
\begin{align}
	I & = \frac{1}{2\pi}\int d^{2}z\:\left[\frac{G_{\mu\nu}+B_{\mu\nu}}{2}\partial X^{\mu}\:\overline{\partial}X^{\nu}+\frac{G_{\mu\nu}-B_{\mu\nu}}{2}\overline{\partial}X^{\mu}\:\partial X^{\nu}\right]\label{eq: String WS action in lc coordinates}
\end{align}
To compare this with Eq. (4.5) of \cite{Chakraborty:2019mdf} we have to use a $4$-dimensional target space, i.e. $X^{\mu}$'s
can be the four coordinates $\phi,y,\gamma,\bar{\gamma}$, whereby we readily read off,
\begin{equation}
	G_{\phi\phi}=k,\quad G_{yy}=\frac{h}{f},\label{eq: 4-dimensional metric components pure}
\end{equation}
and,
\begin{align}
	G_{\gamma\overline{\gamma}}-B_{\gamma\bar{\gamma}}=hk, & \qquad G_{\gamma\overline{\gamma}}+B_{\gamma\bar{\gamma}}=0,\nonumber \\
	G_{y\gamma}+B_{y\gamma}=2\epsilon_{+}h\sqrt{k}, & \qquad G_{y\gamma}-B_{y\gamma}=0,\nonumber \\
	G_{y\overline{\gamma}}-B_{y\overline{\gamma}}=2\epsilon_{-}h\sqrt{k}, & \qquad G_{y\overline{\gamma}}+B_{y\overline{\gamma}}=0.\label{eq: 4-dimensional coefficients from sigmal model comparisons}
\end{align}
Solving these we obtain the $4$ dimensional metric components,
\begin{equation}
	G_{\overline{\gamma}\gamma}=G_{\gamma\overline{\gamma}}=\frac{hk}{2},\qquad G_{y\gamma}=G_{\gamma y}=\epsilon_{+}h\sqrt{k},\qquad G_{\overline{\gamma}y}=G_{y\overline{\gamma}}=\epsilon_{-}h\sqrt{k},\label{eq: 4-dim.  metric components mixed}
\end{equation}
or more conventionally the $4$ dimensional line element.
\begin{equation}
	ds_{4}^{2}=k\;d\phi^{2}+\frac{h}{f}dy^{2}+hk\;d\gamma\:d\overline{\gamma}+2\epsilon_{+}h\sqrt{k}\;dy\:d\gamma+2\epsilon_{-}h\sqrt{k}\:dy\:d\overline{\gamma}.\label{eq: 4 dimensional line element}
\end{equation}
This line element expression is exactly the same as in Eq. (4.9) of
\cite{Chakraborty:2020xyz} with $\alpha'=1$ and modulo the decoupled $T^{3}$ and $S^{3}$ directions. In $U,\gamma,\overline{\gamma}$ coordinates,
\begin{equation}
	ds^{2}=\frac{k}{U^{2}}dU^{2}+\frac{h}{f}dy^{2}+hk\;d\gamma\:d\overline{\gamma}+2\epsilon_{+}h\sqrt{k}\;dy\:d\gamma+2\epsilon_{-}h\sqrt{k}\:dy\:d\overline{\gamma},\label{eq: 4-dimensional line element in U-coordinates}
\end{equation}
while the $4$ dimensional $B$-field components,
\begin{equation}
	B_{\gamma\bar{\gamma}}=-\frac{hk}{2},\qquad B_{y\gamma}=-B_{\gamma y}=\epsilon_{+}h\sqrt{k},\qquad B_{y\overline{\gamma}}=-B_{\overline{\gamma}y}=-\epsilon_{-}h\sqrt{k}.\label{eq: 4-dimensional B-field components}
\end{equation}
or in component-basis form,
\begin{align}
	B & =-\frac{hk}{2}\;d\gamma\wedge d\overline{\gamma}+\epsilon_{+}h\sqrt{k}\;dy\wedge d\gamma-\epsilon_{-}h\sqrt{k}\;d\overline{\gamma}\wedge dy.\label{eq: 4-dim B-field}
\end{align}
The $4$ dimensional field strength, $H$ is thus
\begin{equation}
	H=dB=-\frac{h'k}{2}\;dU\wedge d\gamma\wedge d\overline{\gamma}+\epsilon_{+}h'\sqrt{k}\;dU\wedge dy\wedge d\gamma-\epsilon_{-}h'\sqrt{k}\;dU\wedge dy\wedge d\overline{\gamma}.\label{eq: H-field with basis}
\end{equation}
where $h'=\frac{dh}{dU}$. The components of the $H$ 3-form
field strength tensor,
\begin{equation}
	H_{U\gamma\overline{\gamma}}=-\frac{h'k}{2},\qquad H_{Uy\gamma}=\epsilon_{+}h'\sqrt{k},\qquad H_{Uy\overline{\gamma}}=-\epsilon_{-}h'\sqrt{k}.\label{eq: H-field strength tensor covariant}
\end{equation}
From these we compute that
\begin{equation}
	-H^{2}=\frac{6U^{2}\:h'(U)^{2}}{k\:l_{s}^{2}\:h(U)^{2}}.\label{eq: minus H^2 in 4D}
\end{equation}
Here we have restored factors of $l_{s}$. The $4$ dimensional volume element is,
\begin{equation}
	\sqrt{G}d^{4}x=\frac{hk^{3/2}l_{s}^{3}}{2U}dU\:d\gamma\:d\overline{\gamma}dy.\label{eq: sqrt of dim metric}
\end{equation}
Thus the $4$ dimensional onshell Lagrangian contribution of the three form field strength $H$ is,
\begin{equation}
	\sqrt{G}\:\left(-\frac{1}{12}H^{2}\right)=\frac{l_{s}\sqrt{k}}{4}\left(\frac{Uh'^{2}}{h}\right).\label{eq: LHS of (1.24) of Pope}
\end{equation}
The vanishing of the worldsheet beta functions give the Dilaton, $e^{-2\left(\Phi_{(4)}-\Phi_{0}\right)}=kU^{2}h^{-1},$ and thus the $H$ term contribution to the $4$-dimensional gravity action is,
\begin{equation}
	\frac{1}{16\pi G_{N}^{(4)}}\int\sqrt{G}\:d^{4}x\;e^{-2\left(\Phi_{(4)}-\Phi_{0}\right)}\left(-\frac{1}{12}H^{2}\right)=\frac{1}{16\pi G_{N}}\left(\frac{l_{s}}{\sqrt{k}}\right)\int dU\:\frac{h^{2}}{U^{3}}\int d\gamma\:d\overline{\gamma}\;.\label{eq: H term action in 4d in lc coordinates}
\end{equation}
Here we have already performed the $y$-integration: $\int dy=2\pi R_{y}$,
$R_{y}$ being the radius of the $y$-circle. Here the $3$ dimensional
Newton's constant is defined as,
\begin{equation}
	G_{N}=\frac{G^{(4)}}{2\pi R_{y}}. \label{3d and 4d Newton's constant}
\end{equation}
Next we switch to $X$ and $T$ defined by,
\begin{align*}
	X & =\frac{\sqrt{k}l_{s}}{2\sqrt{\epsilon_{+}\epsilon_{-}}}\left(\epsilon_{+}\gamma+\epsilon_{-}\overline{\gamma}\right),\\
	T & =\frac{\sqrt{k}l_{s}}{2\sqrt{\epsilon_{+}\epsilon_{-}}}\left(\epsilon_{+}\gamma-\epsilon_{-}\overline{\gamma}\right).
\end{align*}
In such case we have
\[
dXdT=\frac{kl_{s}^{2}}{2}d\gamma\:d\overline{\gamma}
\]
and we have the $H$-field contribution (\ref{eq: H term action in 4d in lc coordinates}),
	\begin{align}
		\frac{1}{16\pi G_{N}^{(4)}}\int\sqrt{G}\:d^{4}x\;e^{-2\left(\Phi_{(4)}-\Phi_{0}\right)}\left(-\frac{1}{12}H^{2}\right) & =\frac{1}{16\pi G_{N}}\left(\frac{l_{s}}{\sqrt{k}}\right)\int dU\:\frac{h^{2}}{U^{3}}\left(\frac{2\int dX\:dT}{kl_{s}^{2}}\right)\nonumber \\
		& =\frac{L_{x}}{8\pi G_{N}l_{s}k^{3'2}}\int dU\:\frac{h^{2}}{U^{3}}\:\int dT.\label{eq: H-term contribution in action in X, T}
	\end{align}
\section{Kaluza-Klein reduction on the $y$ circle} \label{KK redone}
Here we repeat the exercise of KK reduction of the $y$-circle  to fill in some of the details, in particular, the KK reduced $3$ dimensional Kalb-Ramond field and the associated KK one-form gauge field expressions were omitted in \cite{Chakraborty:2019mdf}, as well as with the aim to check the equivalence of the $4$ dimensional and the $3$ dimensional (onshell) actions. The equivalence of the $4$ dimensional and $3$ dimensional actions guarantee that the action complexity remains the same before and after the KK reduction. For the KK reduction we closely follow Pope's review \cite{Pope:2000xxx} but departing
from its convention by setting
\begin{equation}
	\alpha=0,\beta=1,
\end{equation}
and calling the KK scalar $\sigma$ following \cite{Chakraborty:2019mdf}. This convention is advantageous because it implies,
\begin{equation}
	\sqrt{-G}\,e^{-2\Phi^{(4)}}=\sqrt{-g}\,e^{-2\Phi}.\label{identity 1}
\end{equation}\\
The $4$ dimensional metric in this convention is split up as,
\[
ds_{4}^{2}=ds_{3}^{2}+e^{2\sigma}\left(dy+\mathcal{A}_{\mu}dx^{\mu}\right)^{2}
\]
from which we can determine the $3$ dimensional metric components,
$g_{\mu\nu}$ and the gauge field components, $A_{\mu}$
\begin{align*}
	G_{yy} & =e^{2\sigma},\\
	G_{y\mu} & =e^{2\sigma}\mathcal{A}_{\mu},\\
	G_{\mu\nu} & =g_{\mu\nu}+e^{2\sigma}\mathcal{A}_{\mu}\mathcal{A}_{\nu}.
\end{align*}
Using (\ref{eq: 4-dimensional metric components pure}) and (\ref{eq: 4-dimensional coefficients from sigmal model comparisons})
we identify,
\[
e^{2\sigma}=\frac{h}{f},
\]
\[
\mathcal{A}_{\gamma}=\epsilon_{+}f\sqrt{k},\qquad\mathcal{A}_{\overline{\gamma}}=\epsilon_{-}f\sqrt{k},\qquad\mathcal{A}_{\phi}=0,
\]
\[
g_{\phi\phi}=k,\quad g_{\gamma\gamma}=-\epsilon_{+}^{2}hfk,\quad g_{\overline{\gamma}\overline{\gamma}}=-\epsilon_{-}^{2}hfk,\quad g_{\gamma\bar{\gamma}}=\frac{hk}{2}\left(1-2\epsilon_{+}\epsilon_{-}f\right).
\]
Thus the $3$ dimensional line element is,
\begin{align}
	ds_{3}^{2} & =k\left[d\phi^{2}-\epsilon_{+}^{2}hfd\gamma^{2}-\epsilon_{-}^{2}hfd\overline{\gamma}^{2}+h\left(1-2\epsilon_{+}\epsilon_{-}f\right)d\gamma d\overline{\gamma}\right]\label{eq: 3d line element in String convention}\\
	& =k\left(d\phi^{2}+h\;d\gamma d\overline{\gamma}-fh\left(\epsilon_{+}d\gamma+\epsilon_{-}d\overline{\gamma}\right)^{2}\right)\nonumber
\end{align}
For this  $3$ dimensional   metric we note that,
\begin{equation}
	\sqrt{-g}=\frac{k^{3/2}\sqrt{fh}}{2}.\label{eq: sqrt 3d metric in string convention}
\end{equation}
With this choice of $\alpha,\beta$ obviously, one needs to change
the Dilaton, such that the action has same normalization for the Ricci
and the c.c. term
\[
\sqrt{-G}e^{-2\Phi^{(4)}}R^{(4)}=\sqrt{-g}e^{-2\Phi}\left(R+...\right)
\]
\[
\sqrt{-G}e^{-2\Phi^{(4)}}\Lambda=\sqrt{-g}e^{-2\Phi}\Lambda
\]
Acccording to (1.14) of Pope,
\[
\sqrt{-G}=e^{\sigma}\sqrt{-g}.
\]
and the unnumbered equation in the passage before (1.11),
\[
\sqrt{-G}R^{(4)}=e^{\sigma}\sqrt{-g}R.
\]
Thus we can consistently choose, $\Phi=\Phi^{(4)}-\frac{\sigma}{2}$
or,
\begin{equation}
	e^{2\Phi}=g_{s}^{2}e^{-2\phi}\sqrt{fh}=g_{s}^{2}\frac{\sqrt{fh}}{kU^{2}}.\label{eq: 3 D dilation using Soumangsu's convention for KK reduction}
\end{equation}
This coincides with eq. (4.8) of \cite{Chakraborty:2019mdf}.\\
\subsection{KK reduction of the Kalb-Ramond field} \label{KK H}
Now we perform the KK reduction of the $4$ dimensional three-form
field strength, $H$ to the $3$ dimensional three-form field strength
$\widetilde{H}$ and two-form field strength $\widetilde{F}$.
But before we do that, we note that in $4$ dimensions, the term in the action is,
\[
\sqrt{-G}e^{-2\left(\Phi^{(4)}-\Phi_{0}\right)}\left(H^{(4)}\right)^{2}.
\]
Under the current convention this term becomes,
\[
\sqrt{-G}e^{-2\left(\Phi^{(4)}-\Phi_{0}\right)}\left(H^{(4)}\right)^{2}=\left(e^{\sigma}\sqrt{-g}\right)\left(e^{-2\left(\Phi-\Phi_{0}\right)}e^{-\sigma}\right)\left(H^{(4)}\right)^{2}=\sqrt{-g}e^{-2\left(\Phi-\Phi_{0}\right)}\left(H^{(4)}\right)^{2}.
\]\\
Now, the first step in the KK reduction for the $H^2$ term as per the recipe of \cite{Pope:2000xxx} is to split up the $4$ dimensional NS-NS $B$-field potential
(\ref{eq: 4-dim B-field}) using Eq.(1.18) of \cite{Pope:2000xxx} :
\begin{align*}
	B & =-\frac{hk}{2}\;d\gamma\wedge d\overline{\gamma}+\epsilon_{+}h\sqrt{k}\;dy\wedge d\gamma-\epsilon_{-}h\sqrt{k}\;dy\wedge d\overline{\gamma}\\
	& =-\frac{hk}{2}\;d\gamma\wedge d\overline{\gamma}+\left(-\epsilon_{+}h\sqrt{k}\:d\gamma+\epsilon_{-}h\sqrt{k}\,d\overline{\gamma}\right)\wedge dy\\
	& =A_{(2)}+A_{(1)}\wedge dy
\end{align*}
where,
\begin{equation}
	A_{(2)}=-\frac{hk}{2}\;d\gamma\wedge d\overline{\gamma},\qquad A_{(1)}=-\epsilon_{+}h\sqrt{k}\;d\gamma+\epsilon_{-}h\sqrt{k}\;d\overline{\gamma}.\label{eq: dimensionally reduced potential fields-1}
\end{equation}
We have previously noted that from the Ricci sector,
\begin{equation}
	\mathcal{A}=f\sqrt{k}\left(\epsilon_{+}\;d\gamma+\epsilon_{-}\;d\overline{\gamma}\right).\label{eq: KK gauge field ricci sector}
\end{equation}
Next step is to substitute (\ref{eq: dimensionally reduced potential fields-1}),
(\ref{eq: KK gauge field-1}) in Eq. (1.21) of \cite{Pope:2000xxx} to get
the $3$-dimensional field strengths, $\widetilde{H}$ and $\widetilde{F}$:
\begin{equation}
	\widetilde{F}=dA_{(1)}=\sqrt{k}h'\left(-\epsilon_{+}dU\wedge d\gamma+\epsilon_{-}\;dU\wedge d\overline{\gamma}\right).\label{3d F KK}
\end{equation}
\begin{align}
	\widetilde{H} & =dA_{(2)}-dA_{(1)}\wedge\mathcal{A}=-\frac{h'fk}{2h}\;dU\wedge d\gamma\wedge d\overline{\gamma} = -\frac{h'fk}{2h}\;\, \tilde{\epsilon}.\label{eq: 3 dimensional 3 form field strength}
\end{align}
where $\tilde{\epsilon}$ is the $3$ dimensional Levi-Civita symbol (nontensor).
In terms of components,
\begin{equation}
	\widetilde{H}_{U\gamma\overline{\gamma}}=-\frac{h'fk}{2h}, \label{3d H components}
\end{equation}
and,
\begin{equation}
	i\widetilde{F}_{\gamma\overline{\gamma}}=0,\qquad i\widetilde{F}_{U\gamma}=-\epsilon_{+}\sqrt{k}h',\qquad i\widetilde{F}_{U\overline{\gamma}}=\epsilon_{-}\sqrt{k}h'. \label{3d F components}
\end{equation}\\
\subsection{Matching the 4d action terms with the 3d action terms} \label{DR action match}
The volume terms in the bulk action \eqref{fullaction} are
\begin{eqnarray}
	S_{4D}=\frac{1}{16\pi G_{N}^{(4)}} \int  d^4 x \sqrt{-G}e^{-2(\Phi^{(4)}-\Phi_0^{(4)})}\left( R^{(4)}+ 4G^{\mu \nu } \partial_\mu \Phi^{(4)} \partial_\nu \Phi^{(4)}-\frac{1}{12}H^2-4\Lambda\right). \label{bulk action 4d}
\end{eqnarray}
Here we demonstrate that this $4$ dimensional action is equal to the following $3$ dimensional action as a consistency check,
\begin{align}
	S_{(3D)}&=\frac{1}{16\pi G_N}\int  d^3 x \sqrt{-g} ~ e^{-2\left(\Phi-\Phi_0\right)}\Big(R-\left(\partial\sigma\right)^{2}-2\square\sigma-\frac{1}{4}e^{2\sigma}\mathcal{F}^{2}-4\Lambda \nonumber\\&\hspace{7cm}+ 4g^{\mu \nu } \partial_\mu \Phi \partial_\nu \Phi+ 4g^{\mu \nu } \partial_\mu \Phi \partial_\nu \sigma  \nonumber\\&\hspace{9cm} -\frac{1}{12}\widetilde{H}^{2} -\frac{1}{4}e^{-2\sigma}\widetilde{F}^{2}\Big). \label{bulk action 3d}
\end{align}
One might wonder why are we keeping total derivative terms like $\square \sigma$ in the $3$ dimensional action upon KK reduction since they do not contribute to the equation of motion. The reason we have to keep these terms is that these total derivative terms \emph{do not vanish on-shell} and in fact make non-trivial contributions to action complexity, including \emph{introducing new surface (GHY) counterterms}. Similar phenomenon was first pointed out in \cite{Goto:2018iay}.\\
By separately considering equality of blocks of terms in the actions, \eqref{eq: 1.24 of Pope-2}, \eqref{Dilaton kinetic term in 3d}, \eqref{3d-4d match of CC} and \eqref{4d-3d match for Ricci}, in the following subsections and upon summing over both sides of those terms, we demonstrate the equality of the two actions \eqref{bulk action 4d} and \eqref{bulk action 3d}.\\
\subsubsection{Matching the contributions of the Kalb-Ramond field sector before and after KK reduction} \label{Kalb Ramond sector match}
From \eqref{eq: 3 dimensional 3 form field strength} we obtain,
\begin{align}
	-\widetilde{H}^{2} & =H_{\lambda\mu\nu}H_{\rho\sigma\tau}g^{\lambda\rho}g^{\mu\sigma}g^{\nu\tau}\nonumber \\
	& =\left(-\frac{h'fk}{2h}\right)^{2}\underbrace{\widetilde{\epsilon}_{\lambda\mu\nu}\widetilde{\epsilon}_{\rho\sigma\tau}g^{\lambda\rho}g^{\mu\sigma}g^{\nu\tau}}_{=3!g^{-1}}\nonumber \\
	& =\left(\frac{h'fk}{2h}\right)^{2}\frac{3!}{g}.\label{eq: H-tilde squared-1}
\end{align}
and from \eqref{3d F KK} we get,
\begin{align}
	-\widetilde{F}^{2} & =i\widetilde{F}_{\mu\nu}i\widetilde{F}_{\rho\sigma}g^{\mu\rho}g^{\nu\sigma}\nonumber \\
	& =\text{tr}\left(\widetilde{F}g^{-1}\widetilde{F}g^{-1}\right)\nonumber \\
	& =\frac{8h'^{2}U^{2}\epsilon_{+}\epsilon_{-}}{hk}.\label{eq: F-tilde squared-1}
\end{align}
So the RHS of (1.24) of \cite{Pope:2000xxx} using eq. (1.23) works out in this case to be, {\small{}
	\begin{align}
		\sqrt{-g}e^{-2\Phi}\left(-\frac{1}{12}\widetilde{H}^{2}-\frac{1}{4}e^{-2\sigma}\widetilde{F}^{2}\right) & =e^{-2\Phi_{(4)}}\left(\left(\frac{h'fk}{2h}\right)^{2}\frac{e^{\sigma}}{2\sqrt{-g}}+\frac{2h'^{2}U^{2}\epsilon_{+}\epsilon_{-}}{hk}e^{-\sigma}\sqrt{-g}\right)\nonumber \\
		& =e^{-2\Phi_{(4)}}\frac{h'^{2}k^{2}f^{2}}{4h^{2}}\frac{\sqrt{\frac{h}{f}}U}{k^{3/2}\sqrt{fh}}+\frac{2e^{-2\Phi_{(4)}}h'^{2}\epsilon_{+}\epsilon_{-}U^{2}}{hk}\left(\sqrt{\frac{f}{h}}\right)\left(\frac{k^{3/2}}{2U}\sqrt{fh}\right)\nonumber \\
		& =\frac{e^{-2\Phi_{(4)}}h'^{2}f\sqrt{k}U}{4h^{2}}+\frac{e^{-2\Phi_{(4)}}h'^{2}f\sqrt{k}U\epsilon_{+}\epsilon_{-}}{h}\nonumber \\
		& =e^{-2\Phi_{(4)}}\frac{h'^{2}f\sqrt{k}U}{4h^{2}}\underbrace{\left(1+4\epsilon_{+}\epsilon_{-}h\right)}_{=h/f}\nonumber \\
		& =e^{-2\Phi_{(4)}}\underbrace{\frac{\sqrt{k}Uh'^{2}}{4h}}_{=-\sqrt{-G}H^{2}/12}\nonumber \\
		& =-\sqrt{-G}e^{-2\Phi_{(4)}}\frac{H^{2}}{12}.\label{eq: 1.24 of Pope-1}
	\end{align}
}Thus we have just verified that the LHS and RHS of Eq. (1.24) of \cite{Pope:2000xxx} are consistent
for this special string background.
From \eqref{eq: 1.24 of Pope-1} of last section and integrating out the $y$-circle leads to the equivlanece of the $H^2$ term in 4d and to the $\widetilde{H}^2, \widetilde{F}^2$ terms in 3d,
\begin{align}
	S_H=\frac{1}{16\pi G_N ^{4}}\int d^3x\, dy \sqrt{-G}e^{-2\left(\Phi_{(4)}-\Phi_0\right)}\left(-\frac{H^{2}}{12} \right) & =\frac{1}{16\pi G_N} \int d^3x \sqrt{-g}e^{-2\left(\Phi-\Phi_0\right)}\left(-\frac{1}{12}\widetilde{H}^{2}-\frac{1}{4}e^{-2\sigma}\widetilde{F}^{2}\right).\label{eq: 1.24 of Pope-2}
\end{align}
Plugging in the background fields and integrating over the (regularized) WdW patch, this contribution can be expressed as a nested integral,
\begin{equation}
	S_H=\frac{L_{X}}{4\pi G_{N}k}\int\frac{dU}{U^{3}}h^{2}(U)\int_{U}^{\frac{l_{s}}{\epsilon}}\frac{dU'}{U'}\frac{1}{\sqrt{h(U')}},\label{H-field contribution}
\end{equation}
where we have used, $\frac{h'}{h}=\frac{2h}{kU^{3}}$. $\mathcal{M}_{3}$
If we set $\epsilon_{\pm}=0$, $h=f$, then the above contribution
becomes,
\begin{equation}
	S_{H}=-\frac{L_{X}}{4\pi G_{N}k}\int_{0}^{\frac{ls}{\epsilon}}\frac{dU}{U^{3}}f^{2}(U)\int_{U}^{\frac{ls}{\epsilon}}\frac{dU'}{U'}\frac{1}{\sqrt{f(U')}}, \nonumber
\end{equation}
which is the exact same expression as Eq. (3.34) of our previous paper \cite{Chakraborty:2020fpt}
on $\mathcal{M}_{3}$ complexity.
\subsubsection{Matching the contributions of the Dilaton sector before and after KK reduction} \label{Dilaton sector match}
Since $\Phi=\Phi(U)$, the Dilaton contribution to the 4d action simplifies to
\begin{equation}
	S_{\Phi}=\frac{1}{16\pi G_{N}^{(4)}}\int_{WdW}d^{4}x\:\sqrt{-G}\:e^{-2\left(\Phi_{(4)}-\Phi_{0}\right)}\;4G^{UU}\:\partial_{U}\Phi_{(4)}\:\partial_{U}\Phi_{(4)}.\label{eq: Dilaton kinetic term in 4d}
\end{equation}
In our conventions, $\sqrt{-G}\:e^{-2\left(\Phi_{(4)}-\Phi_{0}\right)}=\sqrt{-g}\:e^{-2\left(\Phi-\Phi_{0}\right)}$
and $G^{UU}=g^{UU}$ while $\Phi_{(4)}=\Phi+\frac{\sigma}{2}$. Substituting
all this in the $4$ dimensional action (\ref{eq: Dilaton kinetic term in 4d})
and then integrating over the $y$-circle we get the desired 3d action,
\begin{align}
	S_{\Phi} & =\frac{1}{16\pi G_{N}^{(4)}}\int_{WdW}d^{4}x\:\sqrt{-G}\:e^{-2\left(\Phi_{(4)}-\Phi_{0}\right)}\;4G^{UU}\:\partial_{U}\Phi_{(4)}\:\partial_{U}\Phi_{(4)} \nonumber\\ &=\frac{1}{16\pi G_{N}}\int_{WdW}d^{3}x\:e^{-2\left(\Phi-\Phi_{0}\right)}\:\sqrt{-g}\;g^{UU}\:\left(4\partial_{U}\Phi\:\partial_{U}\Phi+4\partial_{U}\Phi\:\partial\sigma+\left(\partial_{U}\sigma\right)^{2}\right).\label{Dilaton kinetic term in 3d}
\end{align}
As a nested integral this is,
\begin{align}
	S_{\Phi} & =\frac{\lambda'^{2}k^{3}L_{X}}{2\pi G_{N}}\int_{0}^{\frac{l_{s}}{\epsilon}}dU\:\frac{U^{5}}{\left(1+\lambda'\:kU^{2}\right)^{2}}\int_{U}^{\frac{l_{s}}{\epsilon}}\frac{dU'}{U'\sqrt{h(U')}}.\label{eq: Dilaton CA contribution as a nested integral}
\end{align}
Again we can explicitly check that setting $\epsilon_{\pm}=0,$ i.e. $\lambda'=\lambda,h=f$ reproduces
the $\mathcal{M}_{3}$ dilaton action-complexity contribution Eq.
(3.28) of our previous paper \cite{Chakraborty:2020fpt}.\\
\subsubsection{Matching the contributions of the Cosmological constant term before and after KK reduction} \label{cc term}
Next consider the contribution to the action from the cosmological constant term. Here we will see again for this term the $3$ and $4$ dimensional calculations match
exactly. Our convention for the change of metric and
Dilaton under KK reduction \eqref{identity 1} implies,
\[
\sqrt{-G}\,e^{-2\left(\Phi^{(4)}-\Phi_0\right)}=\sqrt{-g}\,e^{-2\left(\Phi-\Phi_0\right)}
\]
and hence we get the desired match between cc terms in the the $4$ dimensional and $3$ dimensional actions:
\begin{equation}
	S_\Lambda = \frac{1}{16\pi G_{N}^{(4)}}\int\sqrt{-G}e^{-2\left(\Phi^{(4)}-\Phi_{0}\right)}\left(-4\Lambda\right)=\frac{1}{16\pi G_{N}}\int d^{3}x\sqrt{-g}e^{-2\left(\Phi-\Phi_{0}\right)}\left(-4\Lambda\right). \label{3d-4d match of CC}
\end{equation}
Here we have again integrated out the $y$-circle in going from the LHS to the RHS and set the 3d Newton's constant $G_N = G_N^{4}/2\pi R_y$. In the static coordinates  we get
\[
\sqrt{-g}=\frac{\sqrt{k}l_{s}}{U}f,
\]
and recall that in $3$ dimensions the dilaton factor is
\[
e^{-2\left(\Phi-\Phi_{0}\right)}=\frac{kU^{2}}{\sqrt{fh}}.
\]
Finally using,
\[
\Lambda=-\frac{1}{kl_{s}^{2}}
\]
we get the $3$ dimesional cosmological term contribution as a nested
integral
\begin{align}
	S_\Lambda &  =\frac{L_{X}k}{2\pi G_{N}}\int_{0}^{\frac{l_{s}}{\epsilon}}dU\:U\:\int_{U}^{\frac{l_{s}}{\epsilon}}\frac{dU'}{U'\sqrt{h(U')}}.\label{eq: CC term in action complexity as a nested integral}
\end{align}
$\mathcal{M}_{3}$ limit check: Again setting $\epsilon_{\pm}=0$
and $h=f$ we get,
\[
S_\Lambda=\frac{L_{X}k}{2\pi G_{N}}\int_{0}^{\frac{l_{s}}{\epsilon}}dU\:U\:\int_{U}^{\frac{l_{s}}{\epsilon}}\frac{dU'}{U'\sqrt{f(U')}},
\]
which is the same as Eq. (3.31) of our previous \cite{Chakraborty:2020fpt} paper.
\\
\subsubsection{Matching the contributions of the Ricci scalar sector before and after KK reduction} \label{Ricci sector match}
Finally consider the Ricci scalar term In $4$ dimensions. Upon KK reduction this term gets split up into action term contributions from the $3$ dimensional Ricci scalar, the KK-scalar, $\sigma$ and the KK gauge field, $\mathcal{A}$. To this end recall Equation (1.14) of \cite{Pope:2000xxx}. In our convention, i.e. $\beta=1,\alpha=0$, and for our case, $D=3$, it reduces to
\begin{equation}
	R^{(4)}= R^{(3)}-2\left(\partial\sigma\right)^{2}-2\square\sigma-\frac{1}{4}e^{2\sigma}\mathcal{F}^{2}.\label{eq: correct form of 1.13 of Pope in D=00003D4}
\end{equation}
For the string background under consideration the Ricci scalar is,
\begin{equation}
	R^{(4)}=\frac{-6+8\lambda'\,k\,U^2}{k\,l_s^2\,\left(1+\lambda'\,k\,U^2\right)^2},\label{4d Ricci scalar}
\end{equation} and the dilaton is given by,
\begin{equation}
	\Phi^{(4)} = \Phi_{0}^{(4)} +\frac{1}{2} \log \left(\frac{ h(U)}{k\, U^2}\right)~.
\end{equation}
On the other hand, for the 3 dimensional background, the Ricci scalar works out to be
\begin{equation}
	R=\frac{8\lambda k^{3}\left(\lambda^{2}+8\epsilon_{+}^{2}\epsilon_{-}^{2}-6\lambda\epsilon_{+}\epsilon_{-}\right)U^{6}-2k^{2}\left(-5\lambda^{2}+16\epsilon_{+}^{2}\epsilon_{-}^{2}+20\lambda\epsilon_{+}\epsilon_{-}\right)U^{4}+4k(2\epsilon_{+}\epsilon_{-}-\lambda)U^{2}-6}{k\left(\lambda kU^{2}+1\right)^{2}\left((\lambda-4\epsilon_{+}\epsilon_{-})kU^{2}+1\right)^{2}} \label{3d Ricci Scalar}
\end{equation}
while the  $3$ dimensional dilaton factor is,
\begin{equation}
	e^{-2\left(\Phi-\Phi_{0}\right)}=\sqrt{\lambda kU^{2}+1}\sqrt{\left(\lambda-4\epsilon_{+}\epsilon_{-}\right)kU^{2}+1}, \label{3d dilaton factor}
\end{equation}
the KK-gauge field strength contribution,
\begin{equation}
	\mathcal{F}^2 = \frac{64 k U^4 \epsilon_{+}^2 \epsilon_{-}^2}{\left(  \lambda k U^2+1\right)^4}, \label{KK field strength squared}
\end{equation}
and the KK scalar $\sigma$ contributions,
\begin{eqnarray}
	\left( \partial \sigma \right)^2 = \frac{16 k  U^4 \epsilon_{+}^2 \epsilon_{-}^2}{\left( \lambda k  U^2+1\right)^2 \left((\lambda -4 \epsilon_{+} \epsilon_{-}) k U^2+1\right)^2} \label{del sigma squared}\\
	\square \sigma = \frac{8 U^2 \epsilon_+ \epsilon_{-} \left( k^2 \lambda  U^4 (4 \epsilon_+ \epsilon_{-}-\lambda )+k U^2 (\lambda -2 \epsilon_+ \epsilon_{-})+2\right)}{\left(\lambda k  U^2+1\right)^2 \left((\lambda -4 \epsilon_+ \epsilon_{-})\,k  U^2 +1\right)^2} \label{box sigma}
\end{eqnarray}
From \eqref{4d Ricci scalar}-\eqref{box sigma} one can explicity verify \eqref{eq: correct form of 1.13 of Pope in D=00003D4}. Now that we have checked \eqref{eq: correct form of 1.13 of Pope in D=00003D4}, it is easy to verify (in light of \eqref{identity 1}) that,
\\
\begin{equation}
	S_R = \frac{1}{16\pi G_N^{(4)}} \int d^4 x\sqrt{-G}\,e^{-2\left(\Phi^{4}-\Phi_0\right)} R^{(4)} = \frac{1}{16\pi G_N} \int d^3 x \sqrt{-g}\,e^{-2\left(\Phi-\Phi_0\right)} \left( R^{(3)}-2\left(\partial\sigma\right)^{2}-2\square\sigma-\frac{1}{4}e^{2\sigma}\mathcal{F}^{2} \right) \label{4d-3d match for Ricci}
\end{equation}
after integrating out the $y$-circle. Restricting the integral over the $3$ dimensional WdW patch, this contribution can be expressed as the nested integral,
\begin{equation}
	S_R = \frac{k\,L_X}{4\pi G_N} \int_0^{\frac{l_s}{\epsilon}} dU \,U\,\frac{-3+4\lambda' k U^2}{\left(1+\lambda' k U^2\right)^2} \int_{U}^{\frac{l_s}{\epsilon}} \frac{dU'}{U\sqrt{h(U')}}. \label{Ricci sector contribution to action complexity as a nested integral}
\end{equation}
$\mathcal{M}_{3}$ limit check: On setting $\epsilon_{\pm}=0$
and $h=f, \lambda'=\lambda$, we get,
\[
S_R = \frac{k\,L_X}{4\pi G_N} \int_0^{\frac{l_s}{\epsilon}} dU \,U\,\frac{-3+4\lambda k U^2}{\left(1+\lambda k U^2\right)^2} \int_{U}^{\frac{l_s}{\epsilon}} \frac{dU'}{U\sqrt{f(U')}}
\]
which is the same as Eq. (3.25) of our previous \cite{Chakraborty:2020fpt} paper.\\
Full nonperturbative result for the Ricci sector:
\begin{align}
	S_{R}=\frac{L_{X}}{4\pi G_{N}\sqrt{\lambda'}}\left[-\frac{\pi^{2}}{6}-\frac{7}{2\sqrt{1+\frac{\epsilon^{2}}{\lambda'\:kl_{S}^{2}}}}+\frac{7}{2}\sqrt{\frac{\epsilon^{2}}{\lambda'\:kl_{s}^{2}}+1}-\frac{7\frac{\epsilon^{2}}{\lambda'\:kl_{s}^{2}}}{2\sqrt{1+\frac{\epsilon^{2}}{\lambda'\:kl_{S}^{2}}}}-2\sqrt{1+\frac{\epsilon^{2}}{\lambda'\:kl_{S}^{2}}}\ln\left(1+\frac{\lambda'\:k\:l_{s}^{2}}{\epsilon^{2}}\right)\right.\nonumber \\
	-\frac{7}{2}\sinh^{-1}\left(\frac{\sqrt{\lambda'\:k}\:l_{s}}{\epsilon}\right)+4\ln\left(\sqrt{1+\frac{\lambda'\:k\:l_{s}^{2}}{\epsilon^{2}}}+\frac{\sqrt{\lambda'\:k}\:l_{s}}{\epsilon}\right)+2\ln\left(1+\frac{\lambda'\:k\:l_{s}^{2}}{\epsilon^{2}}\right)\sinh^{-1}\left(\frac{\sqrt{\lambda'\:k}\:l_{s}}{\epsilon}\right)\nonumber \\
	+2\left(\sinh^{-1}\left(\frac{\sqrt{\lambda'\:k}\:l_{s}}{\epsilon}\right)\right)^{2}-4\ln\left(1+\left(\sqrt{1+\frac{\lambda'\:k\:l_{s}^{2}}{\epsilon^{2}}}+\frac{\sqrt{\lambda'\:k}\:l_{s}}{\epsilon}\right)^{2}\right)\sinh^{-1}\left(\frac{\sqrt{\lambda'\:k}\:l_{s}}{\epsilon}\right)\nonumber \\
	\qquad\qquad\qquad\qquad\qquad\qquad\qquad\qquad\qquad\qquad\left.-2\text{Li}_{2}\left(-\left(\sqrt{1+\frac{\lambda'\:k\:l_{s}^{2}}{\epsilon^{2}}}+\frac{\sqrt{\lambda'\:k}\:l_{s}}{\epsilon}\right)^{2}\right)\right].\label{eq: Ricci sector contribution to action complexity for JTbar}
\end{align}\\
\section{GHY type surface terms in 3 dimensions}\label{section:GHY}
Since the $3$ dimensional action has the a second derivative term
from the KK scalar $\sigma$, one will need a boundary term to cancel
its variation. Here we work out that term,{\small{}
	\begin{align*}
		-\frac{1}{8\pi G_{N}}\int_{M}\sqrt{-g}e^{-2\left(\Phi-\Phi_{0}\right)}\square\left(\delta\sigma\right) & =-\frac{1}{8\pi G_{N}}\int_{M}\sqrt{-g}\nabla^{\mu}\left[e^{-2\left(\Phi-\Phi_{0}\right)}\nabla_{\mu}\left(\delta\sigma\right)\right]+\frac{1}{8\pi G_{N}}\int\sqrt{-g}\nabla^{\mu}e^{-2\left(\Phi-\Phi_{0}\right)}\nabla_{\mu}\left(\delta\sigma\right)\\
		& =-\frac{1}{8\pi G_{N}}\int_{\partial M}\sqrt{-\gamma}\:n^{\mu}\left[e^{-2\left(\Phi-\Phi_{0}\right)}\delta\left(\nabla_{\mu}\sigma\right)\right]-\ldots\\
		& =-\frac{1}{8\pi G_{N}}\int_{\partial M}\delta\left(\sqrt{-\gamma}\:n^{\mu}e^{-2\left(\Phi-\Phi_{0}\right)}\nabla_{\mu}\sigma\right)-\ldots.
	\end{align*}
}This first term can be canceled if we add the surface counter term,
\[
I_{GHY;\sigma}=\frac{1}{8\pi G_{N}}\int_{\partial M}\sqrt{-\gamma}\:e^{-2\left(\Phi-\Phi_{0}\right)}\:n^{\mu}\partial_{\mu}\sigma.
\]
In addition we also have the usual GHY term for the string frame metric variation to be well defined,
\[
I_{GHY;g}=\frac{1}{8\pi G_{N}}\int_{\partial M}\sqrt{-\gamma}e^{-2\left(\Phi-\Phi_{0}\right)}K.
\]
Thus the full $GHY$ term is,
\begin{equation}
	I_{GHY}=\frac{1}{8\pi G_{N}}\int_{\partial M}\sqrt{-\gamma}e^{-2\left(\Phi-\Phi_{0}\right)}\left(K+n^{\mu}\partial_{\mu}\sigma\right).\label{eq: Full GHY term after dimensional reduction}
\end{equation}
\\
\section{Holographic Entanglement Entropy}\label{HEE}
The holographic entanglement entropy of WCFT dual to null warped AdS$_3$ (following the prescription for nontrivail dilaton turned on in the bulk \cite{Klebanov:2007ws}) is,
\begin{align}
	S_{A}&=\frac{1}{4G_N}\int e^{-2(\Phi(U)-\Phi_{\infty})} dx \sqrt{\gamma}\\
	&=\frac{1}{4G_N}\int_{U_{0}}^{\frac{l_s}{\epsilon}}dx \sqrt{\frac{kl_s^2}{U^2}\frac{(U^4({1-kU^2\epsilon_-^2})({U^2-V^2-k\epsilon_-^2(U^4-U_{0}^4)})}{U_{0}^2l_s^2(1-kU_{0}^2\epsilon_-^2)}+kU^2(1-kU^2\epsilon_-^2)}
\end{align}
After simplifying and replacing $dx$ as $\frac{dU}{U^{'}}$ the equation for our metric becomes,
\begin{align}
	S_{A}&=\frac{1}{4G_N}\sqrt{k}l_s\int_{U_{0}}^{\frac{l_s}{\epsilon}} dU \frac{\sqrt{1-kU^2\epsilon_-^2}}{\sqrt{U^2-U_{0}^2-k\epsilon_-^2(U^4-U_{0}^4)}}\label{d.3}
\end{align}
\\
For, Warped AdS, after expanding the integrand in equation \ref{d.3} in a Taylor series with respect to $\epsilon_-$, we get the entanglement entropy to be,
\begin{align}
	S_{A}
	&=\frac{\sqrt{k}l_s}{4G_N}\bigg(1+\frac{6kl_s^2}{L^2}\epsilon_-^2\bigg)\ln{\left(\frac{L}{\epsilon}\right)}+O(\epsilon_-^4)
\end{align}
Here, we have used Eq. (\ref{6.4}) to replace $U_{0}$ as a function of $L$. Also, we can see that, putting warping parameter, $\epsilon_-\to 0$ gets us the result back for pure AdS which is, $ S_{A}=\frac{\sqrt{k}l_s}{4G_N}\ln{\left(\frac{L}{\epsilon}\right)}$. This is also can be seen in \cite{Castro:2015csg} from the comparison of equation (3.23) and (3.27).
\\
Through numerical integration of eqn \ref{d.3}, we have found the nature of the Holographic Entanglement Entropy as a function of $L$ for Warped $AdS_3$. We have used parametric plot here and used eqn \ref{6.14} for the expression of $L$.
\begin{figure}[h!]
	\centering
	\includegraphics[width=0.8\textwidth]{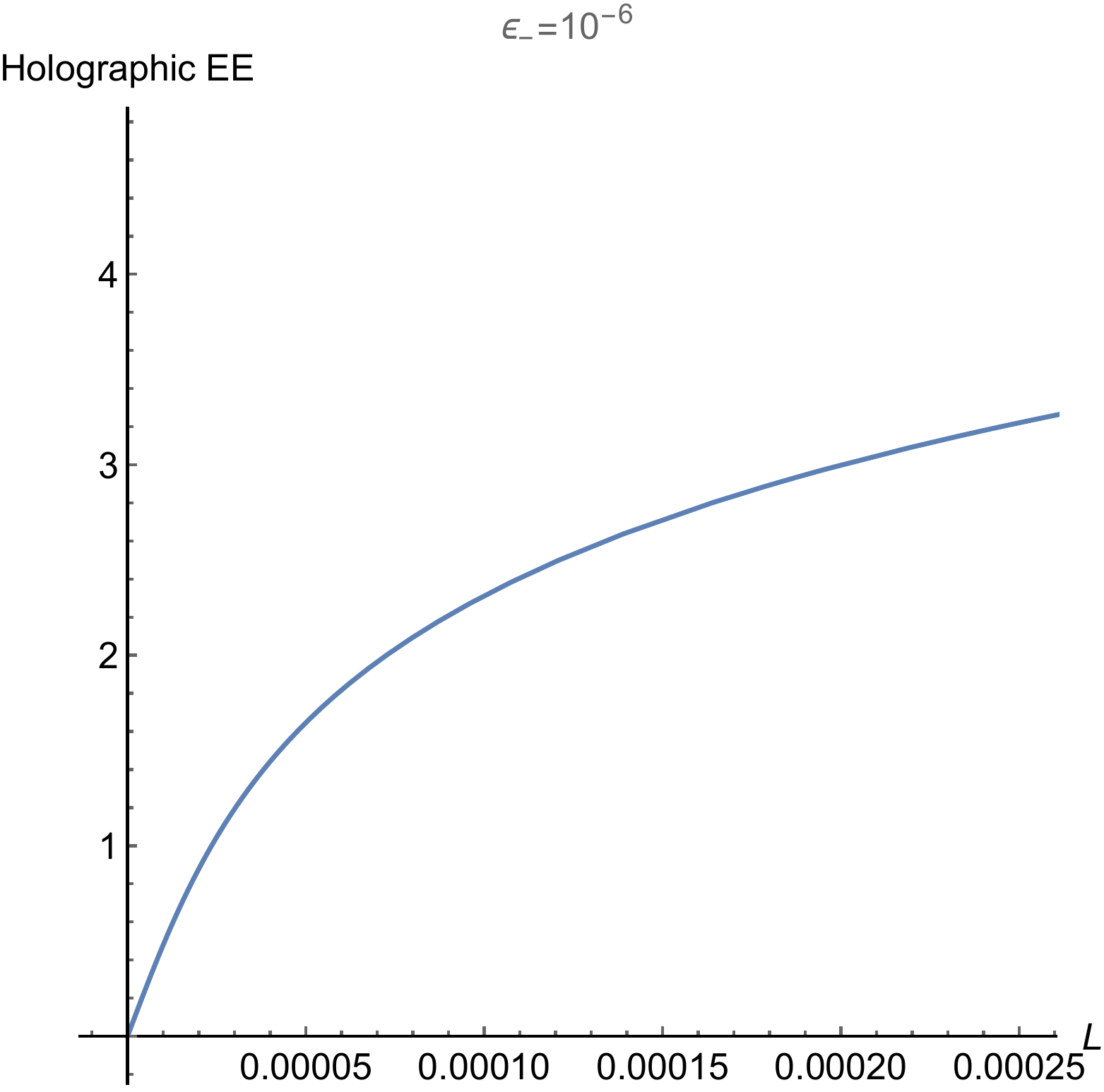}
	\caption{Holographic Entanglement Entropy vs L plot. The values used here are following, $k=10^4$, $l_s=0.01$ and $\epsilon=10^{-5}$}
\end{figure}
Here again, we used the value of the warping factor as, $\epsilon_-=10^{-6}$ for the same reason mentioned at the end of Sec. \ref{s6}.
\\
One curious fact to note is that the holographic entanglement entropy doesn't display any on-locality in terms of the UV divergences appearing - for any value of the warping the only type of UV divergence appearing is the log divergence, much akin to a local field theory like CFT$_2$. This is polar opposite of the pattern of UV divergences appearing in subregion volume complexity (all orders of UV divergences appear there). The fact that the entanglement entropy of a WCFT$_2$, a highly nonlocal and Lorentz boost violating theory, but has the exact same UV divergence structure as that of the entanglement entropy of a local CFT$_2$ has been noted in earlier works \cite{Castro:2015csg, Basanisi:2016hsh}.
\bibliography{ref}\bibliographystyle{JHEP}
\end{document}